\documentclass[hyper(*)]{JHEP3}

\usepackage{epsfig,multicol,bbm}
\usepackage{enumerate}
\usepackage{bibentry}
\usepackage{amsmath}
\title{Higher laminations, webs and $\mathcal{N}=2$ line operators}

\author{Dan Xie

\\ School of Natural Sciences, Institute for Advanced Study \\
Princeton, NJ 08540, USA}

\abstract{A detailed study of  half-BPS line operators of higher rank 4d $\mathcal{N}=2$ theory engineered  from six dimensional
$A_{N-1}$ $(2,0)$ theory on  a bordered Riemann surface with full marked points  is performed. Geometrically, each 4d UV line operator
is represented by an irreducible bipartite web formed by three junctions on Riemann surface, and such web structure is called
higher lamination. Algebraically,
the space of UV  line operators is identified with the integral tropical $a$ coordinates of 
the corresponding $\mbox{PGL}(\text{N,C})$ local system, and the space of IR line operator is identified with the cluster $X$ coordinates of  $\text{SL}(\text{N,C})$ local system.
The expectation value of UV line operator at Coulomb branch parameterized by $X$ coordinates is calculated, and the result is a positive Laurent polynomial in $X$. 
Using the expectation values, we calculate the operator product expansion (OPE) between the line operators, which is then represented geometrically
by  higher rank Skein relations. We also calculate the Poisson brackets of these line operators, and Frenchel-Nielson type coordinates are constructed for  Higher Teichmuller space, etc. }

\begin{document}
\maketitle  

\newpage
\section{Introduction}
The line or loop operators of quantum field theory play a very important role in probing the dynamics of the theory: the expectation value of 
the Wilson or 't Hooft loop of 4d gauge theory can be used to determine the phases of theory; The loop equations
provide an important clue to the discovery of dual string theory \cite{Polyakov:1997tj}, and Wilson loops are very important observables in the study of AdS/CFT correspondence \cite{Maldacena:1998im,Rey:1998ik}; 
The expectation value of certain special Wilson loop is related to the scattering amplitude in 4d $\mathcal{N}=4$ super Yang-Mills theory \cite{Alday:2007hr,CaronHuot:2010ek}, etc.

We are going to study half-BPS line operators of a large class of 4d $\mathcal{N}=2$ quantum field theories engineered using 6d
$A_{N-1}$ $(2,0)$ theory on a Riemann surface $\Sigma$ with various defects \cite{Gaiotto:2009we,Gaiotto:2009hg,Xie:2012hs}.  
A lot of  studies have been done for $A_1$ theory \footnote{By $A_{N-1}$ theory we mean the 4d theories engineered using six dimensional $A_{N-1}$ (2,0) theory.} \cite{Drukker:2009tz,Gaiotto:2010be}, and they have 
a lot of interesting applications:
\begin{itemize}
\item They are important observables of 4d theory.
\item A basis of them gives a nice maximal set of commuting Hamiltonian of the Seiberg-Witten (Hitchin) integrable system, which is 
very useful for the quantization \cite{Nekrasov:2009rc,Nekrasov:2011bc}.
\item They form a very interesting (quantum) algebra, and the Nekrasov instanton partition function is related to the quantization of this algebra \cite{Nekrasov:2010ka,Ito:2011ea,Teschner:2010je,Vartanov:2013ima}. 
\item The framed BPS states can be defined and calculated \cite{Gaiotto:2010be}.
\end{itemize}

The 4d line operators of $A_1$ theory are derived from wrapping half-BPS surface operators of 6d theory\footnote{Six dimensional $A_{N-1}$ $(2,0)$ theory 
are the low energy effective theory of N M5 brane, and the basic surface operator of $(2,0)$ theory comes from M2 brane ending on  M5 branes \cite{Strominger:1995ac,Berenstein:1998ij,Graham:1999pm}.} on
one cycle of Riemann surface, and  the classification should be related to the classification of cycles on $\Sigma$. This picture is confirmed for Lagrangian $A_1$ theory in \cite{Drukker:2009tz}, which shows 
that 4d line operator is classified by non-intersecting closed curves on $\Sigma$. Interestingly, such set of closed curves have been studied in great detail in mathematical
literature and called laminations \cite{fock2005dual}.  Using lamination language, one can extend the classification 
 to non-Lagrangian  $A_1$ theories \cite{Gaiotto:2010be}.

This paper serves as an attempt to do a systematic study about line operators of higher rank theory defined using  Riemann surface $\Sigma$ with full defects (regular and irregular). 
Naively, the generalization is straightforward: one wrap various 6d half-BPS  surface operators on closed curves to get 
4d line operators. This picture seems perfectly right if we only consider the usual Wilson lines built from  weakly coupled gauge group, since 
there is indeed 6d surface operator labeled by arbitrary irreducible representation of $\text{SU}(N)$ gauge group \cite{Lunin:2007ab, Chen:2007ir, Hoker:2007xz}. So the classification using 
6d picture agrees with the result using 
field theory analysis \cite{Kapustin:2005py}.

However, this can not be the whole story for the following two reasons: firstly, unlike $A_1$ theory, there is strongly coupled 
matter system which carries non-trivial
Coulomb branch operators, and there should be other non-trivial line operators charged on them as  
the usual Wilson loop is not suitable; Secondly, it is shown in \cite{Drukker:2010jp} that 
webs on $\Sigma$ can also appear as the 4d line operators using  2d-4d correspondence \cite{Alday:2009aq}.

We will confirm in this paper that the webs are necessary for the classification of  4d line operators. The important 
tool is the operator product expansion (OPE) between ordinary Wilson and 't Hooft line operator, since new line operators must appear
in the OPE if they exist. Remarkably, those new line operators can be represented by webs formed by simple three junctions, and
a complete classification of these three junctions will be given. Using these webs and the cluster algebra tools, we are going to make following main conjectures\footnote{These conjectures are confirmed for some explicit examples.}:
\begin{itemize}
\item The space of 4d UV line operators is identified with the integral tropical $a$ coordinates of the moduli space of decorated $\mbox{PGL}(N,C)$ local system on $\Sigma$;
The cluster $X$ coordinates \footnote{The most important thing about the  definition of cluster coordinate is a quiver, which is actually the BPS quiver.
The tropical $a$ coordinate is just a set of discrete numbers defined on each quiver node, while the cluster
$X$ coordinate is a set of complex number defined on each node.}
 of the framed $\mbox{SL}(N,C)$ local system are identified with the IR line operators, which also parameterize the Coulomb branch.
This is a local description as there are a lot of cluster coordinates systems related by quiver mutations.
\item The expectation value $\text{I}({\cal L})$ of a UV line operator ${\cal L}$ on  Coulomb branch is calculated and expressed as a positive Laurent polynomial
in cluster $X$ coordinate. The leading order term for a UV line operator with  coordinates $(a_1,\ldots, a_n)$ is
\begin{equation}
\text{I}_{A}({\cal L})=\prod X_i^{a_i}+\ldots.
\end{equation}
Using the canonical map, one can define OPE between line operators:
\begin{equation}
\text{I}({\cal L}_1)*\text{I}({\cal L}_2)=\sum_{{\cal L}}\text{C}_{{\cal L}_1 {\cal L}_2}^{\cal L} \mbox{I}({\cal L}),
\end{equation}
where the sum is finite and $\text{C}_{{\cal L}_1 {\cal L}_2}^{\cal L} $ is a positive integer, and $\text{C}_{{\cal L}_1 {\cal L}_2}^{{\cal L}_1+{\cal L}_2} =1$.
\item We find higher rank Skein relations  which give the OPEs in a geometric way.
\item Using the Skein relations, any 4d half-BPS line operator can be represented by an irreducible bipartite 
web on $\Sigma$, and those web structures are called higher laminations. This is a global description, and the space of higher laminations 
is in one-to-one correspondence with the integral tropical $a$ coordinates described in  first statement. 
\end{itemize}

\begin{figure}[htbp]
\small
\centering
\includegraphics[width=8cm]{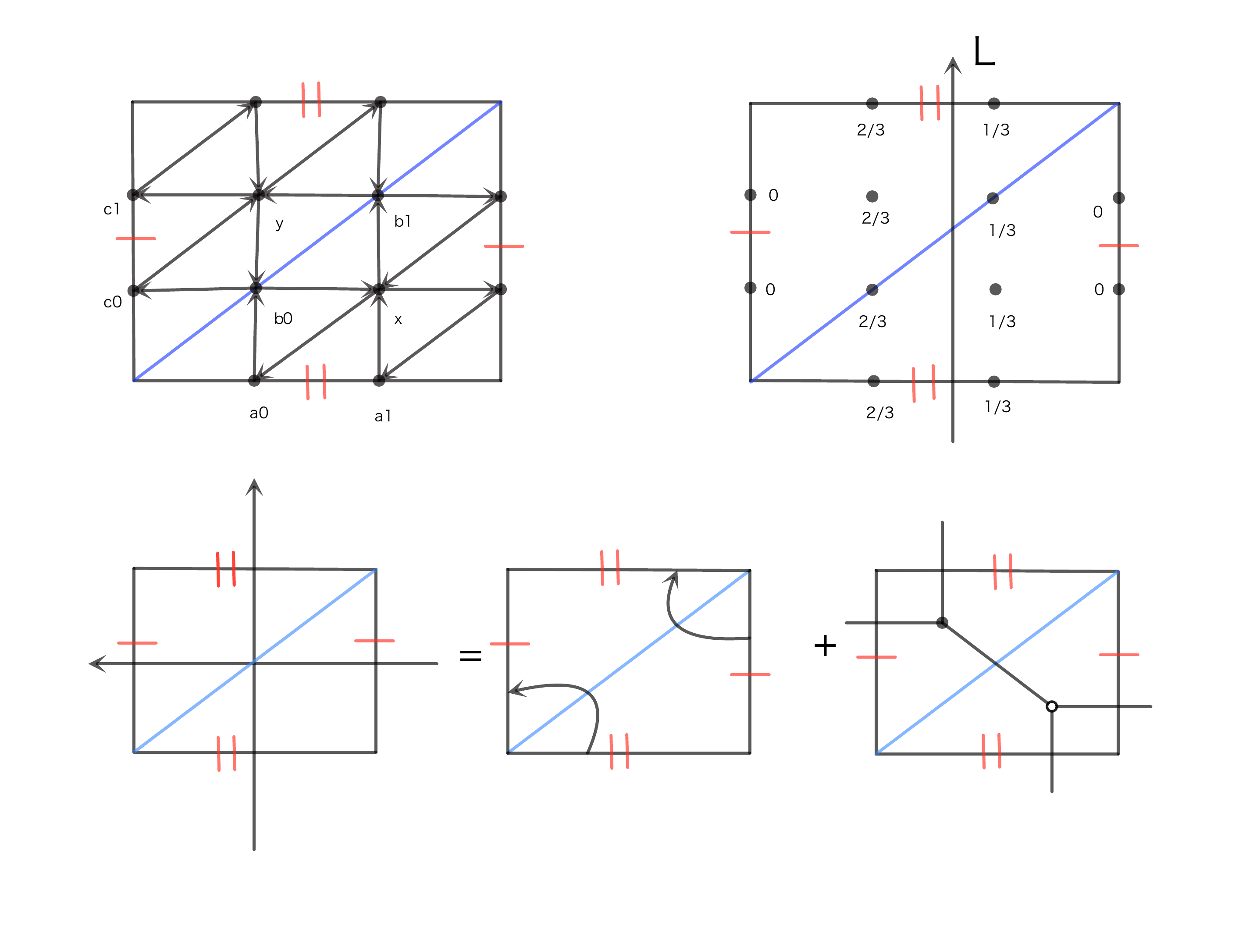}
\caption{Top left: the BPS quiver for $A_2$ theory on once punctured torus, here the quiver nodes on the edges labeled by same red lines are identified. There
is a cluster $X$ variable on each quiver node. Top right:
the tropical $a$ coordinate of a line operator $L$. Bottom: the OPE between Wilson  and 't Hooft line, and  
a bipartite web must appear in the OPE.}
\label{intro}
\end{figure}

Let me explain in some detail what each statement means. As we discussed earlier, the space of UV line operators in the $A_1$ theory is 
identified with the lamination space. On the other hand, it is shown in \cite{fock2005dual} that the lamination space is in one to one correspondence
with the tropical $a$ coordinates of $\mbox{PGL}(2,C)$ local system defined on $\Sigma$. The cluster $X$ coordinate is 
related to the one cycle on  Seiberg-Witten curve, which is then naturally identified with the IR line operator \cite{Gaiotto:2010be}. 
Remarkably, the cluster coordinates (a quiver) for the higher rank theory defined using full defects are found in a seminal paper by Fock and Goncharov \cite{fock-2003}, i.e. 
the quiver for  $A_2$ theory on once punctured torus is shown in figure.~\ref{intro}. Based on the analogy with $A_1$ theory,
it is straightforward to identify the space of UV line operator  as the  tropical $a$ coordinate of the corresponding quiver. 
Here we make the definition much more precise: they actually take values $k/N$ with $k$ integer 
and also satisfy some special constraints (see (\ref{con1}) and (\ref{con2}) for the detailed constraints of $A_2$ theory.). 

The second statement is established by studying line operator represented by closed curve on $\Sigma$: 
one can calculate its monodromy of  $\mbox{SL}(N,C)$ local system using very simple combinatorial method \cite{fock-2003}, 
and the answer is always a positive Laurent polynomial\footnote{We call them Laurent polynomial even though the exponents are fractional.} in cluster $X$ coordinates, 
which  is the expectation value of UV line operator as a direct generalization of what is proposed for $A_1$ case \cite{Gaiotto:2010be} (we also call it canonical map following \cite{fock-2003}.). For the closed curve $\text{L}$ shown in figure. \ref{intro}, the answer is 
\begin{align}
&\mbox{I}(\mbox{L})= {1\over{x^{2/3} y^{1/3} a_0^{1/3} a_1^{2/3} b_0^{1/3} b_1^{2/3}}}+{{b_1^{1/3}}\over{x^{2/3} y^{1/3} a_0^{1/3} a_1^{2/3} b_0^{1/3}}}+\frac{x^{1/3} b_1^{1/3}}{y^{1/3} a_0^{1/3} a_1^{2/3} b_0^{1/3}} 
+\frac{x^{1/3} a_1^{1/3} b_1^{1/3}}{y^{1/3} a_0^{1/3} b_0^{1/3}}+\nonumber\\
&\frac{x^{1/3} b_0^{2/3} b_1^{1/3}}{y^{1/3} a_0^{1/3} a_1^{2/3}}+\frac{x^{1/3} a_1^{1/3} b_0^{2/3} b_1^{1/3}}{y^{1/3} a_0^{1/3}}
+\frac{x^{1/3} y^{2/3} a_1^{1/3} b_0^{2/3} b_1^{1/3}}{a_0^{1/3}}+x^{1/3} y^{2/3} a_0^{2/3} a_1^{1/3} b_0^{2/3} b_1^{1/3}.
\end{align}
These canonical maps
have two remarkable applications: firstly the tropical $a$ coordinates of this line operator can be read from the exponent of leading order term, see figure.~\ref{intro};
Secondly it can be used to define the OPE between  line operators: we simply multiply  canonical maps of two line operators and 
the answer should be expanded into the sum of canonical map of other line operators.
Using the OPE, we find  that the usual Wilson loops are classified by the irreducible representation of the $\text{SU}(N)$ group, and the OPE between
them has the same support as the corresponding tensor product of irreducible representations. This strongly suggests that 
our results are correct.

\begin{figure}[htbp]
\small
\centering
\includegraphics[width=10cm]{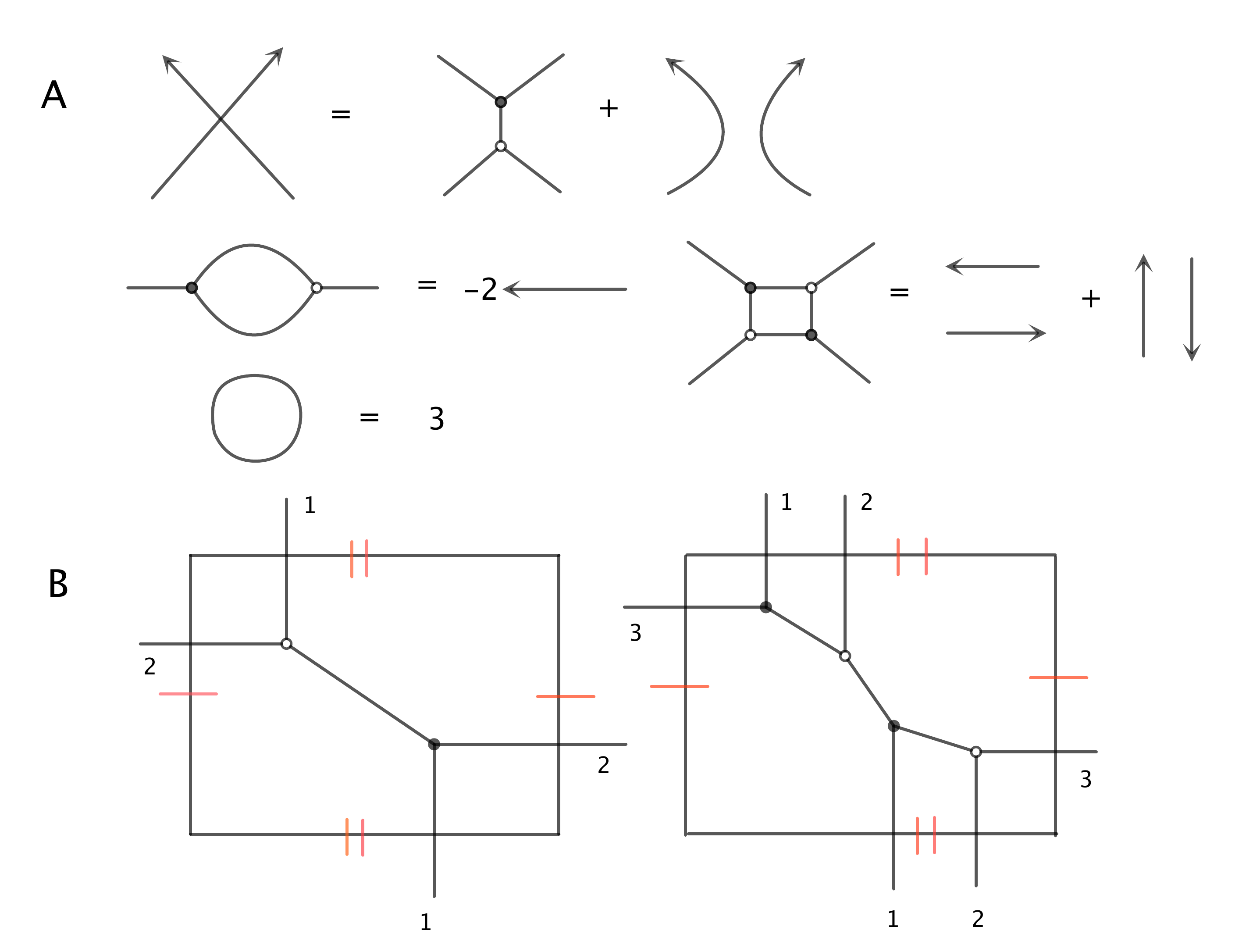}
\caption{A: Skein relations for $A_2$ theory. B. Bipartite webs on once punctured torus of $A_2$ theory, and edges of three junctions with same label are connected.}
\label{intro1}
\end{figure}

Most interestingly, by doing  OPE  between the usual Wilson and 't Hooft line operators, 
we find extra line operator which  is naturally described as a web formed by two three junctions, see figure. \ref{intro}. 
We classify all the three junctions of any higher rank theory, and  any 4d half-BPS line operator
can be represented by the webs formed by three junctions. These
web structures are called higher laminations in  direct analogy with the $A_1$ case;
Some webs on torus are shown in figure. \ref{intro1}. This description is independent of the cluster coordinates system and is the geometric picture for  4d line operators of higher rank theory.
In general, the OPE between any line operators can be found geometrically using local Skein relations, see
figure. \ref{intro1} for  $A_2$ theory. Such Skein relations are also important for the classification of webs: we only consider irreducible webs which can not be reduced by  Skein relations.

There are some compelling evidences supporting our results:  Poisson brackets for the line operators of $A_2$ theory on  once punctured torus are calculated, and 
the answer is in perfect agreement with the underlying integrable system: there are three commuting Hamiltonians, etc. There are 
theories which can be engineered either using  $A_1$  or $A_2$ theory, and the canonical map of  same line operator agrees with each other perfectly!

This paper is organized as follows: in section II, we review some basic facts of line operators of $A_1$ theory,
and we emphasize a coherent picture of geometric and algebraic descriptions. In section III, 
we carry out a detailed study of  line operators for theories defined on once punctured torus, and find 
the basic three junctions, Skein relations, etc. Section IV gives a brief description for line 
operators of theories defined on general Riemann surface. Finally, a conclusion is given in section V. Technical details about
how to calculate monodromy is given in appendix A, and we collect some useful canonical maps in appendix B.

\newpage
\section{ $A_1$ theory: line operator and lamination space}
This section is a review about the line operators for $A_1$ theory. The geometric picture for the classification is the so-called lamination, and the algebraic picture is the tropical cluster algebra \cite{fock2005dual}.
Physically, the geometric picture for the Lagrangian $A_1$ theory is  discovered in \cite{Drukker:2009tz} using field theory , and then extended to the non-Lagrangian theory using lamination in \cite{Gaiotto:2010be}. We are also going to 
discuss the canonical map, OPE, etc, and all of which will be generalized to the higher rank case in the following sections.

Let's first say a couple of words about the four dimensional $\mathcal{N}=2$ theory we have: we start with a six dimensional $A_1$ $(2,0)$ theory and compactify it on a Riemann surface 
 $M_{g,b_i,p_j}$, here $b_i$ denotes $i$th irregular singularity and $p_j$ denotes the $j$th regular singularity
\footnote{The irregular singularity means that the Higgs field of the Hitchin equation has higher order pole, while the 
regular singularity means  first order pole.}.
The IR limit is a four dimensional UV complete theory (conformal or asymptotical free theory).  The UV deformations are encoded in the geometric description: the dimensionless
gauge coupling is identified with the complex structure moduli of the Riemann surface; the mass parameters are the coefficients of the first order pole of regular and irregular
singularity; the dimensional couplings are the coefficients of the higher order pole of the irregular singularity. The IR description can also be easily found from this geometric 
description: a Hitchin equation with specific boundary conditions at the singularity is defined on $M_{g,b_i,p_j}$,
 and the spectral curve of the Hitchin moduli space is identified with the Seiberg-Witten curve of 4d theory.

\begin{figure}[htbp]
\small
\centering
\includegraphics[width=12cm]{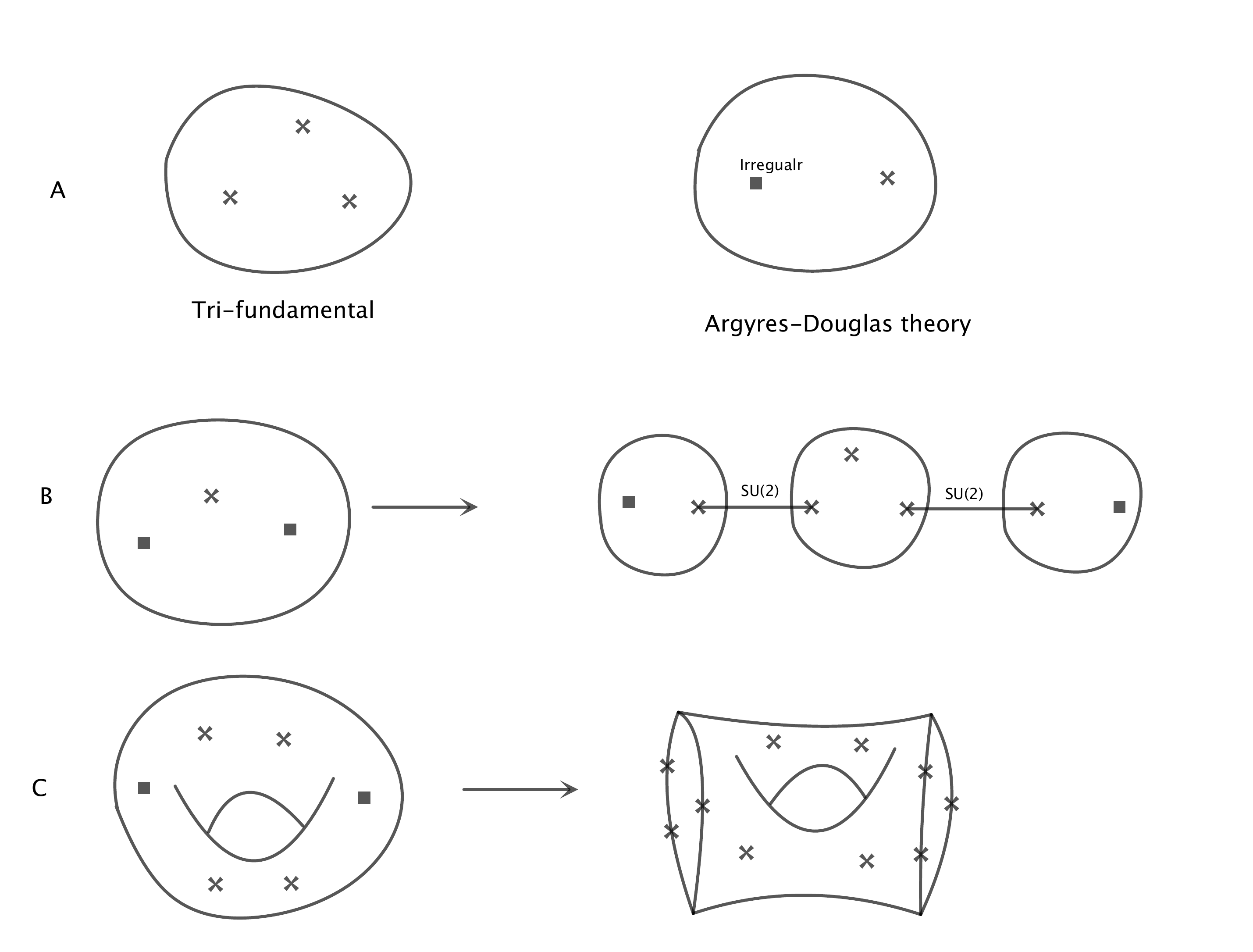}
\caption{A: The two basic matter systems for four dimensional theory built from six dimensional $A_1$ theory: the tri-fundamental represented by a sphere
with three regular singularity and the D type Argyres-Douglas theory represented by a sphere with one irregular and one regular singularity. B: In one weakly coupled gauge group duality frame,
the theory is formed by gluing above two pieces together: physically this is achieved by gauging the diagonal flavor symmetries. C: We replace each irregular singularity with a boundary with marked points.   }
\label{a1}
\end{figure}

The weakly coupled gauge group description \footnote{For the sphere with one irregular singularity or one irregular and one regular singularity, one 
has strongly coupled Argyres-Douglas theory.} is described by 
degeneration limit of the Riemann surface: the long tube in the degeneration limit is associated with the gauge group and other components are the matter system . 
The basic building blocks (matter contents) are:
\begin{itemize}
\item Tri-fundamental represented by  a  sphere with three regular singularities.
\item $\mbox{D}_N$ type\footnote{It is called $\mbox{D}_N$ type theory since $D_N$ quiver appears in the mutation class of its BPS quiver.}  Argyres-Douglas theory represented by an sphere 
with one irregular singularity and one regular singularity .
\end{itemize}
The first piece is free while the second one is a strongly coupled isolated SCFT, see figure. \ref{a1}A.  Each regular singularity carries 
a $\text{SU}(2)$ flavor symmetry, and the weakly coupled gauge group description of the four dimensional theory is derived by gauging these 
$\text{SU}(2)$ flavor symmetries, see figure. \ref{a1}B. Different duality frames correspond to different decompositions of the full
Riemann surface into the above two pieces, which is a remarkably simple description of S duality \cite{Gaiotto:2009we}.

For various applications below, we need to replace an irregular singularity with a boundary plus marked points whose number depends on the type of singularity,
which can be seen by studying the Stokes phenomenon around the irregular singularity, see \cite{Xie:2012jd}.  Therefore, we are 
going to consider a bordered Riemann surface
$\mbox{M}_{g,b_i,n}$, here $g$ is the genus of the Riemann surface, $b_i$ is the number of marked points on 
the $i$th boundary, $n$ is the number of bulk marked points, see figure. \ref{a1}C. The Coulomb branch dimension of the 4d theory is 
\begin{equation}
n_r=n+\sum_i[{b_i-3\over2}]+3g-3,
\end{equation}
where $[{b_i-3\over 2}]$ means taking the integer part of the number inside the bracket.

Each regular puncture contributes a mass parameter (as it carries a $SU(2)$ flavor symmetry.), and each boundary with even number of 
marked points also contributes a mass parameter (since it carries a $U(1)$ flavor symmetry.), and the rank of the charge 
lattice is 
\begin{equation}
R=2n_r+n_f=3n+\sum_i b_i +6g-6.
\end{equation}
We also pick an anti-clockwise orientation for each boundary (the bulk puncture can be thought of as 
a boundary without any marked point.). A bordered Riemann surface with anti-clockwise orientation 
for each boundary is the geometric avatar for our study of line operators.

\newpage
\subsection{UV and IR line operator: geometric picture}
The 4d theory always has a Lagrangian description for bordered Riemann surface $M_{g,0,n}$, so
one can use  field theory tools to study its half-BPS line operator.
For a four dimensional $\mathcal{N} = 2$ gauge theory with gauge group G, the Wilson operators are labeled by irreducible representations R of G and take the form
\begin{equation}
\mbox{W}_{\mbox{R}} =\mbox{Tr}_{\mbox{R}} \mbox{P}\int \exp (iA+\mbox{Re}\phi ds), 
\end{equation}
where $\phi$ is a complex scalar in the vector multiplet and ds is the line element determined by the metric. 
If the integral is taken along a straight line or a circle \footnote{We call these two types as line operators for simplicity, as the space-time shape is
not important for our later purpose.} in flat $R_4$, this Wilson line preserves a half of $\mathcal{N} = 2$ supersymmetry. Equivalently, the Wilson line operator is 
classified by the highest weight in the weight lattice $\Lambda_{G}$ of the gauge group $G$. The 't Hooft operator 
is a disorder operator and defined by specifying the singular behavior of various fields around the loop or line in the space time:
\begin{equation}
F=-{B\over 2}\sin \theta d\theta\wedge d\phi,~~~\phi\sim i {B\over 2r},
\end{equation} 
and the line operator is classified by the highest weight in the coweight lattice in $\Lambda_{\hat{G}}$, or the weight lattice of 
Langlands dual group $\hat{G}$. For general Wilson-'t Hooft line operators, Kapustin found that the line operators are classified by the following data \cite{Kapustin:2005py}:
\begin{equation}
(\mu, \nu)\in \Lambda_{cw}\times \Lambda_{w},~~~(\mu,\nu)\sim (\omega.\mu,~\omega.\nu),~~\omega\in W,
\end{equation}
where $W$ is the Weyl group. The allowed magnetic weight has to satisfy the Dirac quantization condition in the presence of 
matter fields.  Moreover, when the flavor symmetry is available, one can introduce the background gauge field for these
flavor symmetries and consider the line operator carrying magnetic charge on it. 

Let's apply the above general classification to the theory defined by Riemann surface $M_{g,0,n}$, namely a Riemann surface with $n$ regular punctures. There are 
$3g-3+n$ gauge groups and $n$ mass parameters, and a duality frame is described by a pants decomposition of the Riemann surface (since there is only tri-fundamental matter.). The line operator is labeled by 
$(6g-6+3n)$ integers 
\begin{equation}
(p_1,p_2,\ldots,p_{3g-3+2n};q_1,q_2,\ldots,q_{3g-3+n}).
\end{equation}
The Weyl transformation acts as $(p_j,q_j)\rightarrow(-p_j,-q_j)$ for $j=1,2, \ldots,3g-3+n$, and $p_j\rightarrow -p_j$ for $j=3g-2+n,\ldots, 3g-3+2n$. Using Weyl transformation, we may take $p$ to be positive, and 
they are the magnetic charges, so they have to satisfy the condition $p_i+p_j+p_k \in 2Z$ for three gauge groups coupled to a single pants, since the matter from the pants has to satisfy the 
Dirac quantization condition.
\begin{figure}[htbp]
\small
\centering
\includegraphics[width=10cm]{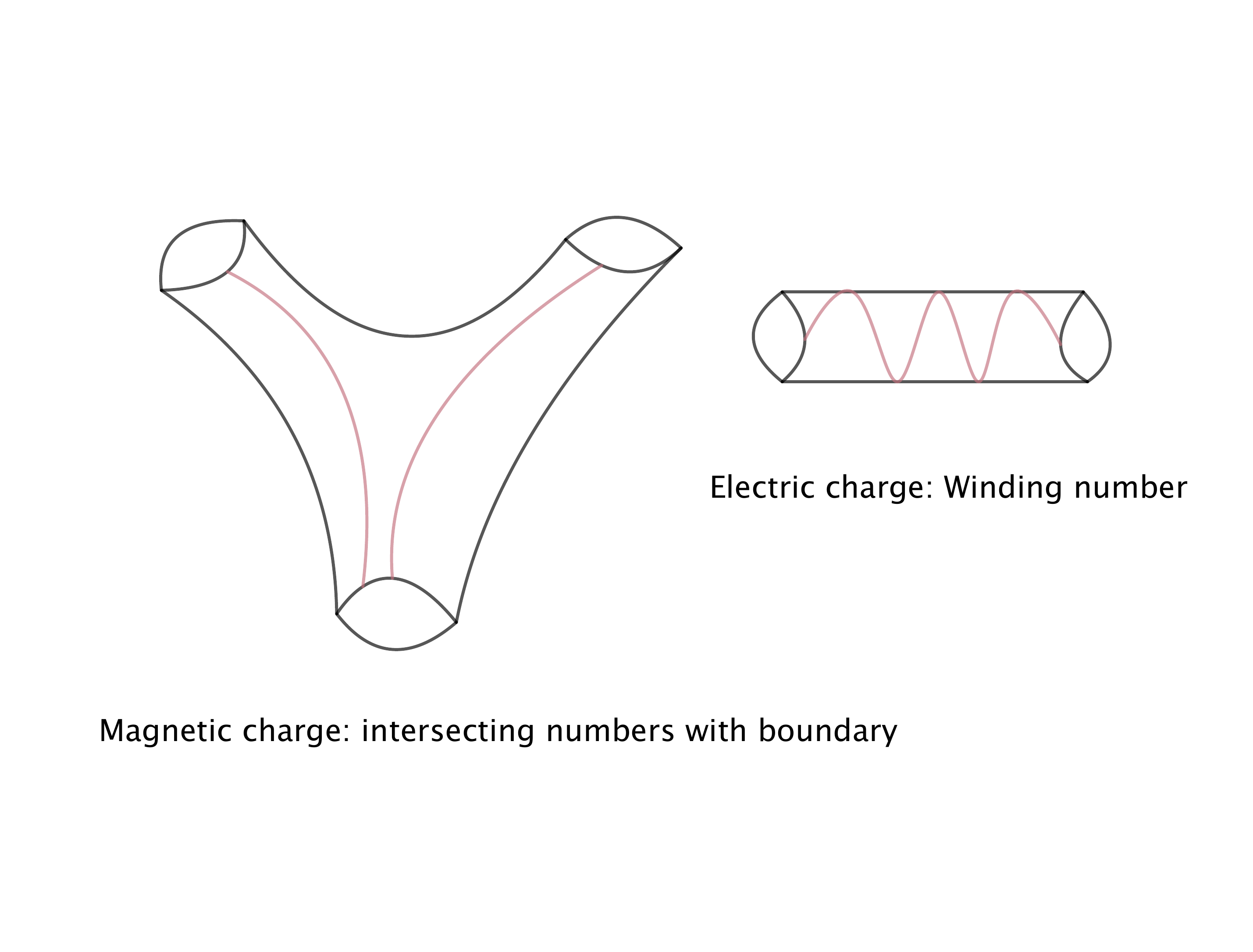}
\caption{The Dehn-Thurston coordinates of a set of closed curves are defined using a pants decomposition: the $p_i$ coordinates are defined by simply counting the weighted
intersecting number, while the $q_i$ coordinates are defined by counting the oriented winding number.}
\label{Dehn}
\end{figure}

Using above field theory results,  it is shown in \cite{Drukker:2009tz} that the space of line operators are identified with the space of self- and mutually nonintersecting unoriented closed curves on $M_{g,0,n}$. 
Therefore, instead of studying the line operator from 4d field theory point of view, we can just study these closed curves on the Riemann surface.
Now let's start with a collection of closed curves (non-intersecting) with positive integer weights \footnote{The label is taken as the charge of that line operator in the duality frame where the closed curve represents a weakly coupled gauge group.},
 and one can construct 
a set of Dehn-Thurston coordinates based on the pants decomposition:  the coordinates $p_i$ (always positive) on a boundary circle of the pants is defined by counting the weighted intersection number, and the $q_i$ coordinates (it can be positive and negative depending on the winding direction) on a tube are defined by counting the winding number of the curves, see figure. \ref{Dehn}. On the other hand, given a set of 
coordinates based on pants decomposition, one can easily find a set of closed curves by doing the inverse construction: one put $p_i$ marked point on the $i$ th boundary of the pants, and 
connect them pairwise without intersection on each pant (which is possible if they satisfy the condition  $p_i+p_j+p_k \in 2Z$ on each pant, and the construction is unique), and also the number of winding on the 
tube is determined by the coordinates $q_i$, and we construct a set of non-intersecting closed curves with weight. Finally we combine the closed curves in same homotopy class, which
 will give back a geometric objects.

The above Dehn-Thurston coordinates are directly related to the field theory description \cite{Drukker:2009tz}, however, it has several disadvantages: firstly,
the transformation of the coordinates  is quite complicated under the change of pants decomposition corresponding to the S duality; Secondly, there is no pants decomposition for
the Riemann surface with boundaries and the above definition of the line operator is not applicable.
In next subsection, we are going to introduce another set of coordinates based on the triangulation of the Riemann surface, and the coordinates has simple 
transformation rule under the change of the triangulation, moreover, the definition can be extended to all bordered Riemann surface. 

\subsubsection{UV line operators: lamination and its coordinates}
The set of the non-intersecting closed curves are called laminations in the mathematics literature \cite{fock2005dual},
so space of  line operator of 4d theory is naturally identified with the lamination space.
Let's now introduce the precise definition of Lamination on bordered Riemann surface. Before that, let's give the definition of special curves:
 a curve is called special if it can be contracted to a bulk marked point or a boundary marked point, and
a curve is contractible if it can shrink to single point which is not a marked point. Now let's spell the definition of the Lamination:

\textbf{Definition}: A integral A-lamination on a bordered Riemann surface is a homotopy class of a collection of 
\textbf{finite} number of self- and mutually nonintersecting unoriented curves either closed or connecting two points of the boundary disjoint from marked points with \textbf{integral} weights and subject to the following conditions and equivalence relations.

(1) Weights of all curves are positive, unless a curve is special.

(2) A lamination containing a curve of weight zero is considered to be equivalent to the lamination with this curve removed.

(3) A lamination containing a contractible curve is considered to be equivalent to the lamination with this curve removed.
 
(4) A lamination containing two homotopy equivalent curves with weights $u$ and $v$ is equivalent to the lamination with one of these curves removed and with the weight $u + v$ on the other.

\begin{figure}[htbp]
\small
\centering
\includegraphics[width=10cm]{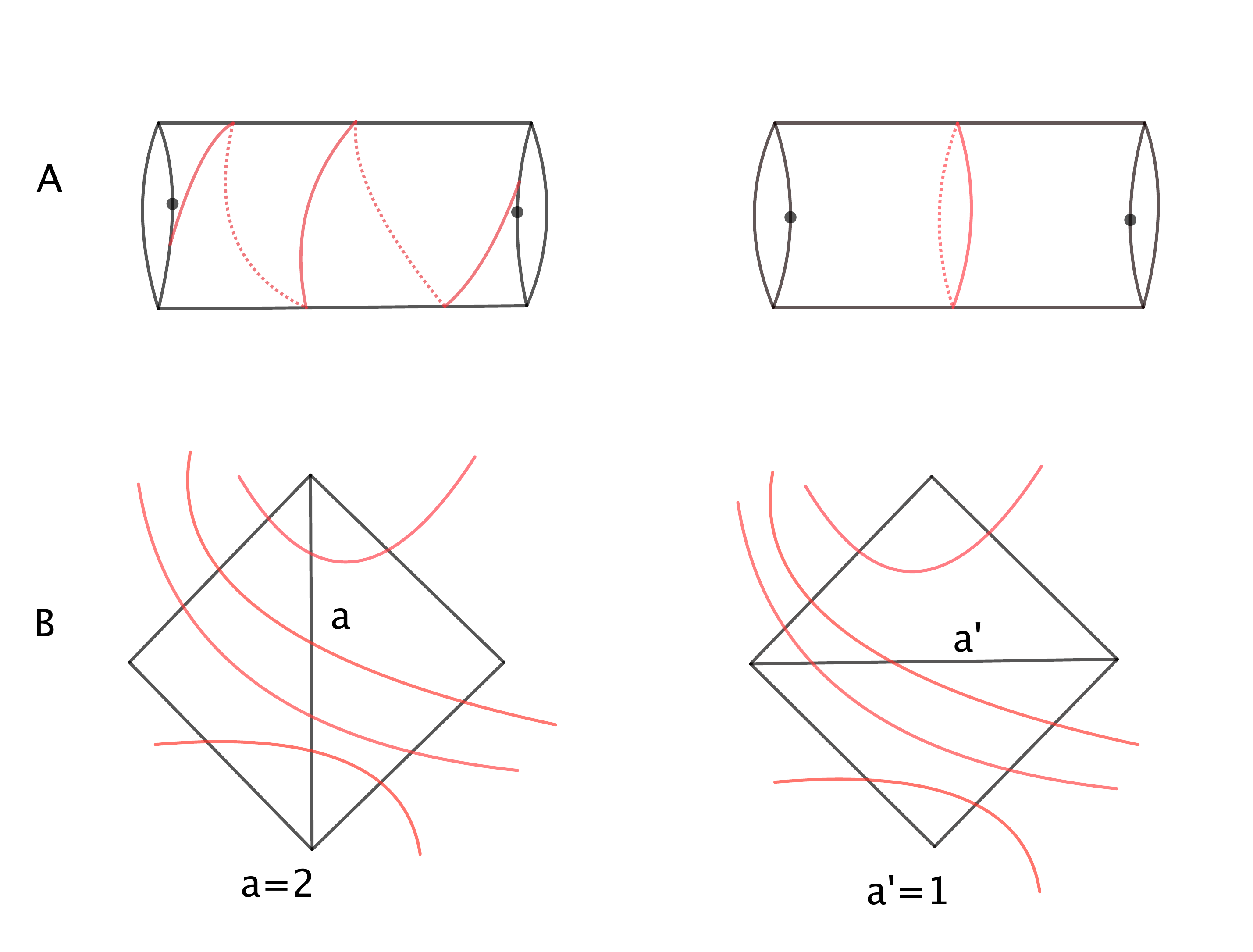}
\caption{A: Laminations on an annulus with one marked points on each boundary, which represent the pure $SU(2)$ theory. B: Constructing coordinates for 
the lamination using the triangulation.}
\label{SU2}
\end{figure}

\begin{figure}[htbp]
\small
\centering
\includegraphics[width=10cm]{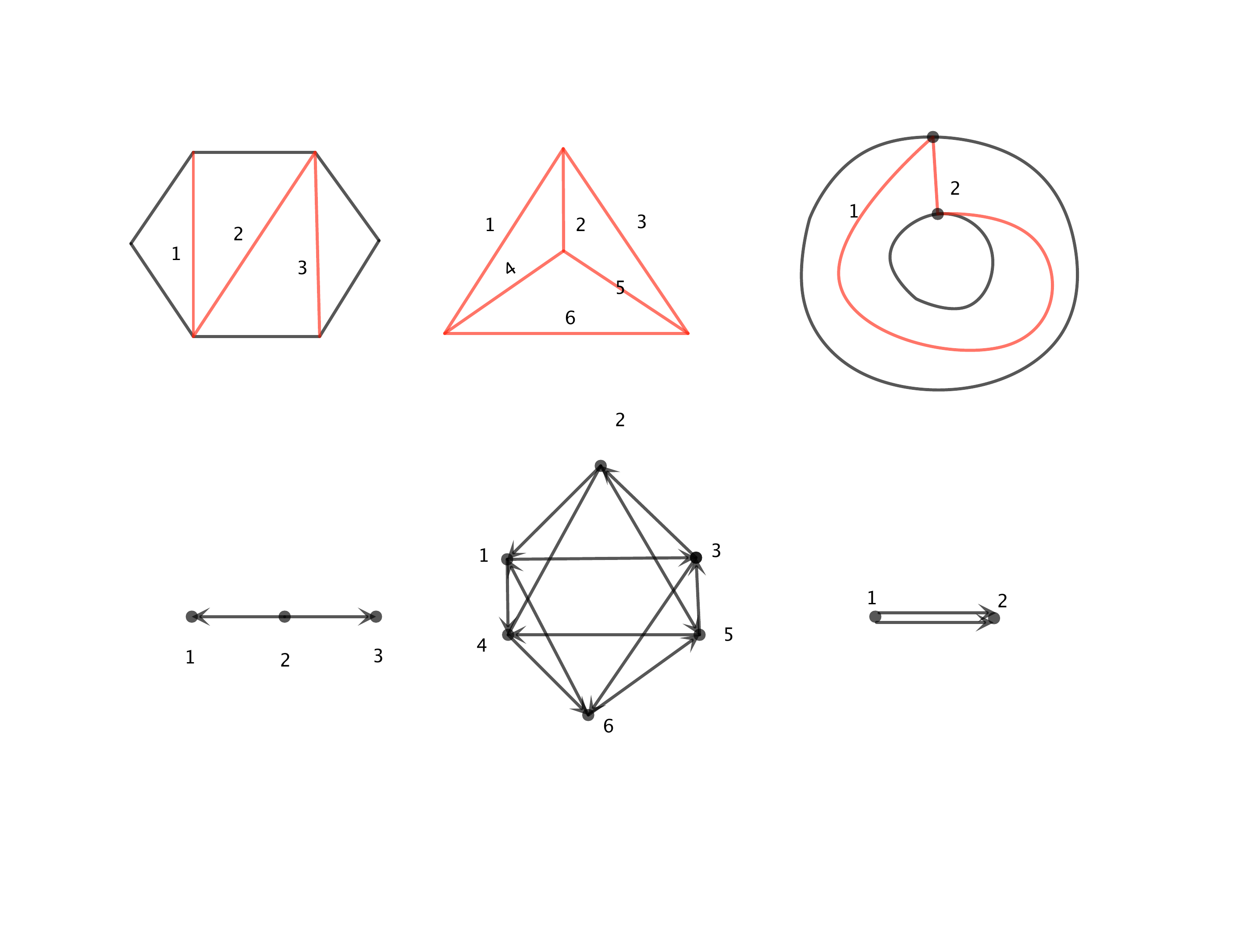}
\caption{Top: The ideal triangulations of bordered Riemann surface: disc with six marked points, sphere with four punctures, and annulus with one marked points on each boundary.
Bottom: the corresponding quiver.}
\label{ideal}
\end{figure}

Notice that the weights for the curves around the marked points can be either positive or negative, while the other curves always have positive weights. 
An example of lamination on an annulus with one point on each boundary is shown in figure. \ref{SU2}. Instead of constructing coordinates 
using the pants decomposition, we use the ideal triangulation whose vertices are the marked points. An ideal triangulation is formed by simple curves whose ending points 
are the marked points, equivalently one can use a quiver to encode the triangulations of the internal edges, and this quiver is actually the BPS quiver \cite{Cecotti:2011rv}. An important fact is that the number of internal edges in the triangulation is $
6g-6+b+3p$ which is exactly the same number of the charge lattice 
We give some example of triangulations and quivers 
for various bordered Riemann surface in figure. \ref{ideal}, more details about the ideal triangulation can be found in \cite{fomin-2008-201}.
Given a lamination and a triangulation, we can construct the coordinates (labeling) for the lamination using the following simple formula:
\begin{equation}
a_{E}={1\over2}\sum_i \omega_i,
\end{equation} 
where the sum is over all the curves intersecting with the curves, see figure. \ref{SU2}B. The coordinates are called tropical $a$ coordinates and taking half-integer values.

On the other hand, given a set of constrained half-integers (the constraints will be spelled later)  on each edge of the triangulation, we can construct a lamination.
First of all, we can assume all the coordinates are positive, since if some of the coordinates of $a_i$ is negative, we can first shift all the coordinates by a positive integers $\omega$ such that  $a_i^{'}=a_i+\omega$ are 
all positive. If we can construct lamination $L^{'}$ for $a_i^{'}$, we can add closed curves with weight $-\omega$ around each marked point to $L^{'}$ (these curves will not intersect with $L^{'}$) to get the lamination $L$ for the coordinates $a_i$.
So let's start with positive half-integer coordinates $\tilde{a}_{E}$ which satisfy the following two conditions for the coordinates on a triangle:
\begin{align}
&|\tilde{a}_{\alpha}-\tilde{a}_{\beta}|<\tilde{a}_{\gamma}<\tilde{a}_{\alpha}+\tilde{a}_{\beta} \nonumber\\
&\tilde{a}_{\alpha}+\tilde{a}_{\beta}+\tilde{a}_{\gamma}\in z.
\label{con}
\end{align}
Now for each edge in the triangle $<ABC>$, we put $2\tilde{a}_{\alpha}$ number of points on it, and for 
a triangle, we connect these points pairwise such that there is no intersections. The construction is unique, since we can divide the marked
points on each edge into the two parts as (and the coordinates on each edge are $(AB,BC,CA)=(  \tilde{a}_{\gamma}, \tilde{a}_{\beta},\tilde{a}_{\alpha})$):
\begin{align}
& a+b=2\tilde{a}_{\gamma}~~\rightarrow~~~~~~~a=\tilde{a}_{\alpha} +\tilde{a}_{\gamma}- \tilde{a}_{\beta} \nonumber\\ 
& a+c=2\tilde{a}_{\alpha}~~~~~~~~~~~~~~b=\tilde{a}_{\beta} +\tilde{a}_{\gamma}- \tilde{a}_{\alpha} \nonumber\\
& b+c=2\tilde{a}_{\beta}~~~~~~~~~~~~~~c=\tilde{a}_{\alpha} +\tilde{a}_{\beta}- \tilde{a}_{\gamma}, 
\end{align}
and the solutions are \footnote{The number of curves around a vertex is equal to the sum of the points on 
two edges connecting this vertex minus the number on the opposite edge.} positive integers due to the constraint (\ref{con}), and they 
determine how to connect the points on the boundary of the triangles: i.e. there are $a$ curves around vertex $A$.
 \begin{figure}[htbp]
\small
\centering
\includegraphics[width=8cm]{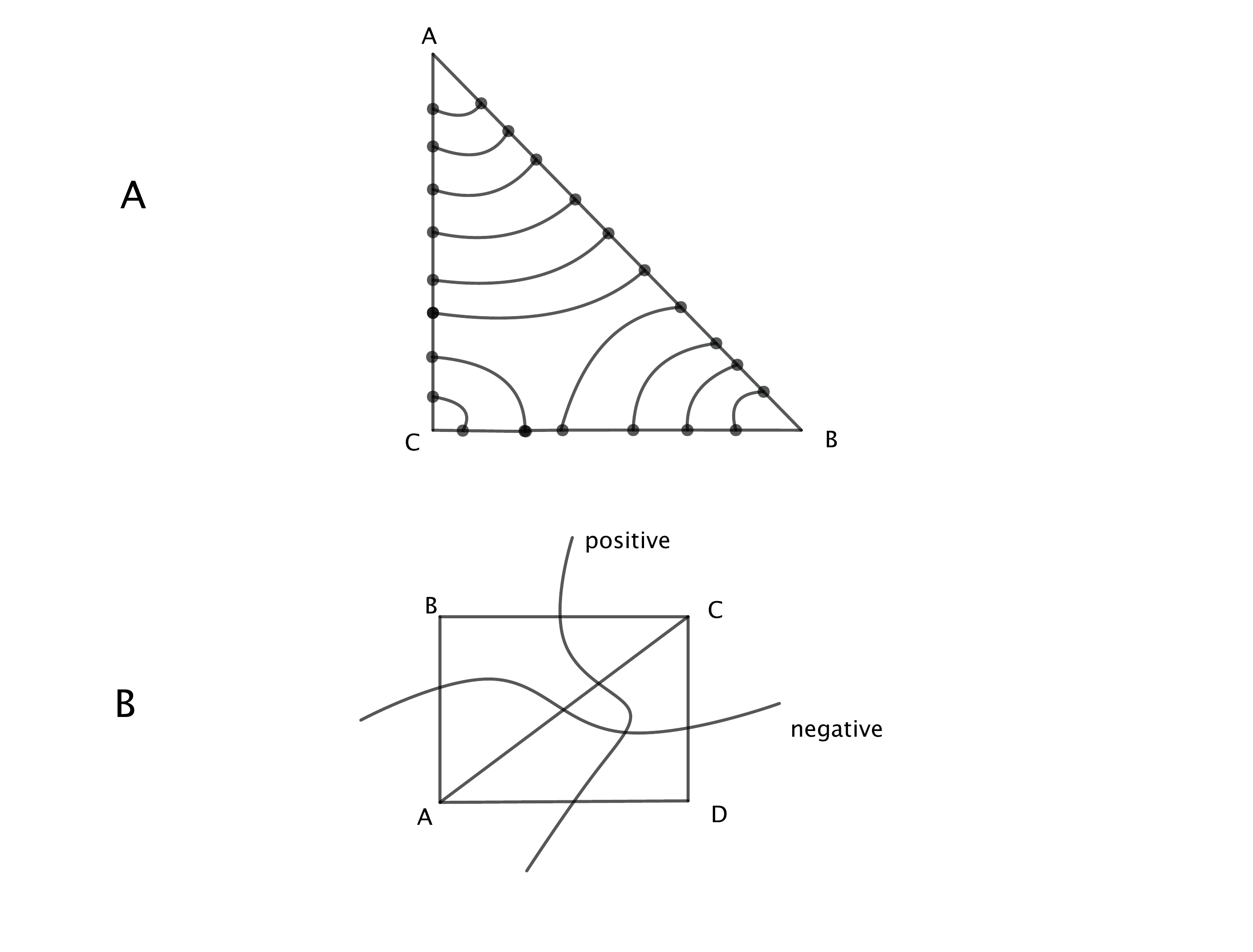}
\caption{A: Given a set of tropical $a$ coordinates on a triangle, one can construct a set of non-intersecting curves, here the coordinates are 
$(AB,BC,CA)=(5,3,4)$. B: The rule for
determining the tropical $x$ coordinate.}
\label{recon}
\end{figure}
The ideal triangulations are not unique, and different triangulations are related by the local moves called flip (see figure. \ref{SU2}). The $a$ coordinates change in a very simple way under the flip:
\begin{equation}
a_e^{'}=-a_{e}+max(a_a+a_c, a_b+a_d).
\end{equation}
If there is a boundary, one need to put further constraints on the $a$ coordinates: 
the coordinates on the boundary edge of the triangulation should be zero. 
The reason for this further constraint is that the external edge is not included in our description of  $\mathcal{N}=2$ theory, as the number of internal edges are 
exactly equal to the rank of the charge lattice.  

One might be curious about the strange constraint (\ref{con}), and the reason is that these coordinates are actually 
the tropical $a$ coordinate of the $\text{PGL}(2,C)$ local system, which is the Langlands dual group of $\text{SL}(2,C)$.
We assume that there is a $\text{SL}(2,C)$ Hitchin system on the Riemann surface for the 
4d field theory description \footnote{If the gauge group of the Hitchin system is $\text{PGL}(2,C)$, then the line operators will take values in the tropical $a$ coordinates of the $\text{SL}(2,C)$ local system and 
will take integer values.}
, and the line operator is taking value on the tropical $a$ coordinates of $\text{PGL}(2,C)$ local system. For more discussion 
on the global issue about the gauge group, see the appendix of \cite{Gaiotto:2009hg}. 

One can construct another set of coordinates for the lamination \footnote{This is called the $X$ lamination, which is also described 
by the curves on Riemann surface, but the precise definition is 
quite different from the $A$ lamination defined earlier. The crucial point is: every $A$ lamination is a also a $X$ lamination, so one can define its tropical $x$ coordinate \cite{fock2005dual}. The 
contrary is not true though: there are $X$ lamination which is not $A$ lamination, namely one can include the curves connecting the punctures.}. 
The definition is the following: pick a vertex of the diagonal edge of the quadrilateral and name it as $A$, then label the other vertices as $B,C,D$ in clockwise direction; if 
a curve goes through $AB$ and $CD$, its coordinate is positive, and the coordinates is negative if it goes through $AD$ and $BC$, see figure. \ref{recon}.
One can check that the $x$  and $a$ coordinates are related by the following simple formula
\begin{equation}
x_i=-\epsilon_{ij}a_j.
\end{equation}
where $\epsilon_{ij}$ is the antisymmetric tensor read from the quiver. These $x$ coordinate is always integer, and can be both positive and negative
even for the non-trivial closed curves.

\subsubsection{IR line operator: one cycle on Seiberg-Witten curve}
The UV theory is described by compactifying 6d theory on a punctured Riemann surface, and the lamination on Riemann surface $\Sigma$ will 
give us the UV line operator. From M theory picture, we have a 4d theory on two M5 branes, and M2 brane ending on M5 branes (wrap on lamination on $\Sigma$) gives 4d line operators. 
 What about the IR line operator? In the IR limit, these multiple M5 branes become a single M5 brane compatified on another
Riemann surface $\Sigma^c$ which is identified with the Seiberg-Witten curve \cite{Witten:1997sc}. It is natural to  expect that M2 brane wrapped on various one cycle of
$\Sigma^c$ will give us IR line operators. 

There is actually a distinguished basis on these one cycles based on 
the ideal triangulation of the Riemann surface $\Sigma$. The construction goes as follows:  first of all, we can 
replace the triangulation by a bipartite network as suggested in \cite{Xie:2012dw}.  Usually one associate a face variable $X_i$ to each face in the original bipartite network, and one can read a quiver from
the network. Given a bipartite network on a Riemann surface, one can construct a
conjugate Riemann surface $\Sigma^c$ using the so-called untwisting procedure, which is 
quite familiar in the dimer study, see \cite{Feng:2005gw}. The details are not important for us, and we want to point out the crucial point: since 
the boundary of the face of network on $\Sigma$ becomes the one cycle on $\Sigma^c$, which will give a IR line operator.
It is then natural  to interpret the 
face variable $X$ as the expectation value of the IR line operators \footnote{Classically, the expectation value of the IR line operator is the length of that one cycle.}. On the other hand,
these one cycles on the SW curve is also used to parameterize the Coulomb branch through the famous Seiberg-Witten construction, so the cluster $X$ coordinates not only 
represent the IR line operator but parameterize the Coulomb branch of 4d theory! 

\subsection{UV and IR line operator: algebraic picture}

Geometrically, the 4d line operator is related to the weighted closed curves on the Riemann surface.  Its tropical $a$ and $x$ coordinates can be constructed using the 
triangulation and their transformation property under the change of triangulation is simple . In fact, the above picture can be put into a general algebraic picture called (tropical) cluster algebra \cite{Fomin2001,fock2005dual}. 

\begin{figure}[htbp]
\small
\centering
\includegraphics[width=8cm]{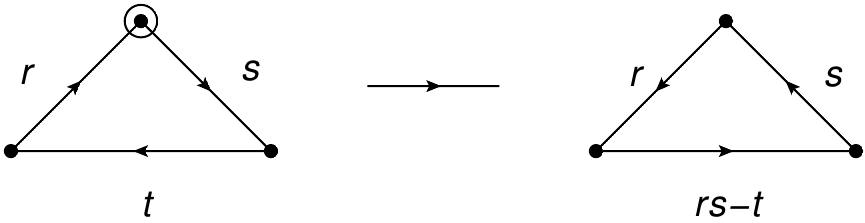}
\caption{The quiver mutation. }
\label{seiberg1}
\end{figure}

The (tropical) cluster algebra consists a family of seeds $(\epsilon_{ij},x_i, a_i)$, here $\epsilon_{ij}$ is 
an antisymmetric matrix which can be represented by a quiver, and $x_i$ and $a_i$ are 
rational, real or integral numbers \footnote{The integral numbers does not necessarily take integer values, but they should be some discrete numbers.} associated with each quiver node.  Different seeds are related by 
the so-called mutations. The mutation acting on the quiver is  
\begin{equation}
\epsilon^{'}_{ij}=\left\{
\begin{array}{c l}
    -\epsilon_{ij}& \text{if}~i=k~or~j=k\\
    \epsilon_{ij}+\text{sgn}(\epsilon_{ik})[\epsilon_{ik}\epsilon_{kj}]_+ & \text{otherwise}
\end{array}\right.
\label{cluster1}
\end{equation}
The quiver mutations can be represented beautifully using the quiver diagram:  A quiver is a directed graph where multiple arrows between two vertices are allowed, which
is derived using $\epsilon_{ij}$ as follows: attach a quiver node for $i=1,\ldots n$, and there are $\epsilon_{ij}$ arrows between node $i$ and node $j$ \footnote{If $\epsilon_{ij}$ is positive, the quiver arrows are
pointing from node $i$ to node $j$; otherwise, they are pointing from node $j$ to node $i$.}.
The quiver mutation for a quiver without one and two cycles (such quiver is called 2-acyclic) is 
 defined as the following: Let Q be a quiver and k a vertex of Q. The mutation $\mu_k(Q)$ is the quiver obtained from Q as follows, see figure. \ref{seiberg1}:

1) for each sub quiver $i\rightarrow k\rightarrow j$, create a new arrow between $ij$ starting from $i$;

2) we reverse all arrows with source or target k;

3) we remove the arrows in a maximal set of pairwise disjoint 2-cycles. 
 
In changing the triangulation, the quiver is changed as quiver mutation (Seiberg duality), and the coordinates 
are changed in following way
\begin{equation}
a^{'}_k= max(a_k{[\epsilon_{ik}]_+},a_i{[-\epsilon_{ik}]_+})-a_k
\end{equation}
\begin{equation}
{x}_j^{'}=\left\{
\begin{array}{c l }
    -{x}_k& if~j=k\\
   x_j+\epsilon_{kj} max(x_k,0) & if~\epsilon_{kj}\geq0\\
       x_j+\epsilon_{kj} max(-x_k, 0)&  if~\epsilon_{kj}\leq0 \\
\end{array}\right.
\end{equation}
The ordinary cluster algebra consists of also  triples $(\epsilon_{ij}, X_i, A_i)$ which are related 
by quiver mutations, where $\epsilon_{ij}$ is the same as the one used above and it has 
the same form as quiver mutations.
The transformation of the $X$ coordinates under mutation is 
\begin{equation}
{X}_j^{'}=\left\{
\begin{array}{c l}
    {X}_k^{-1}& if~j=k\\
    {X}_j(1+{X}_k^{-sgn(\epsilon_{jk})})^{-\epsilon_{jk}} & if~j\neq k,
\end{array}\right.
\label{clusterX}
\end{equation}
and the $A$ coordinates transform in  following way
\begin{equation}
A^{'}_k= {\prod A_i^{[\epsilon_{ik}]_+}+\prod A_i^{[-\epsilon_{ik}]_+} \over A_k}.
\label{clusterA}
\end{equation}
A  two form can be defined on  $A$ space 
\begin{equation}
\omega=\epsilon_{ij}d\log A_i\wedge d \log A_j;
\end{equation}
and a degenerate Poisson structure can be defined on the $X$ space:
\begin{equation}
[X_i,X_j]=\epsilon_{ij}X_iX_j.
\end{equation}
These structures are compatible with the cluster transformation, i.e. if you express $(X_i, \epsilon_{ij})$ in terms of $(X_i^{'}, \epsilon_{ij}^{'})$ using the 
cluster transformation rule, and you will get the same form expressed in terms of $(X_i^{'}, \epsilon_{ij}^{'})$. 

The above two types of coordinates are related by the so-called tropification procedure. Here let's describe a little bit about the idea of tropification, i.e. the triple $(\epsilon_{ij}, x_i, a_i)$ can 
be thought of as the tropical limit of $(\epsilon_{ij}, X_i, A_i)$. Let's explain a little bit of tropification:
the tropification about a polynomial $3a^3+a^2b+1$ is $\text{max}(3a,2a+b,0)$; more generally, for a rational 
function without subtraction, the tropification is 
\begin{equation}
\text{Trop}(f(a_1,a_2,\ldots,a_n))=lim_{\epsilon\rightarrow \infty}\epsilon log(f(e^{\epsilon a_1}, e^{\epsilon a_2},\ldots,e^{\epsilon a_n})). 
\end{equation}
It is then easy to see that $(\epsilon_{ij}, x_i, a_i)$ is a tropification of $(\epsilon_{ij}, X_i, A_i)$ by comparing their transformation property under the mutation.

\subsection{Canonical map: relating UV and IR line operators}
Classically it is natural to identify the expectation value of a UV line operator as the length of the closed curves on the Riemann surface, and 
the length should depend on the Coulomb branch, which is parameterized by the cluster $X$ coordinates. 
There is a remarkably simple method to calculate the length: it is related to the monodromy around the curve in the $\text{SL}(2,C)$ local system, 
and the monodromy can be calculated using  very simple combinatorial methods \cite{fock2005dual}.
Basically, one first construct a dual graph from the ideal triangulation: put a trivalent vertex inside each triangle and glue them together using the triangulation. Then put a square around each 
edge and a hexagon around each vertex, and this structure is called as monodromy graph\footnote{This prescription is different from what is usually written in the literature, and 
we adopt this definition because it is immediately generalized to higher rank.}.
The orientation of edges in monodromy graph is chosen as following: the edges on the cycles around the marked points are taken to be in anti-clockwise 
direction, and the edges in hexagon is  required to be in clockwise direction which can be used to fix the orientation of remaining edges.

\begin{figure}[htbp]
\small
\centering
\includegraphics[width=10cm]{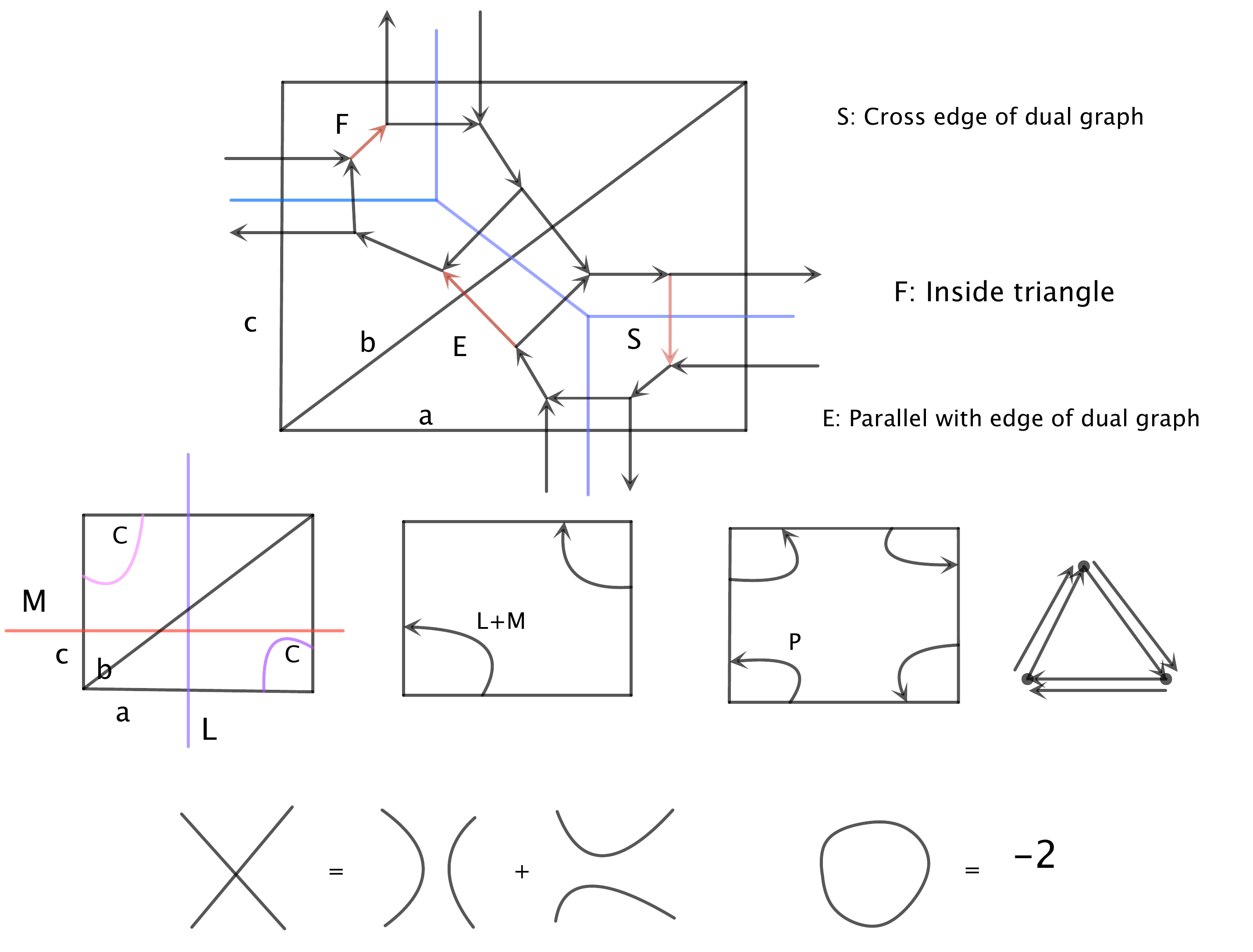}
\caption{Top: the dual graph and the monodromy graph are built from the triangulations. Middle: One triangulation has 
three edges labeled by $a,b,c$, and the opposite edges are identified. We also draw various closed curves, and the quiver. Bottom: the Skein relations.   }
\label{A1}
\end{figure}

There are three types of edges in the monodromy graph: $\text{S}$ edge which intersects with the dual graph, $\text{E}$ edge which is parallel with the dual graph, and finally
the $\text{F}$ edge which is inside the triangle, see figure. \ref{A1}. We associate the following matrices with the three types of edges:
$$
S=\begin{pmatrix}
0 & -1 \\
1 & 0
\end{pmatrix},
E=\begin{pmatrix}
X &  0\\0& 1
\end{pmatrix},
F=\begin{pmatrix}
1 & 0 \\
1& 1
\end{pmatrix}.
$$
If a path is going opposite direction to the prescribed orientation for an edge, we simply use the inverse matrix.
It can be checked that the monodromy around the square and the hexagon is an identity matrix up to a scalar, therefore it defines a 
$\text{PGL}(2)$ local system.  
Now the monodromy of a closed curve $\textbf{C}$ is found as following: find  a closed path on the monodromy graph which is in the same homotopy class as $\textbf{C}$, and then 
multiplying the matrices of  segments in this path which will give the monodromy. 
For our purpose, a $\text{SL}(2)$ local system \footnote{Since we assume the gauge group is SU(2).} is required and the monodromy $M$ is defined as
\begin{equation}
M=|\text{det}(M_0)^{-1/2}M_0|,
\end{equation}
here $M_0$ is the monodromy using the above prescription and we take the absolute value to make all the expressions positive.
Using the monodromy, the expectation value of the UV line operator or canonical map \footnote{Since the expression map  tropical $a$ coordinates to an expression in cluster $X$ coordinates.} for curves in lamination are 
\begin{itemize}
\item For a non-special closed curve $l$ with weight $k$,  the canonical map is $\text{I}(l)=\text{Tr}(\text{M}^k)$.
\item For a special curve $l$ with positive (negative) weight $k$, the canonical map is to take the highest (lowest) eigenvalue of $\text{M}^{|k|}$. 
For example, the eigenvalues of the monodromy matrix have the following form (winding once):
\begin{equation}
(\prod X_i^{1/ 2},\prod X_i^{-{1/2}}),
\end{equation}
and we define the canonical map of curve with weight one as $\text{I}(l^{+})=\prod X_i^{1\over 2}$ and 
the curve with weight minus one as $\text{I}(l^{-})=\prod X_i^{-{1\over 2}}$. 
\item For a curve connecting two points on the boundary, the canonical map is defined as:
\begin{equation}
\text{I}(l)=\begin{pmatrix}
1 &  0\\
\end{pmatrix}*\text{M}*
\begin{pmatrix}
1 \\
0
\end{pmatrix},
\end{equation}
here the definition of $M$ is subtle as there are two choices for the starting and ending edges of the path $l$ in the monodromy graph. Our prescription is that the starting and ending edge of $l$ has the same orientation as the curve.
\item For the line operator with $0$  coordinate, the canonical map is $\text{I}(0)=2$.
\item For a lamination $l$ consists of many curves $l_i$, and the canonical map is defined as 
\begin{equation}
\text{I}(l)=\prod_i \text{I}(l_i).
\end{equation}
\end{itemize}

\textbf{Example}: Let's apply the above discussion to the theory defined by an once punctured torus, which represents the so-called  $\text{SU}(2)$ $\mathcal{N}=2^{*}$  theory. 
The triangulation and the quiver is shown in figure. \ref{A1}, and there are three quiver nodes. There are three non-trivial homology cycles which we call $L, M, C$, and the line operator
associated with this three cycles has tropical $a$ coordinates $(\frac{1}{2},\frac{1}{2},0),(0,\frac{1}{2},\frac{1}{2}),(\frac{1}{2},0,\frac{1}{2})$, and the canonical map is found to be

\begin{align}
&I(L)={ 1\over\sqrt{ab}}+\sqrt{{b\over a}}+\sqrt{ab},~I(M)=     { 1\over\sqrt{bc}}+\sqrt{{c\over b}}+\sqrt{bc}                 \nonumber\\
&I(C)= { 1\over\sqrt{ca}}+\sqrt{{a\over c}}+\sqrt{ac},~I(L+M)= \frac{2 \sqrt{a c}}{a}+\frac{\sqrt{a c}}{a b}+b \sqrt{a c}+\frac{b \sqrt{a c}}{a}+\frac{\sqrt{a c}}{a b c}   \nonumber\\
& I(P^{+})=abc,~~I(P^{-})=a^{-1}b^{-1}c^{-1},
\end{align}
here $L+M$ is the cycle shown in figure. \ref{A1}. One can calculate the OPE using the canonical map: let's multiple the canonical map of two line operators, and expand the answer in terms 
of the canonical map of other line operators. A simple example is
\begin{equation}
I(L)*I(M)=I(L+M)+I(C),
\end{equation}
Geometrically, the OPE can be found using the Skein relation, see figure. \ref{A1}, and 
the above OPE can be easily found using Skein relations. Here are some properties for the canonical map $I(l)$ (for more details, see \cite{fock-2003}):
\begin{itemize}
\item The canonical map is a  Laurent polynomial in $X_i$ with positive integer coefficient, and the
leading order term for a line operator with coordinates $a_i$ is $\prod X_i^{a_i}$. 
\item There is an OPE between line operators  $\text{I}(l_1)\text{I}(l_2)=\sum_l C_{l_1l_2}^l \text{I}(l)$,
here $C_{l_1l_2}^l$ is a positive integer, with $C_{l_1l_2}^{l_1+l_2}=1$.
\item The Poisson structure between  line operators can be easily calculated using the canonical map and the Poisson structure of cluster $X$ coordinates.
\end{itemize}

\subsection{Summary}
Let's summarize the important facts about  $A_1$ line operators (see figure. \ref{lesson}): 
\begin{figure}[htbp]
\small
\centering
\includegraphics[width=8cm]{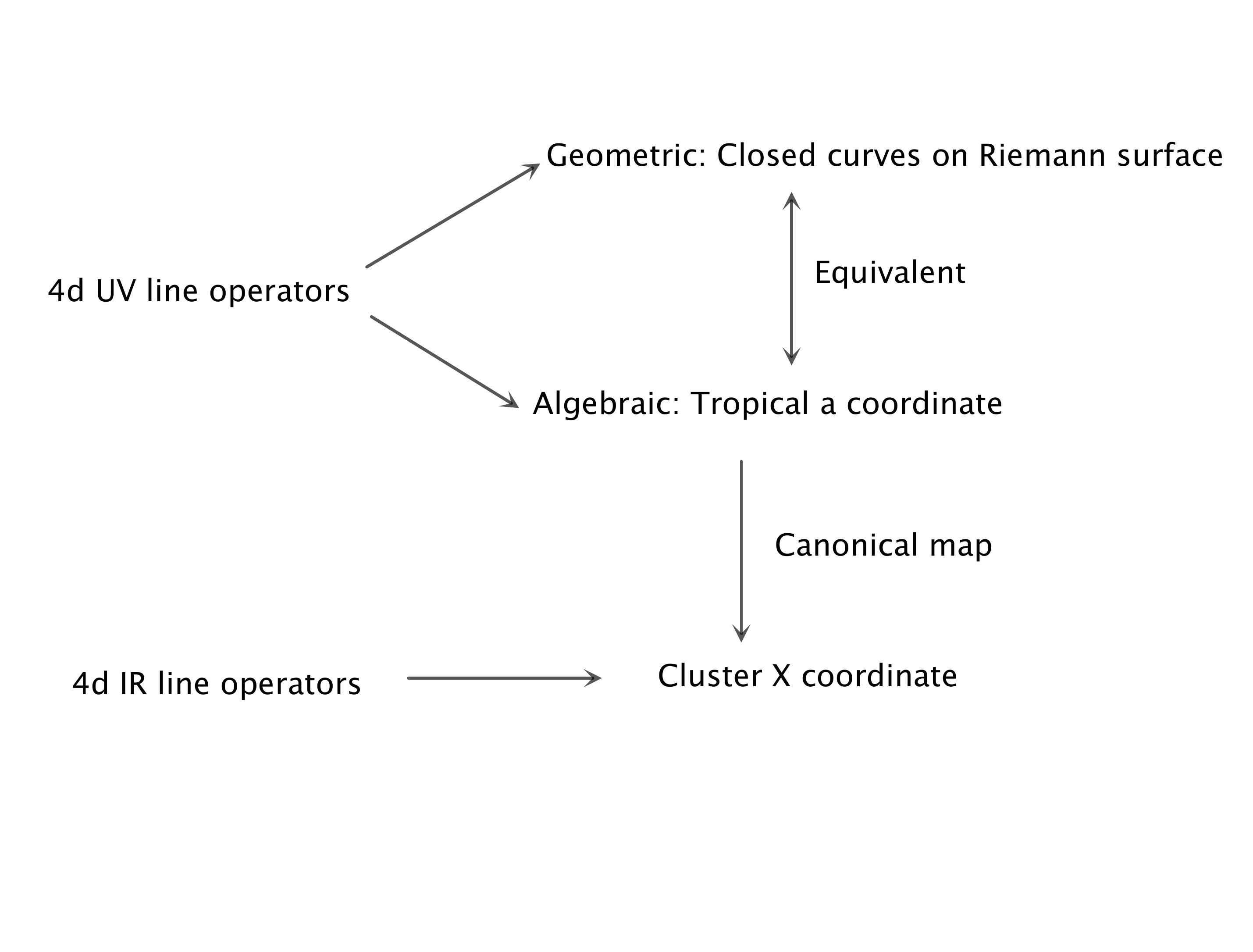}
\caption{The basic picture of $A_1$ line operators. }
\label{lesson}
\end{figure}
\begin{itemize}
\item Each 4d UV line operator is represented by a lamination, which is a global description.
\item  One can parameterize the UV line operator using the cluster algebra and tropical $a$ coordinates: given a lamination, one can construct its coordinates; and given 
a set of coordinates, one can construct a lamination. This description is local as there are a lot of ideal triangulations.
\item The space of IR line operator is identified with the cluster $X$ coordinate.
\item There is a canonical map $\mbox{I}(l)$ between the lamination space and the cluster $X$ coordinates, and there is an OPE between them, which 
has very interesting property.
\end{itemize}

\newpage
\section{$A_{N-1}$ theory: torus with one full puncture}
We are going to generalize the story of  $A_1$  line operators to $A_{N-1}$ theory defined using full regular and irregular singularities\footnote{
In the regular singularity case, the full puncture means the most generic singularity, and the 
specific irregular singularity which will give the full marked points is explained in \cite{Xie:2012jd}.}. Similarly one can replace
the irregular singularity with a boundary with full marked points, and the geometric object is still a bordered Riemann surface (see \cite{Xie:2012jd} for 
the exact correspondence between the bordered Riemann surface and the physical 4d $\mathcal{N}=2$ theory). In the $A_1$ case without the boundary, the 
field theory tools are available since the theory has Lagrangian description, which provides a lot of informations. 
Now although the weakly coupled gauge group description is still valid and can give us some clues,  there are always 
strongly coupled matter and it is clear new phenomenon must appear.

On the other hand, the geometric picture we learned from $A_1$ case is still valid, and closed curves on Riemann surface should still give 
 4d line operators. However, it should be clear that this is not enough as each closed curve give a weakly coupled gauge group, and 
 these line operators can not be used to probe the strongly coupled matter part. 

We do have the cluster algebra tools on hand, so it is straightforward to argue that the UV line operators are parameterized by the 
integral tropical $a$ coordinates of the cluster algebra, and the IR line operator
is given by the cluster $X$ coordinate. 
However, the specific tropical $a$ coordinates (what kind of constraints one need to 
impose on the coordinates) are not known, and given a set of tropical coordinates, we do not have a geometric picture.

Given all these difficulties,  it is probably useful to look at the simplest theory: torus with one full puncture. It turns out that we are going 
to discover the generic features about the line operators of  $A_{N-1}$ theory, and 
remarkably we will establish the same story as the $A_1$ theory.  The key is the canonical map of the line operator associated 
with closed curve and the OPE!

\subsection{$A_2$ theory}
\subsubsection{Closed curves: ordinary Wilson line}
\begin{figure}[htbp]
\small
\centering
\includegraphics[width=10cm]{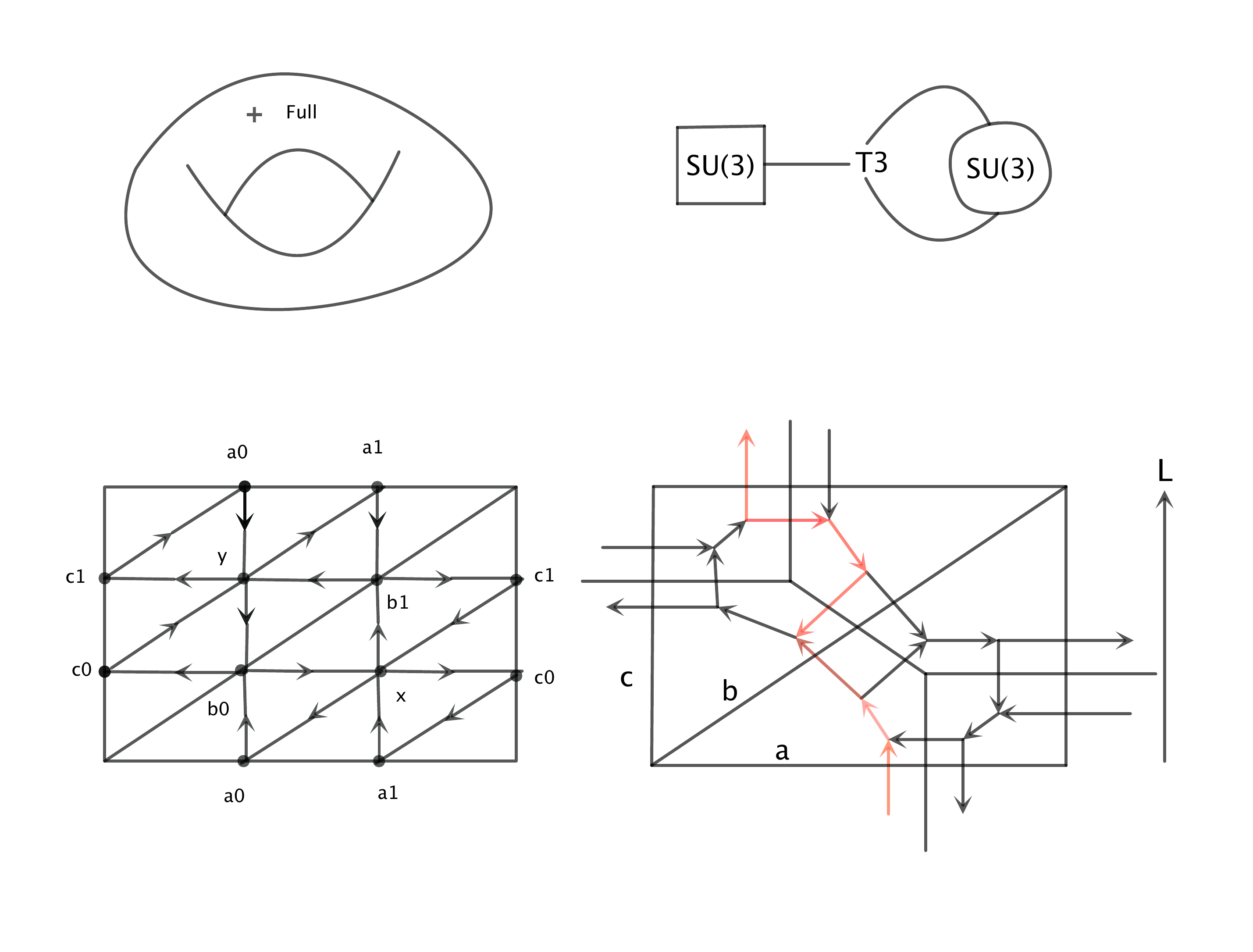}
\caption{Top: The 4d theory defined by putting 6d $A_2$ theory on a torus with one full puncture, and it is derived by gauging the flavor symmetry of $T_3$ theory. 
Bottom Left: The cluster coordinates for $A_2$ theory on once punctured torus; Bottom Right: the closed curve $L$ is represented by the red path in the 
monodromy graph.}
\label{A2}
\end{figure}

Let's consider 4d  $\mathcal{N}=2$ theory defined by compactifying six dimensions $A_2$ theory on a torus with 
a full puncture. The field theory in one duality frame is derived by gauging the diagonal of two
$\mbox{SU}(3)$ flavor symmetries of $\mbox{T}_3$ ($\mbox{E}_6$) theory.  This theory has three Coulomb branch operators: two are carried by \mbox{SU}(3) gauge group,
and one is carried by the $\mbox{T}_3$ theory, see figure.\ref{A2}; there are two mass parameters carried by the full puncture, so the charge lattice has dimension $8$.  

The cluster coordinates (or the quiver) for this theory is constructed as following: one start with a triangulation of  once punctured torus, which is the same as 
the $A_1$ case, and the new features are: there are $2$ nodes on each edge, and one extra nodes inside each triangle.  
The quiver is shown in figure. \ref{A2}.

It is natural to identify the integral tropical $a$ coordinates as the space of half-BPS line operators. Although this statement 
is quite powerful, we actually want to know the geometric counterpart which is more useful in many contexts. Namely, we 
would like to find the geometric representation for each line operator labeled by a tropical $a$ coordinate. 

We do have a natural geometric object on the Riemann surface: the closed curve. Since each closed curve represent a 
weakly coupled gauge group, it is natural to expect that one should get a Wilson loop by wrapping six dimensional surface operator on it, and 
we would like to know its tropical  $a$ coordinates.  According to $A_1$ theory, the tropical $a$ coordinates can be read from the top 
component of the canonical map calculated from monodromy; We expect this fact is also true in the higher rank theory, and luckily the way 
of calculating the monodromy is given in \cite{fock-2003}. We are going to use the monodromy calculation to find the tropical $a$ coordinates for each closed curve.

\begin{figure}[htbp]
\small
\centering
\includegraphics[width=10cm]{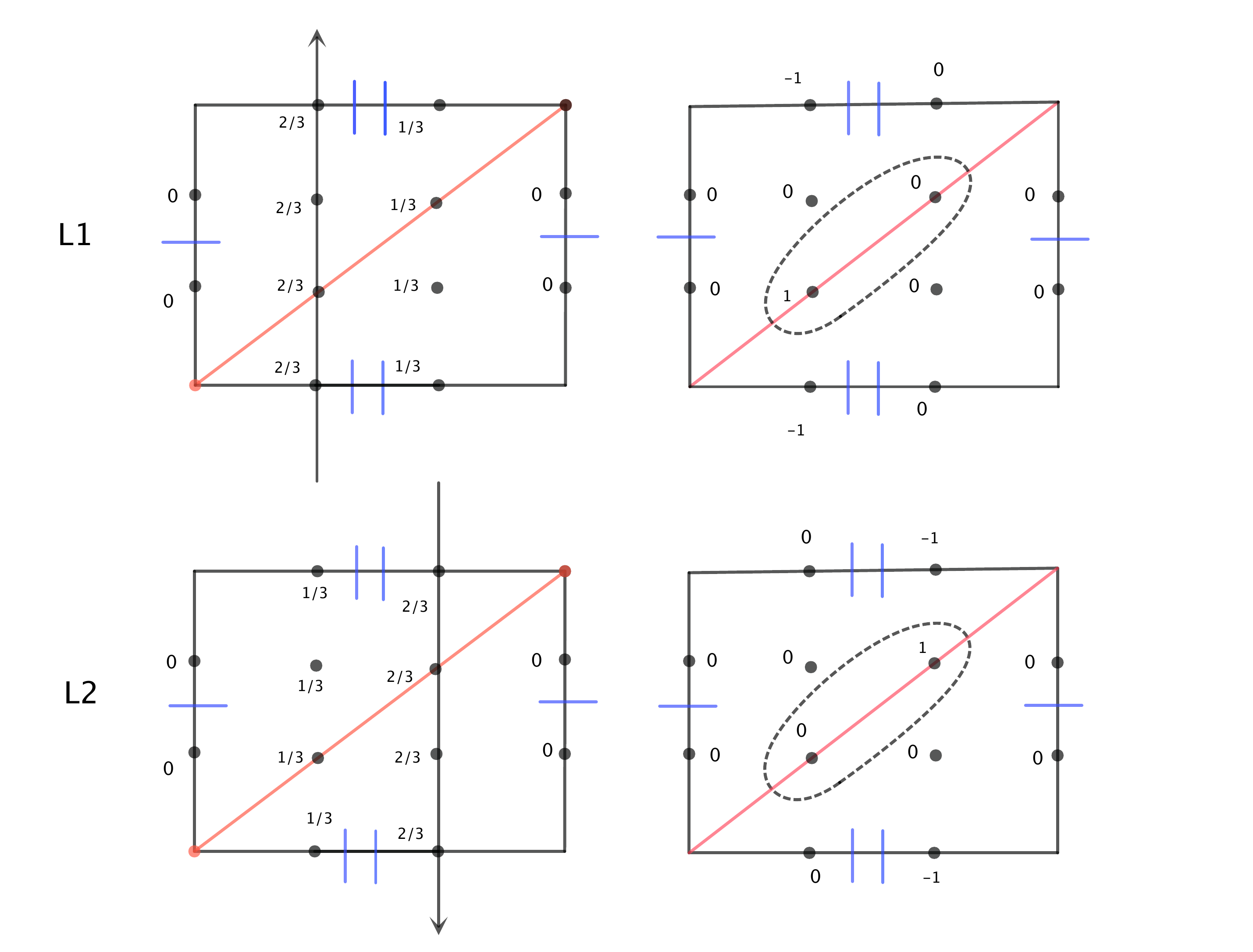}
\caption{Left: the tropical $a$ coordinates for line operators in cycle $L$ with two different orientations. Right: the tropical $x$ coordinate for $L_1$ and $L_2$.}
\label{funda}
\end{figure}

There are three simple nontrivial  cycles which we denote as $\mbox{L},\mbox{M},\mbox{ C}$ which can be identified as three duality frames of the corresponding gauge theory, see figure. \ref{A1}. 
Let's consider one duality frame in which the $\mbox{L}$ cycle represents the $\mbox{SU}(3)$ gauge group. A closed curve wrapping around the 
$\mbox{L}$ cycle should give us a Wilson loop. The monodromy of the closed curves can be similarly found using the following three by three matrices in the same monodromy graph as used for $A_1$ theory:
$$
\mbox{S}[3]=\begin{pmatrix}
0 & 0&-1 \\
0 & 1&0\\
-1&0&0
\end{pmatrix},~~
\mbox{E}[a,3]=\begin{pmatrix}
a_0a_1 &  0&0\\
0& a_1&0\\
0&0&1
\end{pmatrix},~~
\mbox{Face}[x,3]=\begin{pmatrix}
 x & 0 & 0 \\
 x & x& 0 \\
 x & 1+x & 1 \\
\end{pmatrix}.
$$
The above prescription actually gives a $\mbox{PGL(3)}$ matrix and $\mbox{SL}(3)$ canonical map is given by the following formula
\begin{equation}
\mbox{I}(l)=(\mbox{Det}(M))^{-{1\over 3}}\mbox{Tr}(M).
\label{formula}
\end{equation}
The canonical map of $L$ (we call it as $\mbox{L}_1$.) is found using  the path in monodromy graph shown in figure. \ref{A2}; and
 the monodromy matrix is
\begin{equation}
\mbox{M}=\mbox{E}[a,3]*\mbox{Face}[x,3]*\mbox{E}[b,3]*\mbox{S}[3]^{-1}*\mbox{Face}[y,3]^{-1}*\mbox{S}[3]^{-1};
\end{equation}
Using (\ref{formula}), the canonical map for $L_1$ is
\begin{align}
&\mbox{I}(\mbox{L}_1)= {1\over{x^{2/3} y^{1/3} a_0^{1/3} a_1^{2/3} b_0^{1/3} b_1^{2/3}}}+{{b_1^{1/3}}\over{x^{2/3} y^{1/3} a_0^{1/3} a_1^{2/3} b_0^{1/3}}}+\frac{x^{1/3} b_1^{1/3}}{y^{1/3} a_0^{1/3} a_1^{2/3} b_0^{1/3}} 
+\frac{x^{1/3} a_1^{1/3} b_1^{1/3}}{y^{1/3} a_0^{1/3} b_0^{1/3}}+\nonumber\\
&\frac{x^{1/3} b_0^{2/3} b_1^{1/3}}{y^{1/3} a_0^{1/3} a_1^{2/3}}+\frac{x^{1/3} a_1^{1/3} b_0^{2/3} b_1^{1/3}}{y^{1/3} a_0^{1/3}}
+\frac{x^{1/3} y^{2/3} a_1^{1/3} b_0^{2/3} b_1^{1/3}}{a_0^{1/3}}+x^{1/3} y^{2/3} a_0^{2/3} a_1^{1/3} b_0^{2/3} b_1^{1/3}.
\end{align}
This is a positive Laurent polynomial with positive integer coefficient, and the exponents of the leading order term 
should be identified with the tropical $a$ coordinates of $L_1$\footnote{We use the same letter say $a_0$ to denote the cluster $X$ coordinates and the tropical $a$ coordinates, as the 
tropical $a$ coordinates can be read from the exponent of the monomial in cluster $X$ coordinates.}:
$(a_0,a_1,b_0,b_1,c_0,c_1,x,y)=({2\over3},{1\over3},{2\over3},{1\over3},0,0,{1\over3},{2\over3})$.
We now argue that $L_1$ should be identified with the Wilson line in fundamental representation of $\text{SU}(3)$ gauge group by looking at its dual tropical $x$ coordinate:
\begin{equation}
x_i=-a_j\epsilon_{ij},
\end{equation}
and the answer for $L_1$ is $(-1,0,1,0,0,0,0,0)$.  Now take one of the edge (say $b$ edge) of the triangulation, and there are 
$2$ dots on it which can be thought of as the Dynkin nodes for the $A_2$ group, and our line operator has coordinates $(1,0)$ which 
is exactly the Dynkin label for fundamental representation of $\mbox{SU}(3)$ group, so it is natural to identify it as the Wilson line in fundamental representation, see figure. \ref{funda}.

The Wilson loop in anti-fundamental representation is naturally identified with the closed curve with opposite orientation (this is different from 
$A_1$ case in which the orientation does not matter, since the fundamental and anti-fundamental representation of $\mbox{SU}(2)$ group 
is the same.) The canonical map is calculated using the matrix $M^{-1}$, and the answer is
\begin{align}
&\mbox{I}(\mbox{L}_2)=\frac{1}{x^{1/3} y^{2/3} a_0^{2/3} a_1^{1/3} b_0^{2/3} b_1^{1/3}}+\frac{b_0^{1/3}}{x^{1/3} y^{2/3} a_0^{2/3} a_1^{1/3} b_1^{1/3}}+\frac{y^{1/3} b_0^{1/3}}{x^{1/3} a_0^{2/3} a_1^{1/3} b_1^{1/3}}+\frac{y^{1/3} a_0^{1/3} b_0^{1/3}}{x^{1/3} a_1^{1/3} b_1^{1/3}}+\nonumber\\
&\frac{y^{1/3} b_0^{1/3} b_1^{2/3}}{x^{1/3} a_0^{2/3} a_1^{1/3}}+\frac{y^{1/3} a_0^{1/3} b_0^{1/3} b_1^{2/3}}{x^{1/3} a_1^{1/3}}+\frac{x^{2/3} y^{1/3} a_0^{1/3} b_0^{1/3} b_1^{2/3}}{a_1^{1/3}}+x^{2/3} y^{1/3} a_0^{1/3} a_1^{2/3} b_0^{1/3} b_1^{2/3}.
\end{align}
Now  $a$ coordinates for this line operator  is $(a_0,a_1,b_0,b_1,c_0,c_1,x,y)=({1\over3},{2\over3},{1\over3},{2\over3},0,0,{2\over3},{1\over3})$, and 
the dual $x$ coordinates is $(0,-1,0,1,0,0,0,0)$. The $x$ coordinates on edge $b$ is $(0,1)$ which is exactly the  Dynkin label for  anti-fundamental 
representation. 

The other irreducible representations for $A_2$ group are labeled by two positive integers $(m,n)$. In group representation theory, these other representations 
can be found by doing the tensor product of  $m$ fundamental and $n$ anti-fundamental representation, and expand them in terms of irreducible representation in which
$(m,n)$ is the leading order component. 
Here we use  exactly same strategy to find the line operator for the other representations: a line operator with Dynkin label $(m,n)$ can be found 
as leading component in the OPE  of product $\mbox{L}_1^m\mbox{L}_2^n$, since the leading order component in the OPE
 has the tropical $x$ coordinates $(m,n)$ on edge $b$. 

The tropical $a$ coordinates of the line operators in representation $(m,n)$ is $ma(\mbox{L}_1)+na(\mbox{L}_2)$, and the canonical map can be found by carefully doing the OPE. 
A very useful fact is that 
the canonical map for the line operator in $(n,0)$ or $(0,n)$ representations can be found using the monodromy matrix $M^n$ and $M^{-n}$ respectively.  
To find the canonical map of the general representation $(m,n)$, one need to do the OPE expansion of the product $\mbox{L}_1^m \mbox{L}_2^n$. Not quite surprisingly, the OPE is 
closely related to the tensor product of the corresponding representations. More precisely, the support of the  the OPE is 
exactly the same as the corresponding tensor product expansion of the group representation theory, but the coefficient of our OPE is not equal to the multiplicities of the tensor 
product. Here are some of the examples we found:
\begin{align}
&\mbox{I}(\omega_1)*\mbox{I}(\omega_2)=\mbox{I}(\omega_1+\omega_2)+\mbox{I}(0),~~~~~~~~~~~~~~~~~(3\otimes \bar{3}=8+1)  \nonumber\\
&\mbox{I}(\omega_1)*\mbox{I}(\omega_1)=\mbox{I}(2\omega_1)+2\mbox{I}(\omega_2) ,~~~~~~~~~~~~~~~~~~~(3\otimes 3=6+\bar{3})       \nonumber\\
&\mbox{I}(\omega_1+\omega_2)*\mbox{I}(\omega_1)=\mbox{I}(2\omega_1+\omega_2)+2\mbox{I}(2\omega_2)+2\mbox{I}(\omega_1),~(8\otimes 3= 15+6+3).
\end{align}
Here we use $\omega_1$ to denote the line operator in fundamental representation ($\mbox{L}_1$) and $\omega_2$ to denote the anti-fundamental representation line operator ($\mbox{L}_2$), and 
the trivial line operator (with zero $a$ coordinates) give a number $3$ \footnote{Since a trivial loop gives an identity matrix, and taking the trace would 
give us a number $3$.} .  Let's now prove the second OPE in above formula. 
The canonical map for $(2,0)$ line operator can be found using the monodromy matrix $M^2$, and the 
result is given in the appendix B.
Using the canonical map of $\mbox{I}(\omega_1)$ for $L_1$ and $\mbox{I}(\omega_2)$ for $L_2$, one can verify the second OPE in above formula.
These wonderful OPE suggests that we got the right answer for the Wilson loops: the Wilson loop is classified by the irreducible representation 
of the gauge group, and the OPE is related to the tensor product of the corresponding representations. These facts play an important role
in the physical approach to Geometric Langlands program in \cite{Kapustin:2006pk}.

Let's now calculate the monodromy around the puncture, and the closed curve is shown in figure. \ref{A1}. By taking a closed path 
representing the above closed curve in the monodromy graph, we have the following representation for the monodromy: 
\begin{align}
&M_p=\mbox{Edge}[e, 3].\mbox{Face}[y, 3].\mbox{Edge}[a, 3].\mbox{Face}[x, 3].\mbox{Edge}[b, 3].\mbox{Face}[y,  3]. \nonumber\\
&\mbox{Edge}[c, 3].\mbox{Face}[x, 3].\mbox{Edge}[f, 3].\mbox{Face}[y, 3].\mbox{Edge}[d, 3].\mbox{Face}[x, 3]
\end{align}
Notice that we use the different labels for the same edges ($(e=c, a=f, b=d)$), and at the end we should identify the coordinates of the same edge in the following way, say
\begin{equation}
e_0=c_1,e_1=c_0.
\end{equation}
we find a upper triangular matrices whose three diagonal terms (after making the matrix to have determinant one) are 
\begin{equation}
\gamma_{11}=a_0a_1b_0b_1c_0c_1xy,~~\gamma_{22}=xy,~~~\gamma_{33}={1\over a_0a_1b_0b_1c_0c_1x^2y^2}.
\end{equation}
Each diagonal term is the central element in the Poisson structure, and it is natural to regard the first two terms as the 
independent mass terms:
\begin{equation}
m_1=a_0a_1b_0b_1c_0c_1xy,~~m_2=xy.
\end{equation}
We take the leading order term as the canonical map for this line operator with positive weights (as motivated by the rule in the $A_1$ case): 
\begin{equation}
\text{I}(\omega_1^{+})=a_0a_1b_0b_1c_0c_1xy.
\end{equation}
Here $\omega_1^{+}=(1,0)$, and we use this rule following the study of the Wilson line.  We can also take the least leading order term as the canonical map for the opposite label:
\begin{equation}
\text{I}(\omega_1^{-})={1\over a_0a_1b_0b_1c_0c_1x^2y^2},
\end{equation}
here $\omega_1^{-1}=(-1,0)$. Similarly, by calculating the monodromy using the opposite orientation and take the leading order term, we have the canonical
map
\begin{equation}
\text{I}(\omega_2^{+})=a_0a_1b_0b_1c_0c_1x^2y^2,~~\text{I}(\omega_2^{-})={1\over a_0a_1b_0b_1c_0c_1xy},
\end{equation}
with $\omega_2^{+}=(0,1), \omega_2^{-}=(0,-1)$. The tropical $a$ coordinates can be easily read from the canonical map.
More generally, a closed curve around the puncture is labeled by  a pair of  integers $(m,n)$, and the tropical $a$ coordinates are
\begin{equation}
\text{Trop}((m,n))=\text{Trop}(\omega_1^{\text{sgn}(m)})+\text{Trop}(\omega_2^{\text{sgn}(n)}).
\end{equation}
and the canonical map is just a single monomial in  cluster $X$ coordinates with the exponent determined by  tropical $a$ coordinates.

\subsubsection{The missing line operator}
Completely same analysis  can be applied to other simple closed curves, say $\mbox{M}$ and $\mbox{C}$ cycles. The conclusion is that the line operators supported on 
these closed curves are generated by the line operators in the fundamental and anti-fundamental representations, and they are labeled by
the irreducible representations of $\text{SU}(3)$ group. In the torus case, 
we have the generators $\mbox{L}_1, \mbox{L}_2, \mbox{M}_1, \mbox{M}_2, \mbox{C}_1, \mbox{C}_2$. If we take a duality frame where $L$ cycle
represents the weakly coupled gauge group, then $\mbox{M}_i$ can be regarded as the 't Hooft loop operator. 

The above story seems to be in parallel with the $A_1$ theory.
However, something is  missing here: first of all, all the line operators we found do not have the charges (tropical $x$ coordinates) under
the internal quiver nodes, and secondly there are only six generators for the loop coordinates and one should have at least $8$ to match the 
total dimension of the cluster coordinates. It is natural that 
we could not find the extra ones by just looking at the closed curves since they should only carry the charge of the gauge group, and they 
are not able to probe the internal Coulomb branch operator carried by the $\mbox{T}_3$ theory. 
\begin{figure}[htbp]
\small
\centering
\includegraphics[width=8cm]{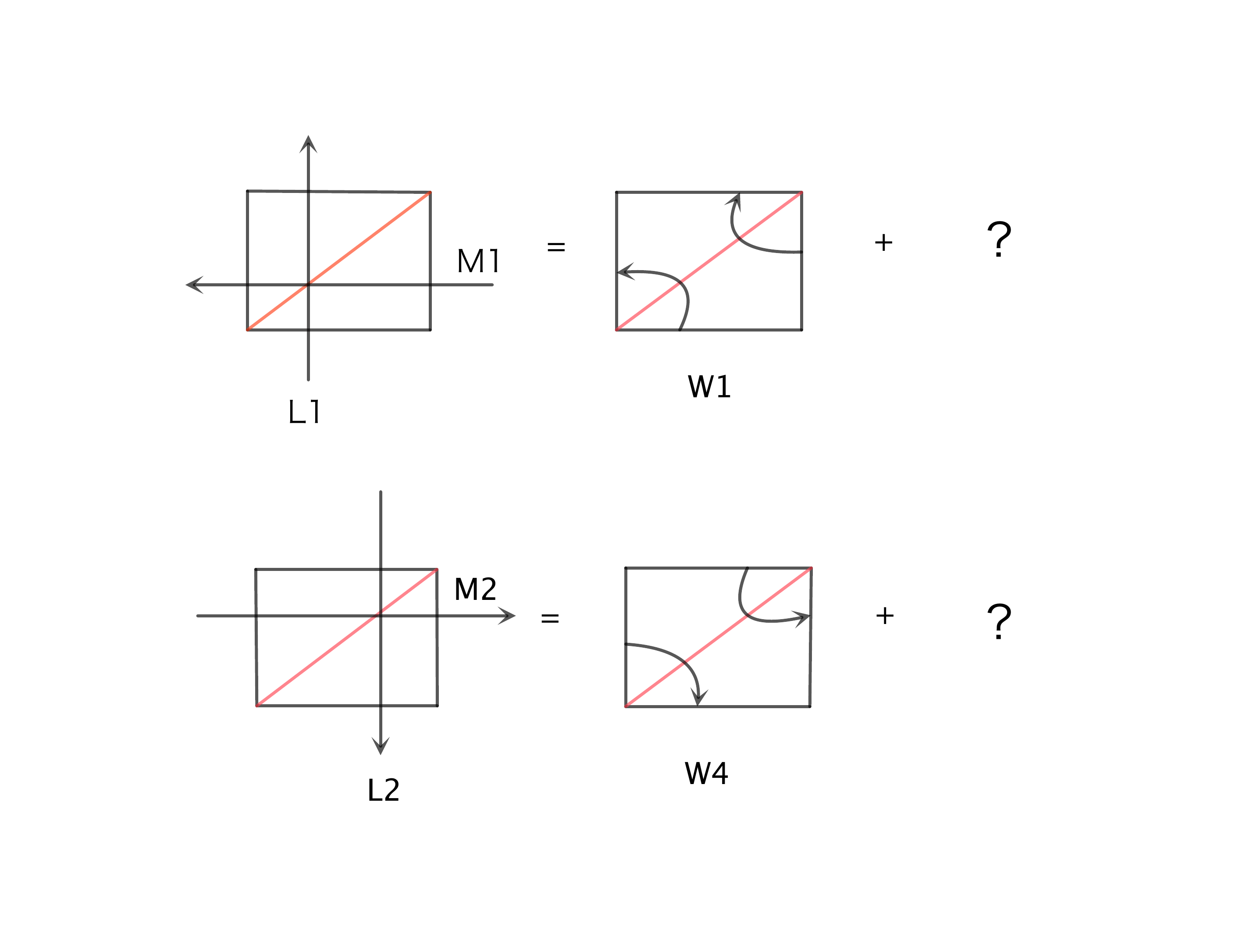}
\caption{The OPE between the Wilson and t'Hooft line operators.}
\label{OPE}
\end{figure}

Although a geometric picture for the missing ones is not available yet (we will turn to this later), 
we should be able to find  them by doing the OPE of these usual Wilson and 't Hooft line operators as 
they form a closed algebra.
Let's do the OPE for $\mbox{L}_1$ and $\mbox{M}_1$ (see the appendix for the canonical map of $\mbox{M}_1$), see figure. \ref{OPE}.
 In the OPE, it is natural to expect that 
the leading order component might be represented by the closed curve $W_1$ shown in figure. \ref{OPE}, whose 
 canonical map  can be found using monodromy representation. We calculate the canonical map for $W_1$  and subtract it from the product of $\text{I}(\text{L}_1) *\text{I}(\text{M}_1)$,
and indeed we get a positive polynomial which means that it is indeed the top component of 
the OPE. We are left with a positive Laurent polynomial whose leading order term is (see appendix for the full Laurent polynomial):
\begin{equation}
\mbox{I}(I_1)=a_0^{2/3} a_1^{1/3} b_0^{1/3} b_1^{2/3} c_0^{2/3} c_1^{1/3} x y+\ldots
\end{equation}
This line operator is actually the missing one!  The leading order monomial should give the tropical $a$ coordinates for this new line operator. As usual, Let's find its tropical $x$ coordinates, and the answer is 
\begin{equation}
(0,0,0,0,0,0,-1,1),
\end{equation}
so it indeed carries the charge under the internal quiver nodes, which strongly suggests that this is the missing line operator we are looking for.
Similarly, one can find another internal line operator $I_2$ by doing the OPE of $L_2$ and $M_2$, and  after subtracting the canonical map of $W_4$,
we find  the canonical map $\text{I}(\text{I}_2)$ whose leading order term is
\begin{equation}
\mbox{I}(I_2)= a_0^{1/3} a_1^{2/3} b_0^{2/3} b_1^{1/3} c_0^{1/3} c_1^{2/3}x y+\ldots\ldots
\end{equation}
and it has the tropical $x$ coordinates
\begin{equation}
(0,0,0,0,0,0,1,-1).
\end{equation}
\begin{figure}[htbp]
\small
\centering
\includegraphics[width=8cm]{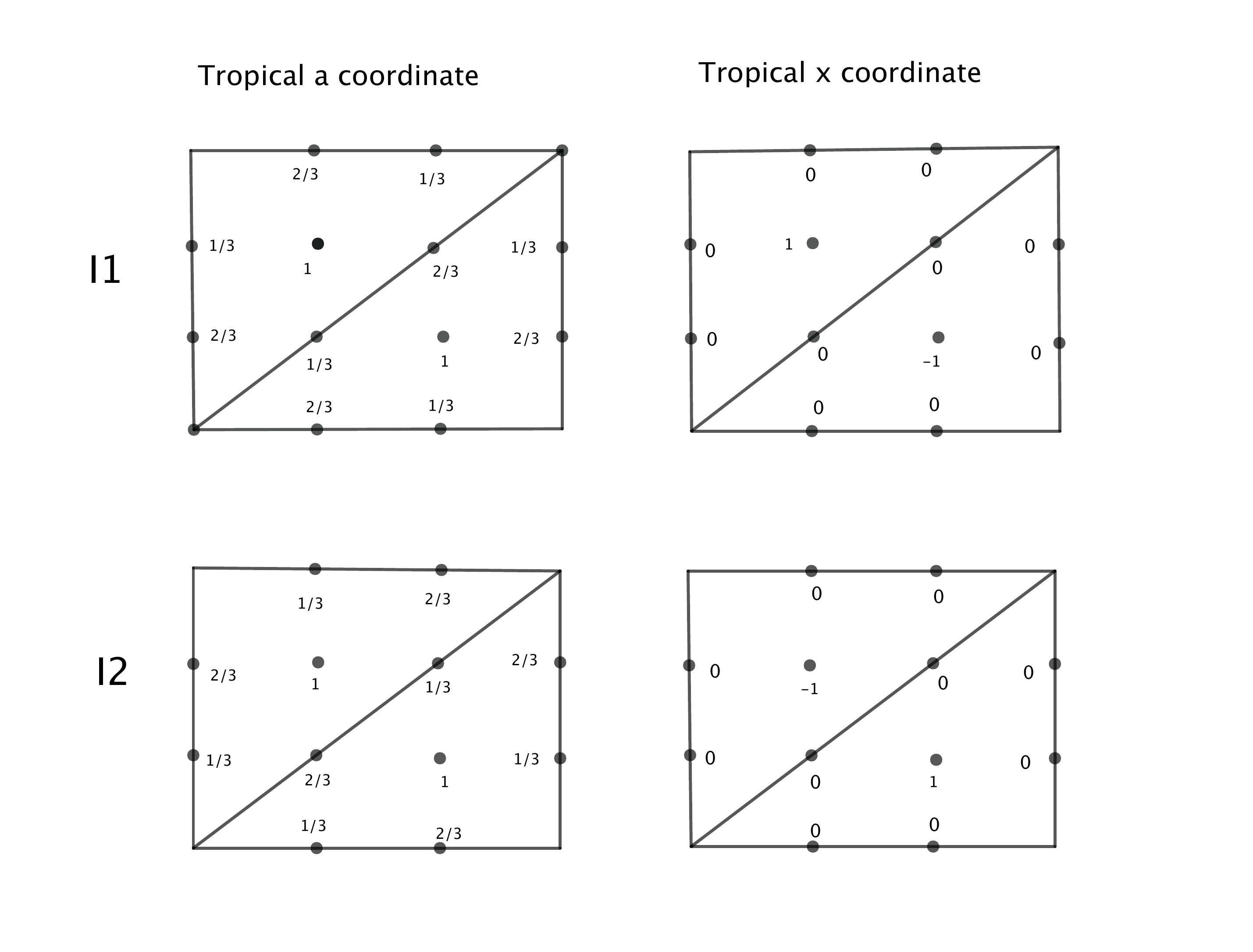}
\caption{The tropical $a$ and tropical $x$ coordinates for $I_1$ and $I_2$ which are found from OPE between the usual Wilson and 't Hooft line operators.}
\label{internal}
\end{figure}
The coordinates of two new line operators are shown in figure. \ref{internal}.
With these two new line operators, we have the following conjecture regarding the generators of the line operator:

\textbf{Conjecture}: The line operators of $A_2$ theory on torus with one full puncture are generated by the following eight nontrivial elements:
\begin{equation}
\mbox{L}_1, \mbox{L}_2, \mbox{M}_1, \mbox{M}_2, \mbox{C}_1, \mbox{C}_2, \mbox{I}_1, \mbox{I}_2;
\end{equation}
By generators, I mean every  nontrivial line operator with positive \footnote{If the initial tropical coordinates is negative, one can add the line operator with label $(n,n)$
around the puncture to shift the $a$ coordinates by a common positive constant,  this will make the new $a$ coordinates positive. 
So the non-trivial part is the line operator with positive tropical $a$ coordinates, which is the same
as $A_1$ case. }
 tropical $a$ coordinates can be formed by taking 
the product of them and take the leading order term in the OPE:
\begin{equation}
l=\prod \mbox{G}_i^{n_i},
\end{equation}
where $G_i$ are the above generators.  The tropical data for these line operators are listed in table \ref{gene}.

\begin{table}
\begin{center}
\begin{tabular}{| l | c | r| }   
  \hline                        
   ~& Tropical a coordinates & Tropical x coordinates \\ \hline
  $L_1$ & $ a_0^{2/3} a_1^{1/3} b_0^{2/3} b_1^{1/3}x^{1/3} y^{2/3}$ & $(-1,0,1,0,0,0,0,0)$ \\ \hline
  $L_2$ & $a_0^{1/3} a_1^{2/3} b_0^{1/3} b_1^{2/3}x^{2/3} y^{1/3} $ & $(0,-1,0,1,0,0,0,0)$ \\ \hline
    $M_1$ & $ b_0^{2/3} b_1^{1/3} c_0^{2/3} c_1^{1/3}x^{2/3} y^{1/3}$ & $(0,0,-1,0,1,0,0,0)$ \\ \hline
  $M_2$ & $ b_0^{1/3} b_1^{2/3} c_0^{1/3} c_1^{2/3}x^{1/3} y^{2/3}$ & $(0,0,0,-1,0,1,0,0)$ \\ \hline
  $C_1$ & $ a_0^{2/3} a_1^{1/3} c_0^{1/3} c_1^{2/3}x^{2/3} y^{1/3}$ & $(1,0,0,0,0,-1,0,0)$ \\ \hline
  $C_2$ & $ a_0^{1/3} a_1^{2/3} c_0^{2/3} c_1^{1/3}x^{1/3} y^{2/3}$ & $(0,1,0,0,-1,0,0,0)$ \\ \hline
  $I_1$ & $a_0^{2/3} a_1^{1/3} b_0^{1/3} b_1^{2/3} c_0^{2/3} c_1^{1/3} x y$ & $(0,0,0,0,0,0,-1,1)$ \\ \hline
  $I_2$ & $a_0^{1/3} a_1^{2/3} b_0^{2/3} b_1^{1/3} c_0^{1/3} c_1^{2/3}x y $ & $(0,0,0,0,0,0,1,-1)$ \\ \hline
  \hline  
\end{tabular}
\end{center}
  \caption{The tropical $a$ and $x$ coordinates for eight generators.}
    \label{gene}
\end{table}

\newpage
\subsubsection{Webs and Skein relations}
We found these extra line operators by doing  OPE, which is purely algebraic as we use the canonical map.  One may wonder if there is 
a geometric structure for them as line operators of 4d $\mathcal{N}=2$ theory
should be represented by M2 branes wrapped on one skeleton on Riemann surface. 
We have already used all the closed curves, so new structure is needed. 

\begin{figure}[htbp]
\small
\centering
\includegraphics[width=8cm]{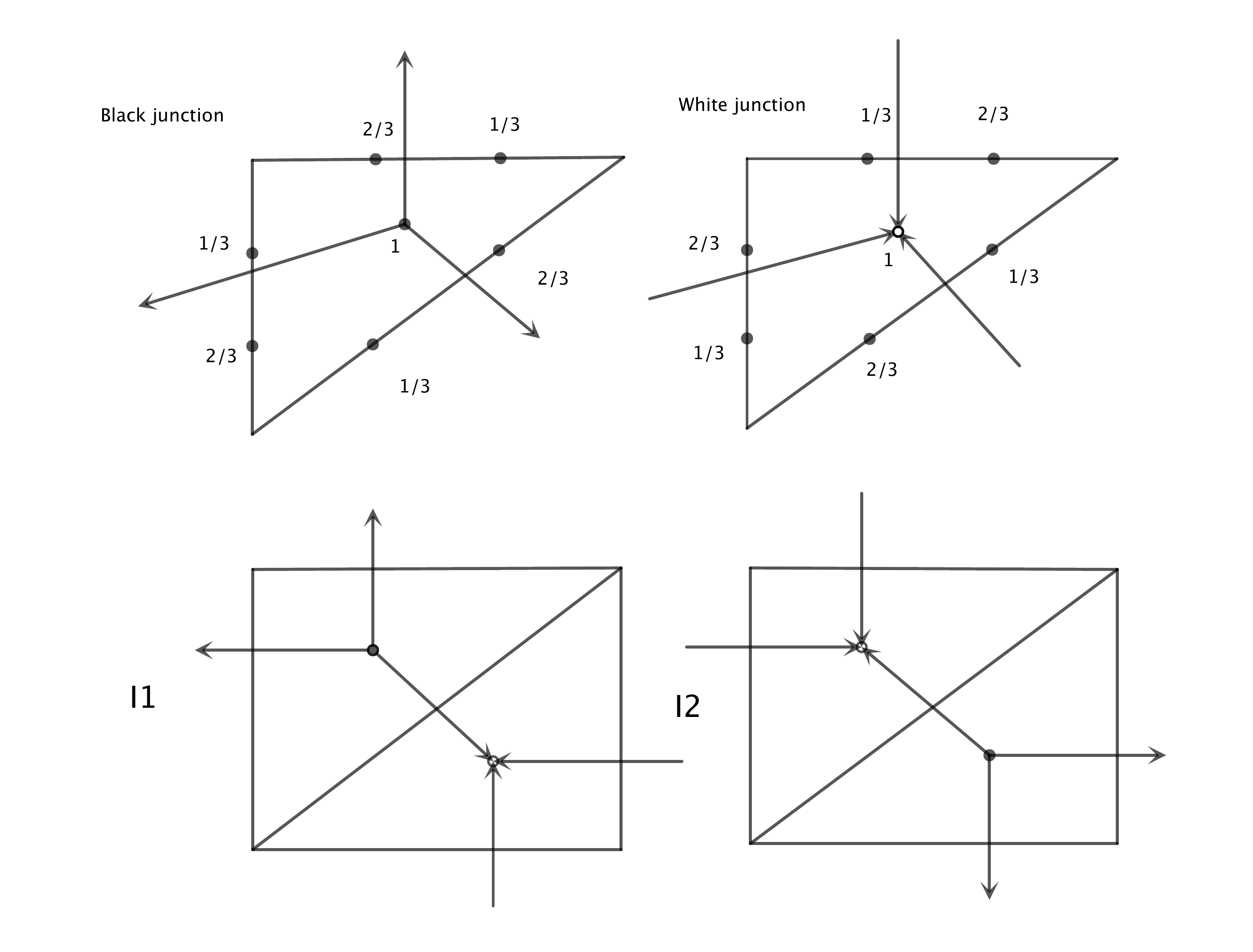}
\caption{Top: the junction and its tropical $a$ coordinates on a dual triangle. Bottom: webs representing $\text{I}_1$ and $\text{I}_2$.}
\label{junction}
\end{figure}

It seems the new structures are webs formed by  three junctions, i.e. the $\text{I}_1$ is formed by connecting two three junctions. 
Since the extra line operator has nontrivial tropical $x$ coordinate on the internal quiver node, it is natural to put the
junction at the internal quiver nodes, and it is useful to 
label the junction as black if the $x$ coordinate is positive and white if the $x$ coordinate is negative. 
Based on the tropical $a$ coordinates of the line operator $\text{I}_1$ and $\text{I}_2$,  it is then straightforward to find the $a$ coordinate for the dual triangle  
of these two junctions, and the result is shown in figure. \ref{junction}.

\begin{figure}[htbp]
\small
\centering
\includegraphics[width=10cm]{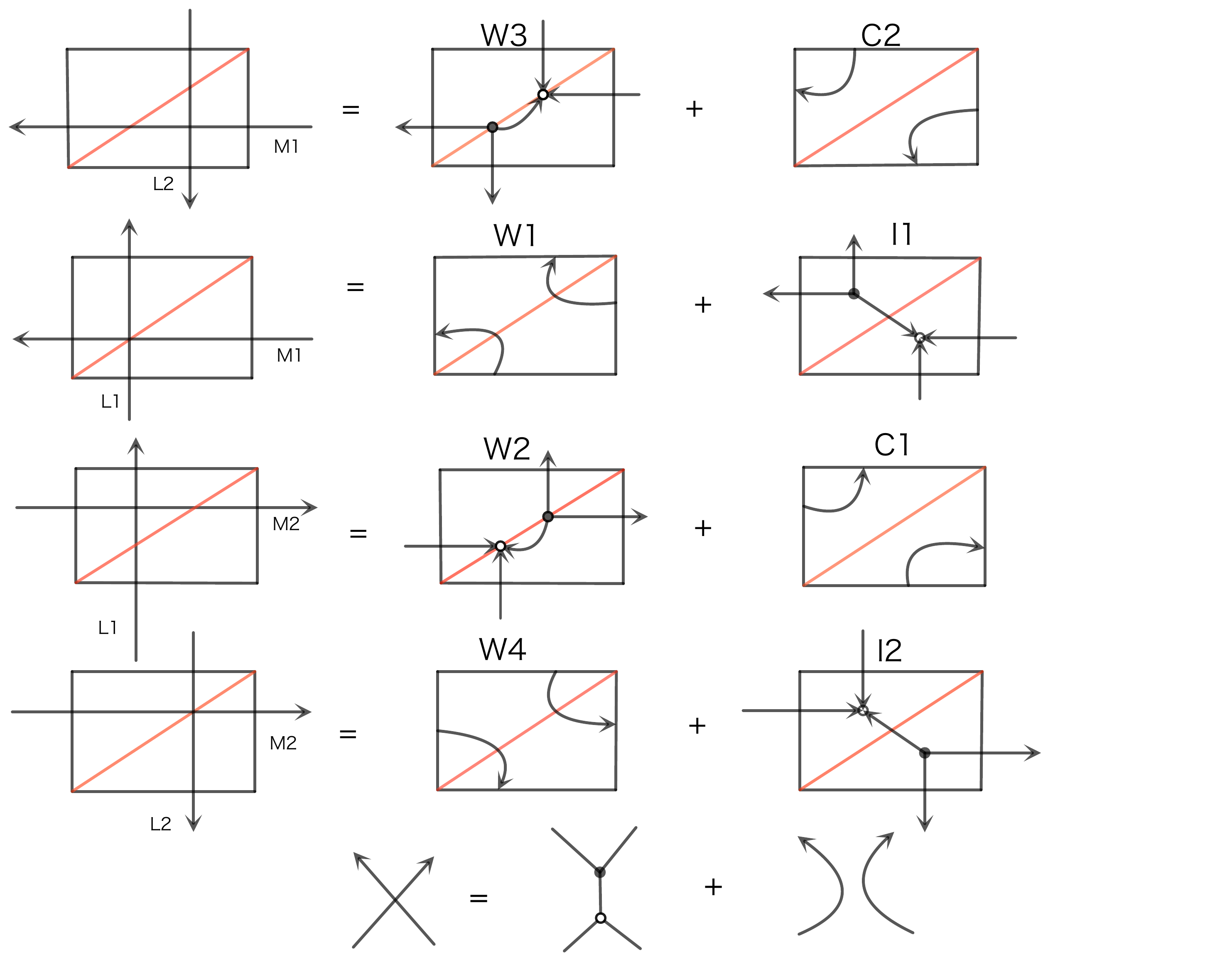}
\caption{Top: The geometric representation for the OPE between the Wilson and 't Hooft line operators. Bottom: the above OPE can be
derived by using this simple Skein relation. }
\label{OPE2}
\end{figure}

Up to now, the junctions and webs are just a good way of representing the new extra line operator. To make it as something really fundamental, we 
are going to do OPE for  other line operators represented by the closed curves, and the result is shown in figure. \ref{OPE2}. 
$\text{W}_2$ (as well as $\text{W}_3$) is represented by bipartite webs, and its tropical $a$ coordinates and canonical map can be 
easily found:
\begin{equation}
\text{I}(\text{W}_2)=a_0^{1/3} a_1^{1/3} b_0 b_1 c_0^{1/3} c_1^{2/3}x^{2/3} y^{4/3}+\ldots
\label{w2}
\end{equation}

\begin{figure}[htbp]
\small
\centering
\includegraphics[width=8cm]{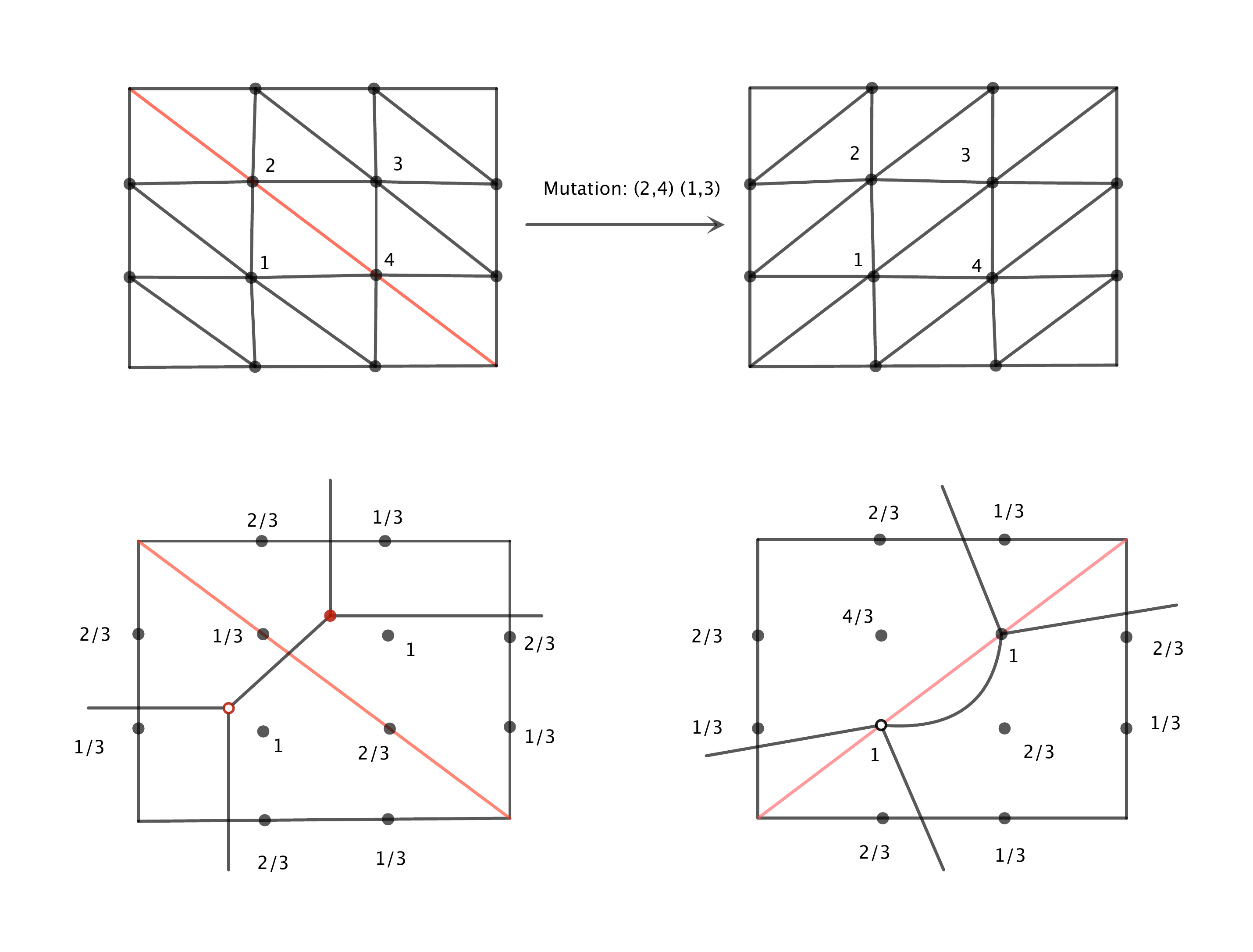}
\caption{Top: 
the two quivers of two different triangulations are related by the quiver mutations $(\mu_2, \mu_4), (\mu_1, \mu_3)$.
Bottom: The transformation of the tropical $a$ coordinates of the $W_2$ from one triangulation to another triangulation. }
\label{check}
\end{figure}

Now a very strong check of the web structure is  following: we can use another coordinate system by doing mutation on edge $b$ and write down 
the tropical $a$ coordinates of $\text{W}_2$ using the junction rule indicated in figure. \ref{junction}, 
then we mutate back using the 
mutation sequences $(\mu_2, \mu_4), (\mu_1, \mu_3)$, see figure. \ref{check}.  The $a$ coordinates of $W_2$ in  original coordinate system can be found using the mutation formula, 
and remarkably, the answer is the same as given by (\ref{w2}).
One can check the transformation of the canonical map and they are actually related by the mutations!

\begin{figure}[htbp]
\small
\centering
\includegraphics[width=12cm]{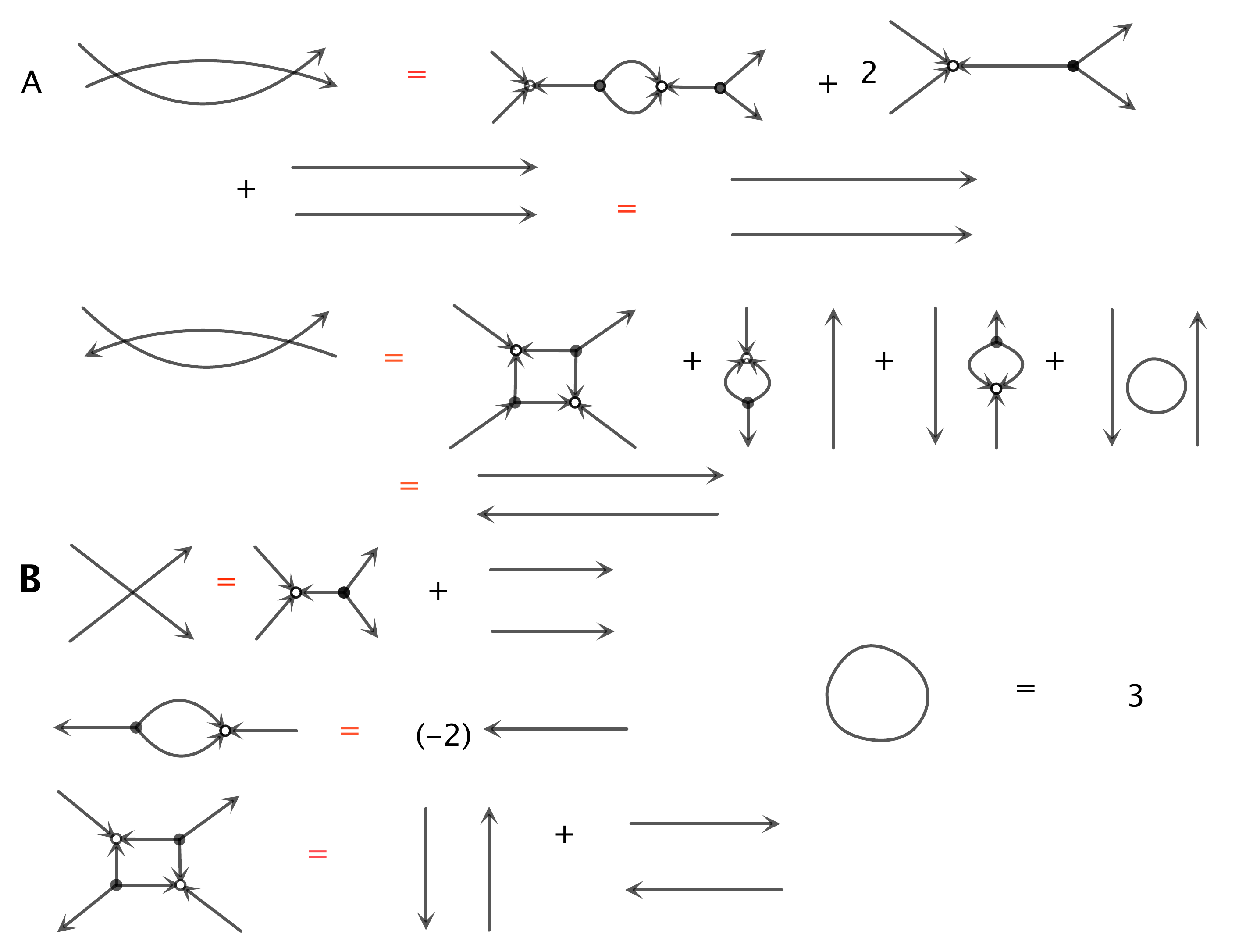}
\caption{A: The derivation of the bubble reduction and the square reduction. B: The full Skein relations of $A_2$ theory.}
\label{sk2}
\end{figure}

By examining the OPE shown in figure. \ref{OPE2}, it is not hard to realize that they can be derived from a local Skein relation shown at the bottom of figure. \ref{OPE2}. 
One might think that  the OPE for any line operators can be found using this Skein relation. This is reasonable as one can 
do following: first draw the closed curves (or webs) for the line 
operators appearing in the operator product, and use the Skein relation  to eliminate all four crossings, etc.
However, the above story is not entirely correct, since there is a very crucial thing we want to address: there are trivial crossings between two closed curves, we'd better make sure 
such configurations are trivial and this would imposed the equivalence condition on the webs. By analyzing the equivalence conditions shown in figure. \ref{sk2},
 we get the Skein relation in figure. \ref{sk2}, which we call as  bubble reduction and square reduction respectively.  So an irreducible web diagram has no two cycles and 
four cycles, and we conjecture that these are the full Skein relations.

In the following, we will ignore the orientation of the edges in the web diagram, and it is easy to recover the orientations using the rules: the arrow is going from black junction to white junction. 
Using the above Skein relation, we have the following geometric picture for line operators for $A_2$ theory
on once punctured torus: all the line operators can be realized as the bipartite webs without two cycles and four cycles on the torus\footnote{Of course, if there is a marked point inside  two or four cycles,
they are allowed; we can only remove those contractible two and four cycles.}.
Let's look at two simple examples of using Skein relations to find  OPE.

\begin{figure}[htbp]
\small
\centering
\includegraphics[width=10cm]{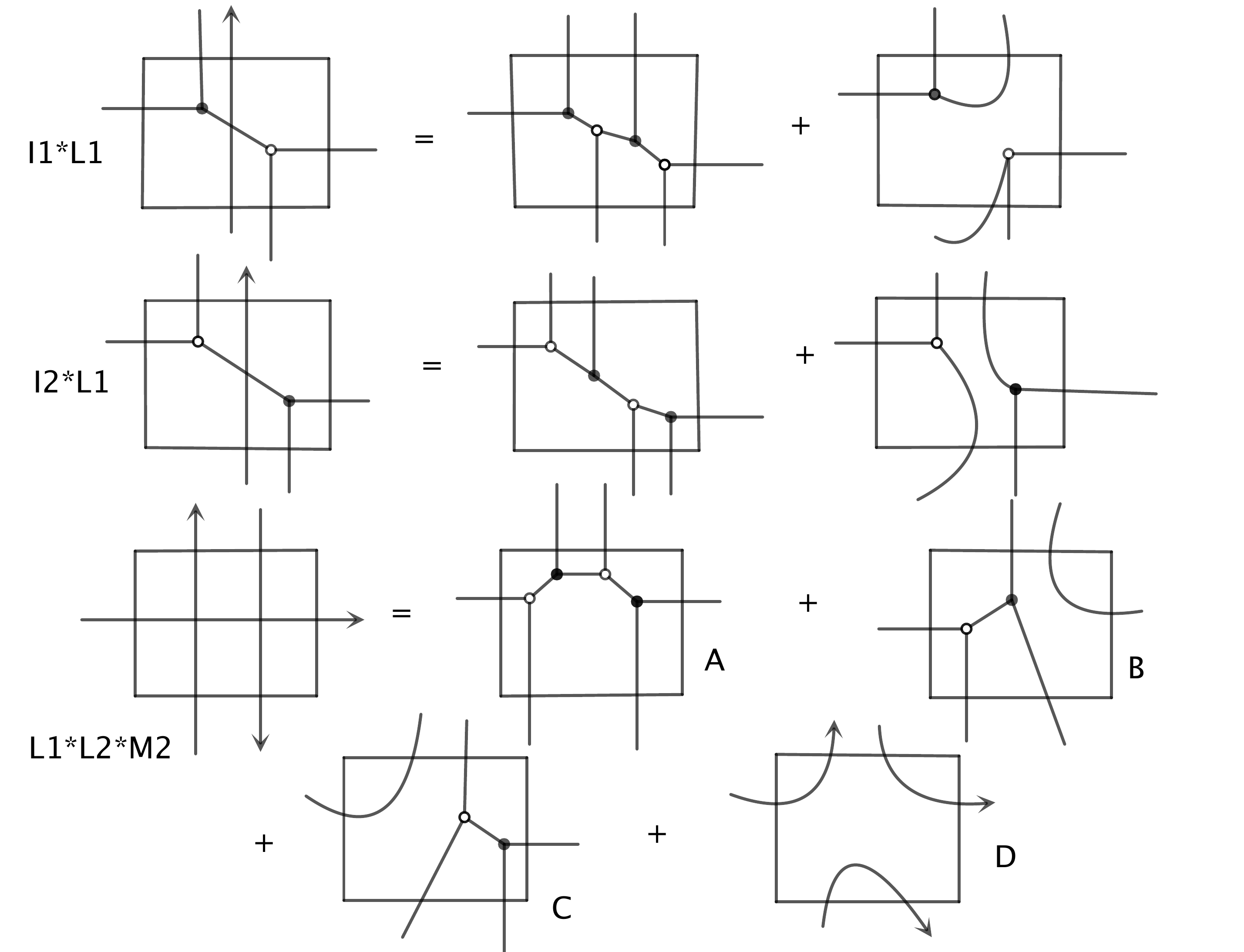}
\caption{Upper: OPE of $\mbox{I}(I_1)*\text{I}(L_1)$. Middle: OPE of $\text{I}(I_2)*\text{I}(L_1)$. Bottom: The OPE of  $\mbox{I}(L_1)*\mbox{I}(L_2)*\mbox{I}(M_2)$.}
\label{exam1}
\end{figure}

\textbf{Example 1}: Let's consider the OPE of   $\mbox{I}_1$ and $\mbox{L}_1$, and    $\mbox{I}_2$ and $\mbox{L}_1$. Use 
the Skein relation,  one have two line operators in the OPE, see figure. \ref{exam1}.

\textbf{Example 2}: Let's study the OPE of $L_1L_2M_2$ using Skein relation, and the result is shown in figure. \ref{exam1}:
\begin{equation}
\mbox{I}(\text{L}_1)*\mbox{I}(\text{L}_2)*\mbox{I}(\text{M}_2)=\mbox{I}(\text{A})+\mbox{I}(\text{B})+\mbox{I}(\text{C})+3\mbox{I}(\mbox{M}_2),
\end{equation}
here $A$ is the top component of $\mbox{I}(\text{I}_2)*\text{I}(\mbox{L}_1)$, B is the top component of $\mbox{I}(\text{L}_1)*\mbox{I}(\text{L}_2)*\mbox{I}(\text{M}_2)$, C is the top component of 
$\text{I}(\mbox{L}_2)*\text{I}(\mbox{C}_1)$, and D is just $\text{M}_2$. Notice that the coefficient $3$ before ${\text{M}}_2$ is invisible from using the Skein relation. 

The above two examples are in perfect agreement with the  calculation using the explicit canonical map. It is amazing that the simple Skein relation reproduces the correct 
OPE, which strongly suggest that the geometric picture we found for the general line operators are correct!

\subsubsection{Construction and reconstruction}
The web construction is based on the assumption that the line operators are built from eight generators (the leading component in OPE). The construction of 
web from $a$ coordinates is rather indirect: we need to first find the combinations of  generators and then do OPE. Moreover, we do not know what kind of 
constraints on $a$ coordinates are needed to have a geometric web representation.  So it is important to find a way of constructing webs directly 
from tropical $a$ coordinate, which is also very useful for dealing with line operators on general Riemann surface. Indeed, we find  similar reconstruction procedure as  
$A_1$ case. 

Let's consider a triangle in the triangulation and assume the tropical $a$ coordinates on the quiver nodes as $\tilde{a}_i$; 
We further  assume that the dual $x$ coordinates of the internal 
quiver nodes to be positive.  We multiple each tropical coordinates by 3, i.e. $a_0=3\tilde{a}_0$, such that all the coordinates are integers now.

\begin{figure}[htbp]
\small
\centering
\includegraphics[width=10cm]{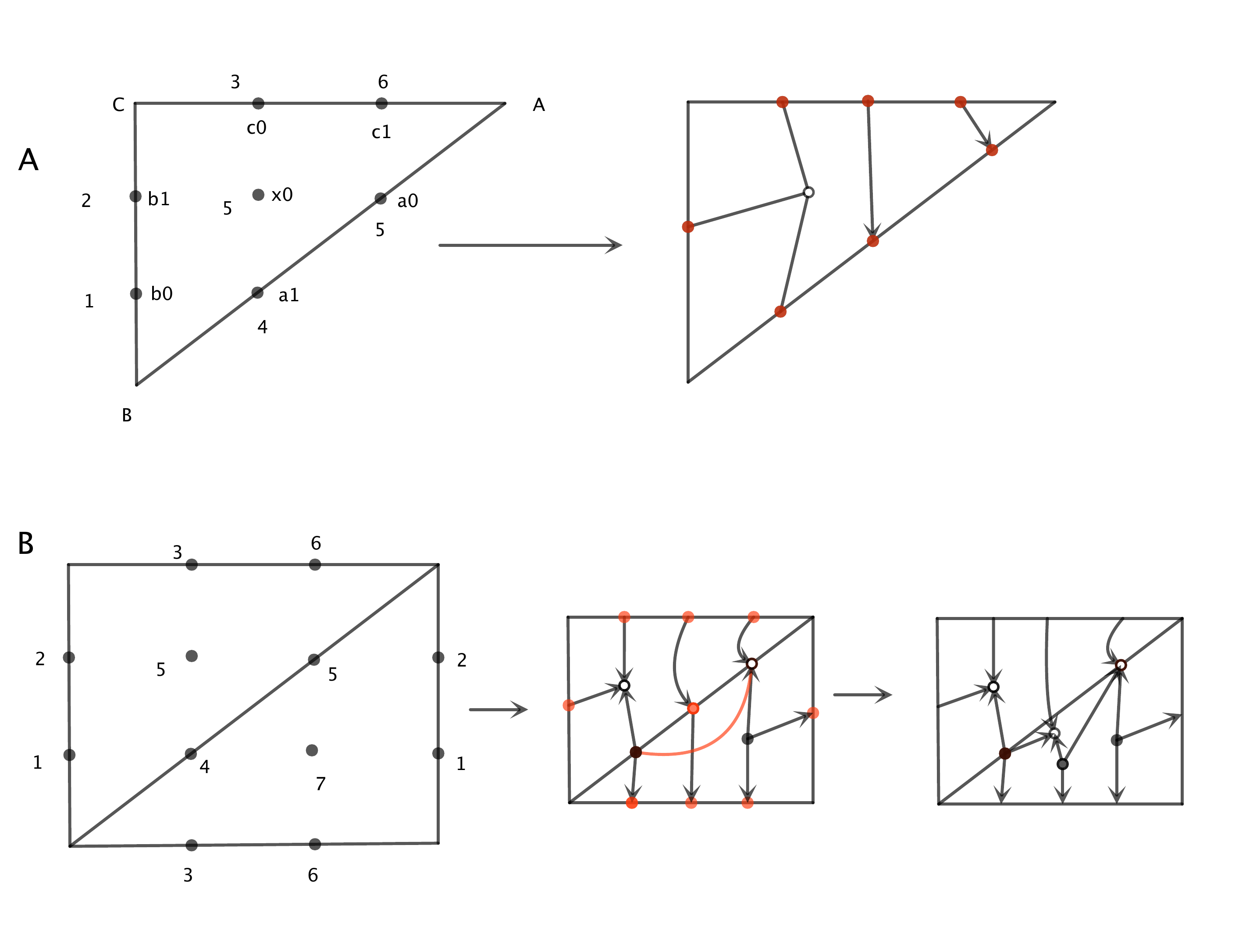}
\caption{Top: construct curves from tropical $a$ coordinates on a triangle. Bottom: construct webs on a torus from tropical $a$ coordinates.}
\label{a2recon}
\end{figure}

The idea of reconstruction is to construct set of non-intersecting curves (junction) inside the triangle such that they give the desired $a$ coordinates. There 
are actually seven kinds of curves: two curves around each vertex and one junction (only one type of junctions is allowed as we will see shortly). So
let's assume there are $n_1$ ($n_2$) anti-clockwise (clockwise) curves around vertex $A$, and we use the similar notation for other two vertices;
finally let's assume there are $J$ black junctions inside the triangle. Summing up the contribution of each curve to the tropical $a$ coordinate on seven quiver nodes, we have
the following equations:
\begin{align}
& 2n_1+n_2+m_1+2m_2+2J=a_0,~~n_1+2n_2+2m_1+m_2+J=a_1,    \nonumber\\
& 2m_1+m_2+l_1+2l_2+2J=b_0,~~m_1+2m_2+2l_1+l_2+J=b_1,   \nonumber\\
& 2l_1+l_2+n_1+2n_2+2J=c_0,~~~~~l_1+2l_2+2n_1+n_2+J=c_1, \nonumber\\
& n_1+2n_2+m_1+2m_2+l_1+2l_2+3J=x_0.
\end{align}
The solution is uniquely fixed by the tropical $a$ coordinates:
\begin{align}
&n_1= \frac{1}{3} (a_0+c_1- x_0),~
n_2= \frac{1}{3} (-a_0+a_1-b_0+x_0),~ \nonumber\\
&m_1= \frac{1}{3} (a_1 + b_0 - x_0),~
m_2= \frac{1}{3} (-b_0 + b_1 - c_0 + x_0),\nonumber\\
&l_1=\frac{1}{3} (b_1 + c_0 - x_0),~~
l_2= \frac{1}{3} (-a_0 - c_0 + c_1 + x_0),~J={1\over 3}({a_0-a_1+b_0-b_1+c_0-c_1}).
\label{sol1}
\end{align}
Interestingly, $J$ is the tropical $x$ coordinates of the internal quiver nodes and therefore is a positive integer, which should be the case since
only the junction can carry the tropical $x$ charge of the internal quiver nodes.
Notice that the number of marked points on one edge $AB$ is equal to 
 \begin{equation}
 n_1+n_2+J+m_1+m_2={1\over3}(a_0+a_1)=\tilde{a}_1+\tilde{a}_0,
\end{equation}
which actually only depends on the coordinate on the edge $AB$. Moreover, the number of curves coming into  $AB$ are $n_1+m_2+J=2\tilde{a}_0-\tilde{a}_1$,
and the number of outgoing curves is $n_2+m_1=2\tilde{a}_1-\tilde{a}_0$, which also only depend on the coordinates on the edge.
To make all these numbers to be positive integer, we need to put the following constraints on the tropical $a$ coordinates on the edge:
\begin{align}
&\tilde{a}_0+\tilde{a}_1\in z \nonumber\\
&\tilde{a}_0\leq2\tilde{a}_1,~~\tilde{a}_1\leq2\tilde{a}_0.
\label{con1}
\end{align}
Moreover, we need the following constraints:
\begin{equation}
\tilde{\alpha}+\tilde{\beta}+2\tilde{\gamma}\in z,~\tilde{\gamma}\leq \tilde{\alpha}+\tilde{\beta}
\label{con2}
\end{equation}
where $\tilde{\gamma}$ is the coordinate on  internal quiver node, and $\tilde{\alpha}, \tilde{\beta}$ are  quiver nodes
in the same quiver triangle as $\tilde{\gamma}$, say $(\tilde{\alpha},\tilde{\beta},\tilde{\gamma})=(a_0,c_1, x_0)$ in figure. \ref{a2recon}.
One can easily see that above three constraints would ensure 
the solutions in (\ref{sol1}) are all positive integers! 
Similarly, if the tropical $x$ coordinate of the internal quiver node is negative, we have the solution:
\begin{align}
&n_1= \frac{1}{3} (a_0+c_1- x_0),~
n_2= \frac{1}{3} (c_0-c_1-b_1+x_0),~ \nonumber\\
&m_1= \frac{1}{3} (a_1 + b_0 - x_0),~
m_2= \frac{1}{3} (a_0 - a_1 - c_1 + x_0),\nonumber\\
&l_1=\frac{1}{3} (b_1 + c_0 - x_0),~~
l_2= \frac{1}{3} (-a_1 + b_0 - b_1 + x_0),~J={1\over 3}({-a_0+a_1-b_0+b_1-c_0+c_1}),
\label{sol2}
\end{align}
and $J$ is equal to the minus of the internal tropical $x$ coordinate and therefore is positive. The constraints are the same as the above case.

\textbf{Example}: Let's look at an example, the coordinates on the triangle are
\begin{equation}
(a_0,a_1,b_0,b_1,c_0,c_1,x_0)=(5,4,1,2,3,6,5),  
\end{equation}
see figure. \ref{a2recon}, so the marked points on each edge are $(AB,BC,CA)=(3,1,3)$. Since the tropical $x$ coordinates
of internal quiver node is negative, we need to use solution (\ref{sol2}). 
Substitute the coordinates  into the  formula, we have
\begin{equation}
(n_1,n_2,m_1,m_2,l_1,l_2,J)=(2,0,0,0,0,0,1),
\end{equation}
so there is one white junction inside triangle and two clockwise curves around vertex $A$.

\begin{figure}[htbp]
\small
\centering
\includegraphics[width=10cm]{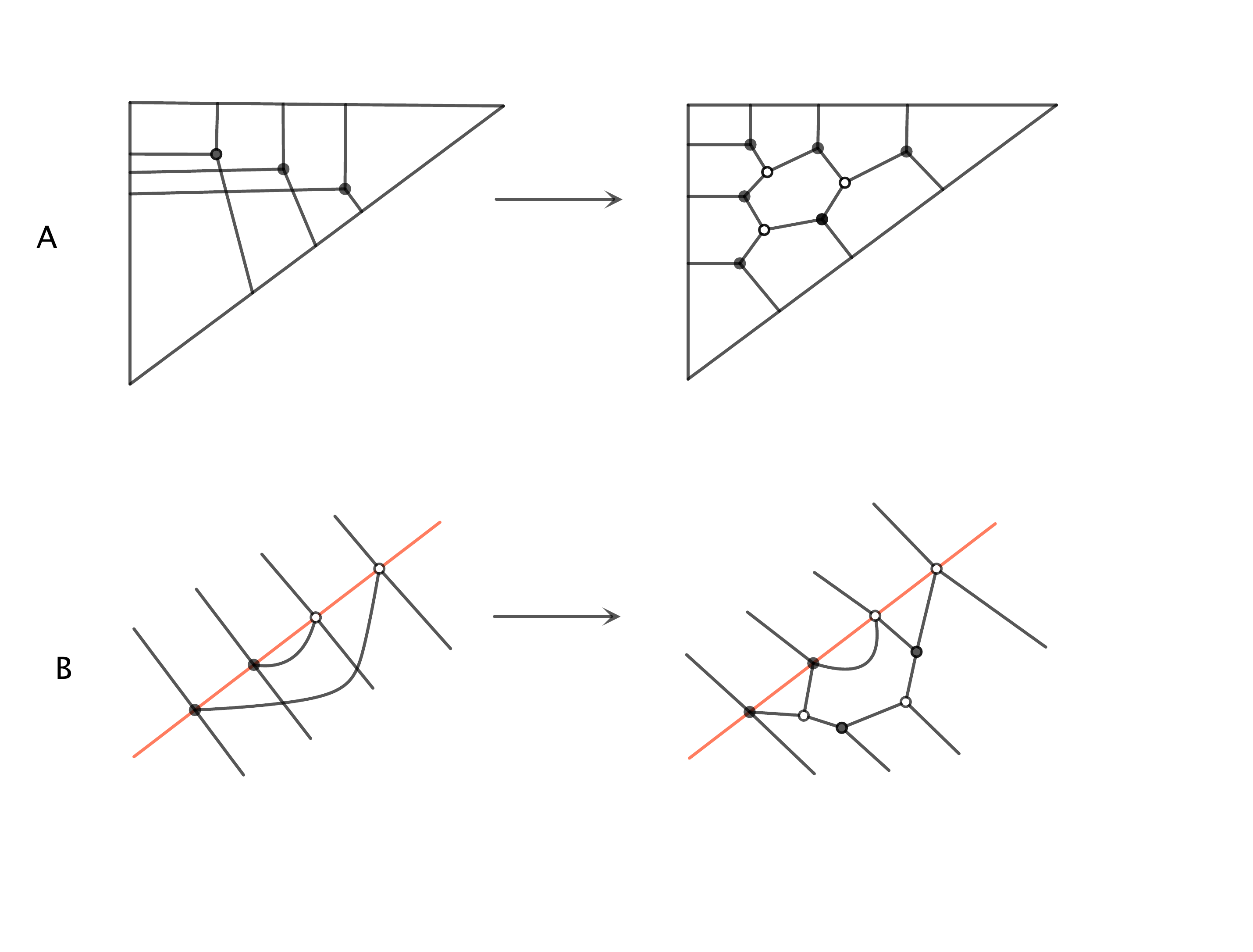}
\caption{Top: resolve the crossing of the three junctions inside the triangle. Bottom: resolve the crossing around the edges.}
\label{resolution}
\end{figure}

Now given a set of $a$ coordinates, we do the above reconstruction for each triangle, then 
we can form a set of closed non-intersecting webs and closed curves on  Riemann surface.
 However, there are two complications we need to address: firstly the web has crossings if there are more than one three junctions inside
the triangle; and secondly the marked points on the edge can be source (two arrows come out) or sinks (two arrows coming in). 
The first complication is easy to address: one replace each crossing by a piece with two junctions using the Skein relation, see an example in figure. \ref{resolution}. One 
always produce an irreducible sub web in this step as all the closed surfaces are hexagons.

The second complication is a little bit difficult to deal with. First, let's color a marked points as black (white) if it is source (sink), and 
it is not hard to observe the following fact: the number of black and white vertices are the same. We can then connect the white
and black vertices in pairs and produce a lot of crossings. The resolution of the crossings are different based on the origin of two crossing lines:
\begin{itemize}
\item  If the two lines coming from the vertices with different color, we keep the second term (two curves) in the Skein relation.
\item If the two lines coming from the vertices with same color, we keep the first term (two junctions) in the Skein relation.
\item If one of two crossing line is a ordinary line connecting two internal junctions,  we keep the second term in the Skein relation; otherwise, we 
keep the second term. 
\end{itemize}
Using the above rules, we can produce an irreducible bipartite web around the edges. However, we do not know 
whether this procedure will produce an irreducible web in gluing the triangles together. This is true for all the examples we check, and 
it would be nice to have a proof though.

\begin{figure}[htbp]
\small
\centering
\includegraphics[width=10cm]{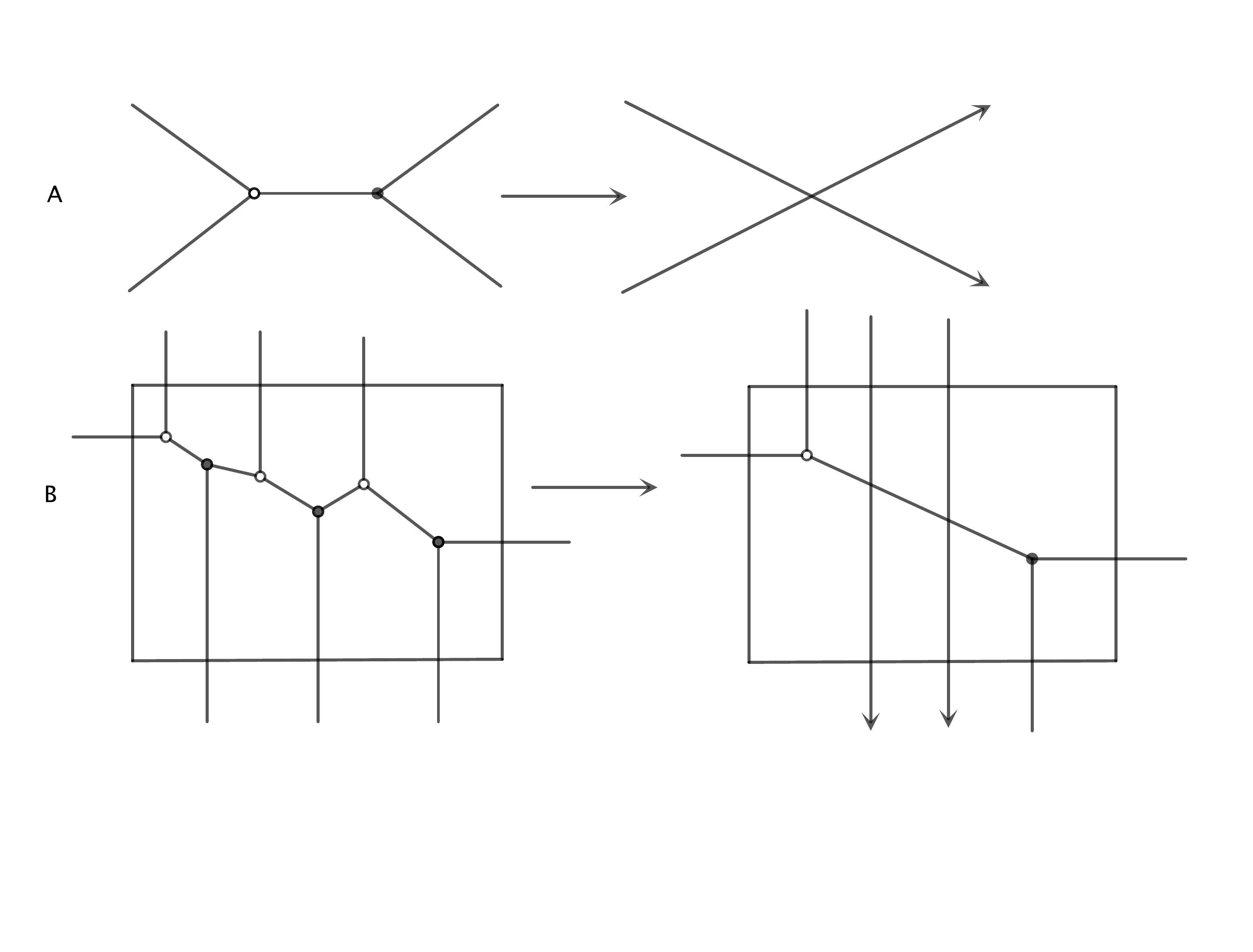}
\caption{A: replace two junctions with the crossing. B: We replace a web with a set of generators from which the $a$ coordinates of the initial web is read.}
\label{A2con}
\end{figure}

In the above consideration, we start with a set of tropical $a$ coordinates and then construct an irreducible web. On the other hand, if we start with a web, we would like to find its
tropical $a$ coordinates. This can be found by doing reverse Skein relation, namely one replace two junctions with a four crossings, see figure. \ref{A2con}\footnote{Although the ordering of dissolving junctions 
does not matter, in practice it is better to start with the junction at the corner to avoid producing some weird diagram.}. 
The above procedure stops if four legs of the two junctions are on the boundary. At the end, we are left with the basic building blocks: closed curves and $\text{I}_1,\text{I}_2$. 
By taking a triangulation, one know the cluster coordinates for each individual component, and the coordinates are just the sum of that 
of components. 

For example, according to above method, the web diagram in figure. \ref{A2con} is reduced to three generators $(\text{L}_2,\text{L}_2, \text{I}_2)$. So the original web
 is the top component of the product $\mbox{I}_2 *\mbox{L}_2 *\mbox{L}_2$ and 
 its $a$ coordinates are equal to $\text{Trop}(\mbox{I}_2 )+\text{Trop}(\mbox{L}_2 )+\text{Trop}(\mbox{L}_2)$, which actually equal to the $a$ coordinates we 
 used in last example. 
The result is in perfect agreement with the one shown in figure. \ref{a2recon}, in which we construct the same web from  same tropical $a$ coordinates. 
 Moreover, once the set of generators are found, one can find the canonical map of  initial web  by doing OPE.
Now we have established a one-to-one correspondence between the constrained tropical $a$ coordinates (see  [\ref{con1}, [\ref{con2}]) and
irreducible bipartite webs. This description is exactly the same 
as the $A_1$ case.

It is worth to remark that  bipartite network on bordered Riemann surface appears in many different physical contexts: 4d $\mathcal{N}=1$ superconformal field theory \cite{Xie:2012mr,Franco:2012mm},
BPS quiver for 4d $\mathcal{N}=2$ theory \cite{Xie:2012dw,Xie:2012jd}, and scattering amplitude of $\mathcal{N}=4$ SYM \cite{arkani2012scattering}. The web diagram we found here is of a different kind;
the equivalence conditions are different: we do not have square move, but we have bubble and square reduction.

\subsubsection{A further check: Poisson brackets}

Let's now verify that the basic line operators (eight generators) are indeed the coordinates for the cluster variety associated with 
the moduli space of the local system on a torus with one full puncture. The most
convincing check is to find the explicit Poisson brackets of these line operators, and show that 
there are indeed three commuting Hamiltonian as suggested by the underlying Hitchin integrable system.

The Poisson bracket of line operators can be easily found due to the remarkably simple Poisson structure for the cluster $X$ coordinates, which takes
the following simple form:
\begin{equation}
[X_i,X_j]=\epsilon_{ij}X_iX_j,
\end{equation}
here $\epsilon_{ij}$ is the antisymmetric tensor read from the quiver.
It is very easy to  calculate the Poisson bracket for two loop operators using the canonical map which is just a positive Laurent polynomial in $X_i$. 
First, for two monomials $A$ and $B$,  the Poisson bracket is simply 
\begin{equation}
[A,B]=(\sum_ia_{A}^ix_{B}^i)A*B,
\end{equation}
here $a_{A}$ is the tropical $a$ coordinates for the monomial $A$, and $B$ is the tropical $x$ coordinates 
for $B$. Using the above noted relation, we could easily calculate the Poisson structure between the 
eight generators using the explicit canonical map (See appendix B for their canonical maps.), and the Poisson brackets are
indeed closed:
\begin{align}
&[L_1,M_1]=-{2\over3}L_1M_1+I_1,~~~~[L_1,M_2]=-{1\over3}L_1M_2+C_1\nonumber\\
&[L_1,C_1]={2\over3}L_1C_1-I_2,~~~~~~~~[L_1,C_2]={1\over3}L_1C_2-M_1\nonumber\\
&[L_1,I_1]={1\over3}L_1I_1-L_2M_1+C_2,~~~~[L_1,I_2]=-{1\over3}L_1I_2+L_2C_1-M_2
\end{align}
\begin{align}
&[L_2,M_2]=-{2\over3}L_2M_2+I_2,~~~~[L_2,M_1]=-{1\over3}L_2M_1+C_2\nonumber\\
&[L_2,C_2]={2\over3}L_2C_2-I_1,~~~~~~~~[L_2,C_1]={1\over3}L_2C_1-M_2\nonumber\\
&[L_2,I_2]={1\over3}L_2I_2-L_1M_2+C_1,~~~~[L_2,I_1]=-{1\over3}L_2I_1+L_1C_2-M_1
\end{align}
The other brackets are very similar
\begin{align}
&[M_1, C_1]=-{1\over3} M_1 C_1+L_1,~~~~~[M_1,C_2]=-{2\over3} M_2C_2+I_2 \nonumber\\
&[M_1, I_1]=-{1\over 3}M_1L_1+L_1M_2-C_1,~~~~[M_1, I_2]={1\over3}M_1I_2-M_2C_2+L_2\nonumber\\
&[M_2,C_1]=-{2\over3}M_2C_1+I_2,~~~[M_2,C_2]=-{1\over3}M_2C_2+L_2\nonumber\\
&[M_2,I_1]={1\over3}M_2I_1-M_1C_1+L_1,~~[M_2,I_2]=-{1\over3}M_2I_2+L_2M_1-C_2
\end{align}
and
\begin{align}
&[C_1,I_1]=-{1\over3}C_1I_1+C_2M_2-L_2,~~[C_1, I_2]={1\over 3}C_1I_2-C_2L_1+M_1 \nonumber\\
&[C_2,I_1]={1\over3}C_2I_1-C_1L_2+M_1,~~~~~[C_2,I_2]=-{1\over3}C_2I_2+C_1M_1-L_1 \nonumber\\
&[I_1,I_2]=L_2M_1C_1-L_1M_2C_2.
\end{align}

There are actually two central elements which could be found from the above Poisson structure. One of them is 
\begin{align}
&I_1I_2-L_1 M_2 C_2-L_2 M_1 C_1 +C_1C_2+M_1M_2+L_1L_2 \nonumber\\
&= 3+m_1+m_2+m_1m_2+{1\over m_1}+{1\over m_2}+{1\over m_1m_2},
\end{align}
and $m_1, m_2$ is expressed as the Fock-Goncharov coordinates as
\begin{equation}
m_1=xy,~~~m_2=a_0 a_1 b_0 b_1 c_0 c_1 xy.
\end{equation}
Since $m_1$ and $m_2$ are the central elements of the Poisson bracket, and the above polynomial is also a central element.
The other central element is much more complicated:
\begin{align}
&L_1L_2M_1M_2C_1C_2+ L_1^3+L_2^3+M_1^3+M_2^3+C_1^3+C_2^3-I_1^3-I_2^3+3I_1 C_1 L_1 \nonumber\\
&+3I_2 C_2L_2+3I_1 L_2M_2+3I_2L_1M_1+3I_1M_2C_2+3I_2 M_1C_1-3L_1M_2C_2-3L_2M_1C_1\nonumber\\
&-2 C_2^2 L_2 M_1 - 2 C_1^2 L_1 M_2 - 2 C_1 L_1^2 M_1 - 2 C_2 L_1 M_1^2 - 2 C_2 L_2^2 M_2 - 2 C_1 L_2 M_2^2 \nonumber\\
&-C_2^2 I_1 L_1 - C_2^2 I_2 M_2 - C_1^2 I_2 L_2 - C_1^2 I_1 M_1 - L_1^2 C_2 I_2 -  L_1^2 I_1  M_2 - L_2^2 C_1 I_1 - \nonumber\\
&L_2^2 I_2  M_1 - M_1^2 C_1 I_2 - M_1^2 I_1 L_2 -   M_2^2 C_2 I_1 - M_2^2 I_2 L_1+I_1^2 L_1 M_1 + C_1 I_1^2 M_2 + C_2 I_1^2 L_2 + C_2 I_2^2 M_1 \nonumber\\
& + I_2^2 L_2 M_2 +  C_1 I_2^2 L_1 +C_1 C_2 L_1 L_2 + C_1 C_2 M_1 M_2 + L_1 L_2 M_1 M_2 + I_1 I_2 M_1 M_2 + I_1 I_2 L_1 L_2 + C_1 C_2 I_1 I_2 \nonumber\\
&C_2 L_2^2 M_1^2 + C_1 L_1^2 M_2^2 + C_1^2 L_2^2 M_2 + C_2^2 L_1^2 M_1 +  C_2^2 L_2 M_2^2 + C_1^2 L_1 M_1^2 \nonumber\\
&-C_1 C_2 I_2 L_1 M_1 - C_1 C_2 I_1 L_2 M_2 - C_1 I_2 L_1 L_2 M_2 - C_1 I_1 L_1 M_1 M_2 - 
 C_2 I_2 L_2 M_1 M_2-L_1 L_2 C_2 I_1 M_1 \nonumber\\
 &=12+\frac{6}{x y}+6 x y+\frac{1}{x^3 y^3 a_0^2 a_1^2 b_0^2 b_1^2 c_0^2 c_1^2}+\frac{1}{a_0 a_1 b_0 b_1 c_0 c_1}+\frac{1}{x^3 y^3 a_0 a_1 b_0 b_1 c_0 c_1}+
 \frac{6}{x^2 y^2 a_0 a_1 b_0 b_1 c_0 c_1}+\nonumber\\
&\frac{6}{x y a_0 a_1 b_0 b_1 c_0 c_1}+a_0 a_1 b_0 b_1 c_0 c_1+6 x y a_0 a_1 b_0 b_1 c_0 c_1+6 x^2 y^2 a_0 a_1 b_0 b_1 c_0 c_1\nonumber\\
&+x^3 y^3 a_0 a_1 b_0 b_1 c_0 c_1+x^3 y^3 a_0^2 a_1^2 b_0^2 b_1^2 c_0^2 c_1^2=\nonumber\\
&12+6({1\over m_1}+m_1+{1\over m_2}+m_2+{m_1\over m_2}+{m_2\over m_1})+{1\over m_2^2m_1}+m_2^2m_1+{1\over m_2^2m_1}+
{m_1\over m_2}+m_1^2m_2.
\end{align}
Although the above formula seems to be very complicated, it is actually very easy to find using the canonical map. The key observation is the following: the central element 
is represented by the closed curves around the puncture and their dual $x$ coordinate is zero! So if we can find the combinations of the simple generators such 
that their tropical $x$ coordinates are vanishing, then its leading component is definitely a central element! We actually have only two independent choices $I_1I_2$ and $L_1L_2M_1M_2C_1C_2$, which
give the leading term in the above central element.
The full expression is found by doing OPE of $\text{I}(I_1)*\text{I}(I_2)$ and the OPE has the following form:
\begin{equation}
\text{I}(I_1)*\text{I}(I_2)=\text{Central~terms}+\sum \prod_i G_i^{n_i};
\end{equation}
and the polynomial $(I_1*I_2-\sum \prod_iG_i^{n_i})$ is the central element.
 We also find the following three commuting Hamiltonians:
\begin{align}
& H_1=L_1*L_2, \nonumber\\
&H_2= L_1^3+L_2^3, \nonumber\\
& H_3=L_1^2C_1M_1 +L_2^2 C_2 M_2+I_1 I_2 +M_1 M_2 +C_1 C_2+L_1L_2 \nonumber\\
&-(I_2 C_2 L_2 + M_2 L_2 I_1+I_1C_1L_1+M_1L_1I_2 ).
\end{align}
In fact, one can choose any linear combinations of $L_1$ and $L_2$ since they both commute with $H_3$. 
The Poisson bracket for eight coordinates $L_i, M_i, C_i, I_i$ are closed, so they are good  coordinates for
the Poisson manifold.  We have found two central elements and three commuting Hamiltonians, therefore it is an integrable 
system as it should be. The action variable can be found such that the symplectic structure has the canonical form
\begin{equation}
\omega=\sum_i dH_i \wedge d\theta_i.
\end{equation}
The detailed form for $\theta$ coordinates will be presented elsewhere. 
These explicit  calculations provide very strong evidence that we have the right answer for
4d $\mathcal{N}=2$ line operators!

\subsection{General N}
\subsubsection{Basic building blocks}
Let's consider line operators of 4d theory defined using general $A_{N-1}$ $(2,0)$ theory on once punctured torus.
The $\mathcal{N}=2$ theory in one duality frame is described by gauging the diagonal of 
 two $\mbox{SU}(N)$ flavor symmetries of 
$\mbox{T}_N$ theory. The $\mbox{T}_N$ theory has $(N-1)(N-2)\over 2$ Coulomb branch operators, and the $\text{SU}(N)$ gauge group has $(N-1)$ Coulomb 
branch operators;  There are also $(N-1)$ mass parameters carried by the full puncture,
so the rank of the charge lattice is 
\begin{equation}
R=(N-1)(N-2)+2(N-1)+(N-1)=N^2-1.
\end{equation}

The cluster coordinates are  found by Fock and Goncharov: first take an ideal triangulation of the once punctured torus
and then put  $(N-1)(N-2)\over 2$ more quiver nodes inside each triangle.
The total number of quiver nodes is equal to the rank of the charge lattice of 4d theory, for more details, see \cite{Xie:2012dw,Xie:2012jd}.

The ordinary Wilson lines  are represented by the closed curves on the Riemann surface. The tropical $a$ coordinates for Wilson loop 
in fundamental representation $\omega_1$ and anti-fundamental representation $\omega_{N-1}$ 
are found by calculating the monodromy $\text{M}$ and $\text{M}^{-1}$ around the simple closed  curve using the similar combinatorial method as $A_1$ and $A_2$ case, 
say  appendix A for more details. 

Once the canonical map (the positive Laurent polynomial) is found, its tropical $a$ coordinates are read from the leading order term.
There are other fundamental representations whose tropical $a$ coordinates and the canonical map can be
 found by doing the OPEs of the two basic fundamentals.  A useful fact is that the canonical map of the representation $(n,0,\ldots,0)$ and $(0,\ldots,0,n)$ can be found by taking the trace 
of the monodromy matrices $\text{M}^n$ and $\text{M}^{-n}$.   For example, we can do the product $\text{I}(\omega_1)*\text{I}(\omega_1)$, where $\omega_1$ is
the fundamental representation, and the same tensor product from the group theory tells us that the OPE has two components $(2,0,0\ldots,0)$ and $(0,1,\ldots,0)$, since
we know the canonical map for representation $(2,0,0\ldots,0)$, we can easily find the canonical map for the second fundamental representation and its $a$ coordinates. 
Using OPE, one can find the tropical $a$ coordinates for all fundamental representation $\omega_i$.
\begin{figure}[htbp]
\small
\centering
\includegraphics[width=10cm]{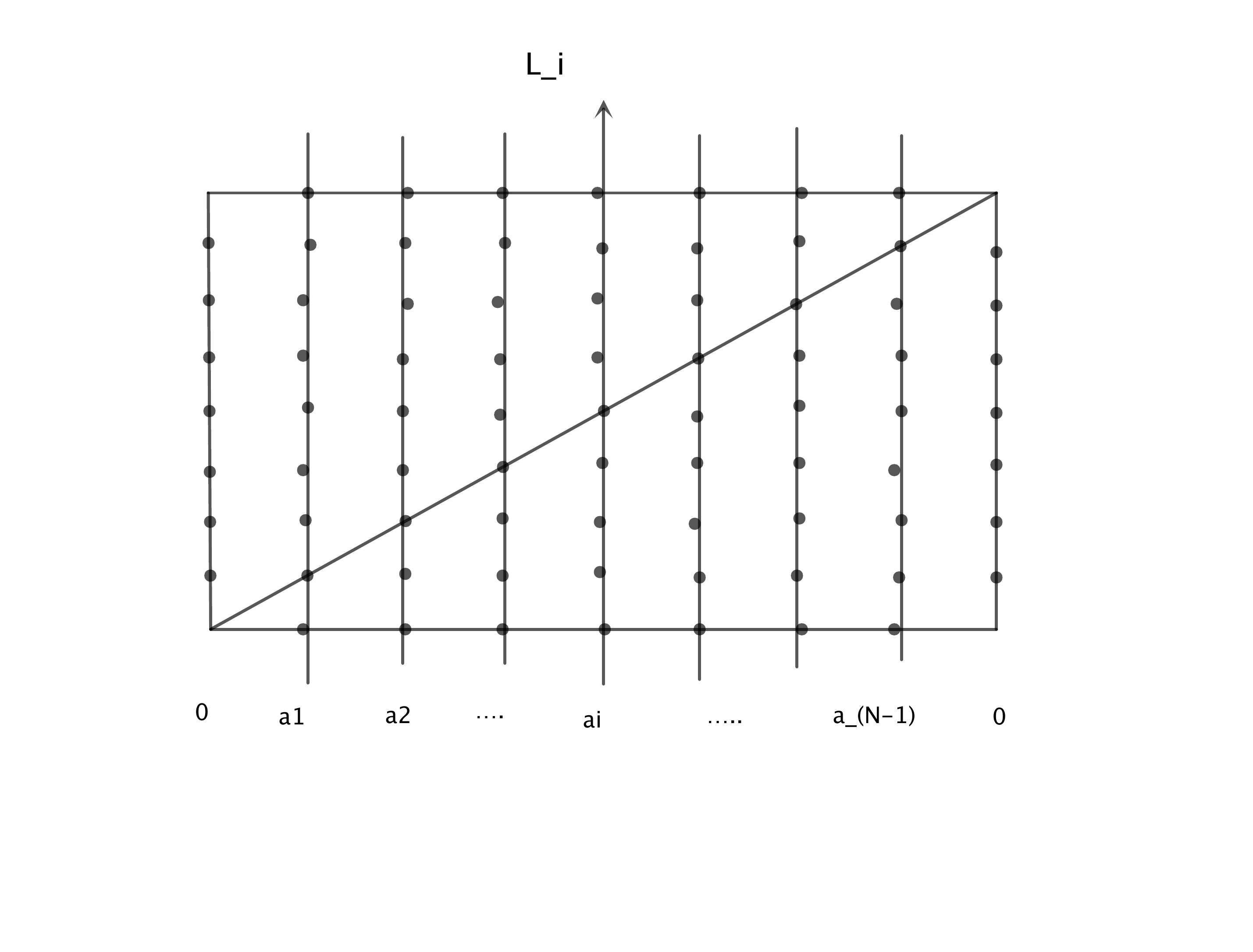}
\caption{The tropical $a$ coordinates for the Wilson line in the $i$th fundamental representation: the quiver nodes
in each straight line have the same indicated tropical $a$ coordinates.  }
\label{WSN}
\end{figure}

Instead of using the OPE to find the tropical $a$ coordinates of $\omega_i$, one can actually find the answer using the following requirement:
the tropical $x$ coordinates on the $i$th quiver  node of $a$ and $b$ edge should be $\pm1$ while  all the other $x$ coordinates are vanishing.
The answer is:
\begin{equation}
a_1={N-i\over N}, a_2={2(N-i)\over N},\ldots, a_i={i(N-i)\over N}, \ldots, a_{N-2}={2i\over N}, a_{N-1}={i\over N};
\end{equation}
namely, the $a$ coordinates take the maximal value at the $i$th line, and decreases along the left and right direction, see figure. \ref{WSN}.
So it is natural to represent such line operators as the closed curve going through the ith line in the quiver diagram.

After finding the tropical $a$ coordinates and the canonical map of the fundamental representations, one can find all other representations by doing 
the OPE, which has the same support as the tensor product of the group theory. The conclusion is: the Wilson line is classified by the irreducible 
representation of $\text{SU}(N)$ group.

Again, these ordinary Wilson operators around three one cycles are not enough. 
The extra line operators can be found by doing the OPE between the Wilson line operators $L_i$ and 't Hooft operators $M_j$ \footnote{The convention of $L_i$ has 
maximal $a$ coordinates on $i$th quiver node counting from left to right, while $M_i$ is based on the counting from bottom to top, see figure. \ref{WSN}.}, and the extra line operator
is found as the sub-leading component in the OPE. There are a total of $(N-1)^2$ pairs and 
the $N-1$ pairs ($L_i$ and  $M_{N-i}$) which will give the line operators in the C cycle, therefore we have $(N-1)(N-2)$ number of extra line operators which exactly match 
the number of internal quiver nodes inside two triangles. The canonical map of the extra line operator can be found using the Poisson bracket: One calculate the 
Poisson bracket of $L_i$ and $M_j$, and the result would have two parts with different coefficients, and these two parts are the canonical map of two line operators in the OPE.

These extra line operators can be represented geometrically using three junctions. By doing the explicit OPE, we find that
 there are ${(N-1)(N-2)\over2}$ type of three junctions which 
labeled by three positive integers satisfying the following conditions:
\begin{equation}
i+j+k=N.
\end{equation}
It is easy to see that the total number of solutions are ${(N-1)(N-2)\over2}$ which equals 
to the number of internal quiver nodes inside one triangle. For each solution, there are two types of 
junctions such that the internal quiver nodes has positive (negative) tropical $x$ coordinate. 

\begin{figure}[htbp]
\small
\centering
\includegraphics[width=8cm]{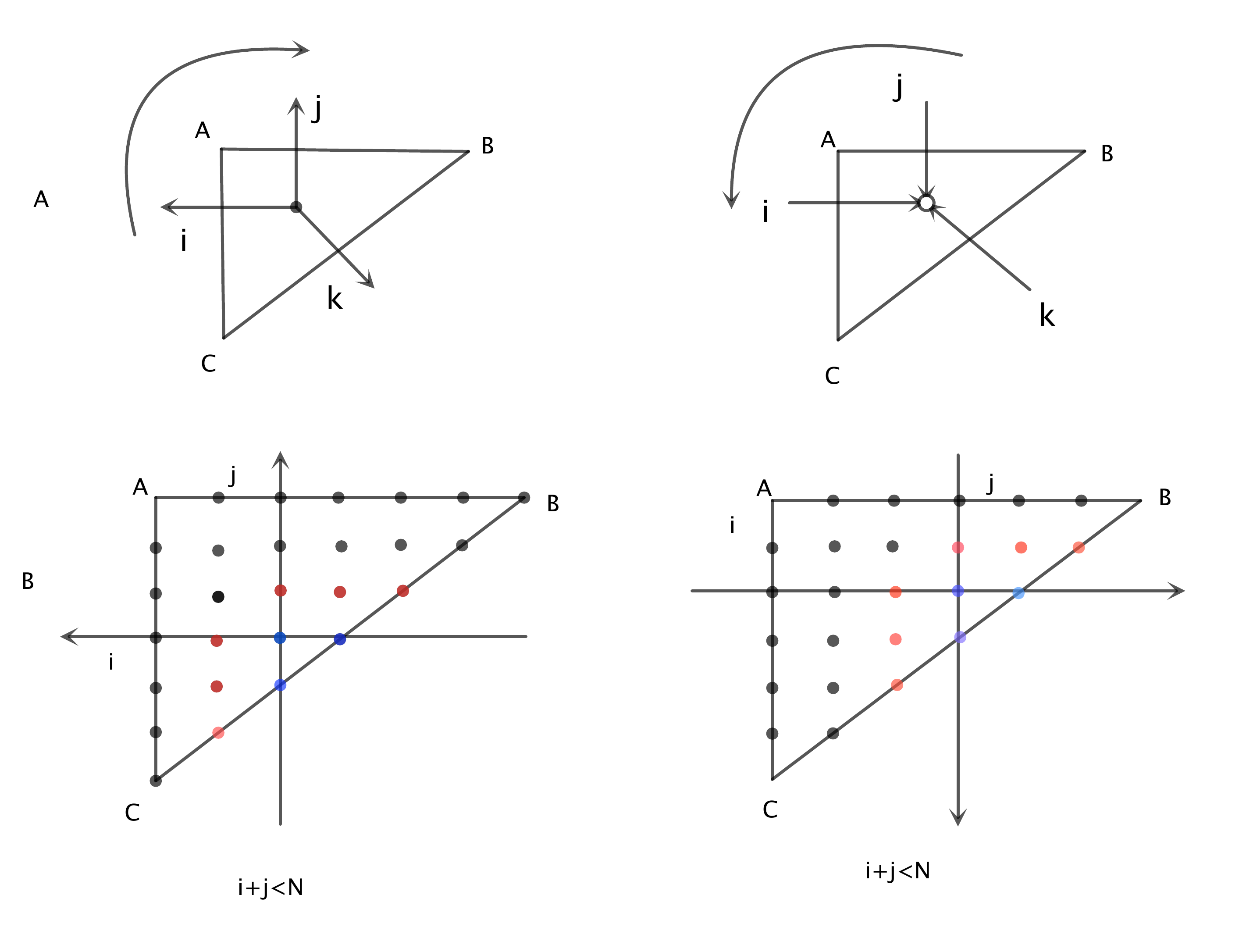}
\caption{Top: the black and white junction with labels $(i,j,k)$, and the meaning of the label is that the tropical $a$ coordinates 
is maximal on $i$th node counting in clockwise direction for the black junction, and anti-clockwise direction for white junction. Bottom: the rule of finding
the tropical $a$ coordinates from using fundamental representation $\omega_i$ and $\omega_j$: first sum up the $a$ coordinates of $\omega_i$
and $\omega_j$ and then subtract $2$ for the green nodes, and $1$ on the red nodes. }
\label{Njunction}
\end{figure}

For the black junction, the arrows are coming out of the junction for all three legs while 
the arrows are coming into the junction for white junction, and we need to label the junction by $(i,j,k)$, see figure. \ref{Njunction}. 
In the $N=3$ case, there is only one solution for the above equation, therefore there is no need to specify the 
label for the legs.

Let's now describe the tropical $a$ coordinates on the quiver nodes of the dual triangle of black and white junctions. Again, the rules are found by looking 
the OPE of the line operators $L_j$ and $M_i$.  For the black junction,  the tropical $a$ coordinates of the internal quiver nodes and the third edge are found 
as following: first let's draw two lines $L_j$ and and $M_i$ in the $i$ and $j$ directions of the three junction; secondly,
 sum up the tropical $a$ coordinates of the line operators $L_j$ and $M_i$, finally we need to  subtract appropriate integer number for the above tropical $a$ on edge $BC$
 such that they are the tropical coordinates of the $k$th fundamental representation (this can always be achieved), moreover, one need to 
 subtract integer numbers from tropical $a$ coordinates of some of the internal quiver nodes, the pattern is summarized in figure. \ref{Njunction}.
 Similarly procedure can be done for the white junction.

The dual tropical $x$ coordinate of the junction would have non-vanishing value on only one internal quiver node. Using the above explicitly described tropical $a$ coordinate, we find 
the position of the quiver node which the junction lives:
\begin{equation}
\begin{cases}
&(a,b,c)=(i, j, k),~~\text{Black junction}\\
&(a,b,c)=(j, k, i),~~\text{White junction}. 
\end{cases}
\end{equation}
One new feature is that one can have $M$ junction with labels $i_1,i_2,\ldots,i_M$ such that 
\begin{equation}
i_1+i_2+\ldots+i_M=N;
\end{equation}
Such general junction can be formed using a sequence of three junctions by un-merging  junction into a 
sequence of three junctions, so there is essentially nothing new. 

We oriented all the Wilson line in the same orientations as the first fundamental representation. 
For the following purpose, it is natural to choose a different rule for the orientation:
 the $i$th fundamental representation with $1\leq i< {N\over2}$ has the same orientation as $\omega_1$, otherwise it is oriented in 
the orientation as $\omega_{N-1}$ and the label is changed to $i^{'}=N-i$.
 If $i={N\over 2}$ and is an integer, the corresponding representation is real, and we do not assign any orientation to it.
With this convention, all the curves have the label with value less or equal than ${N\over 2}$.  

\begin{figure}[htbp]
\small
\centering
\includegraphics[width=10cm]{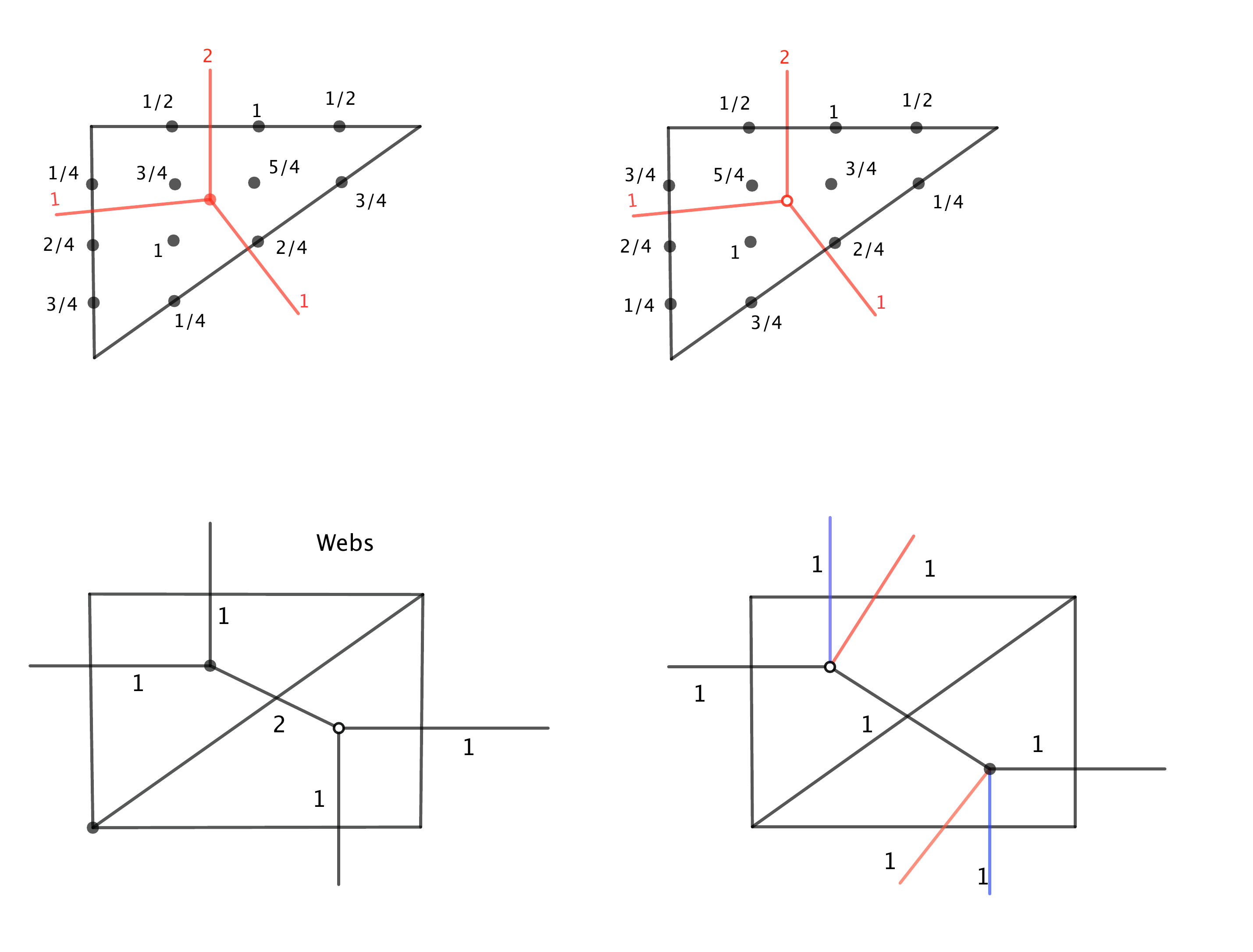}
\caption{Top: the basic junctions and its dual tropical $a$ coordinates of $A_3$ theory. Bottom: some webs of $A_3$ theory.}
\label{sl3}
\end{figure}

Accordingly, one need to change the orientation and the label for  three junctions. The rules are actually very simple: because of the constraint $i+j+k=N$, there is at most
one edge with label bigger than $N/2$, and if there is such type of edge, one simply change its orientation and change the label to $N-i$. The two representations have different 
advantages: the first one is good for constructing the complicated webs using the basic junctions, while the second one is good for the Skein relations we are going to study 
in the following.  It is not hard to transform between these two representations. The basic junctions and its tropical $a$ coordinates for $A_3$ theory are given in figure. \ref{sl3},
and we build some webs using the basic junctions in figure. \ref{sl3}.

\newpage
\subsubsection{Skein relations}
After describing the basic geometric building blocks for the webs, it is straightforward to develop the Skein relations and  have a geometric representation for the OPE.
In the mean time, any 4d line operator would be represented by the bipartite webs on the corresponding Riemann surface.
For the purpose of Skein relations, it is useful to use second labeling for the three junctions, i.e. all the labels have value less than or equal ${N\over 2}$.
Two legs of two different junctions with same label and opposite orientations (including the legs without orientation) can be connected, and 
one can have multi-junction, which is different from the $\text{SU}(3)$ case in which one can 
only have the three junction.
\begin{figure}[htbp]
\small
\centering
\includegraphics[width=12cm]{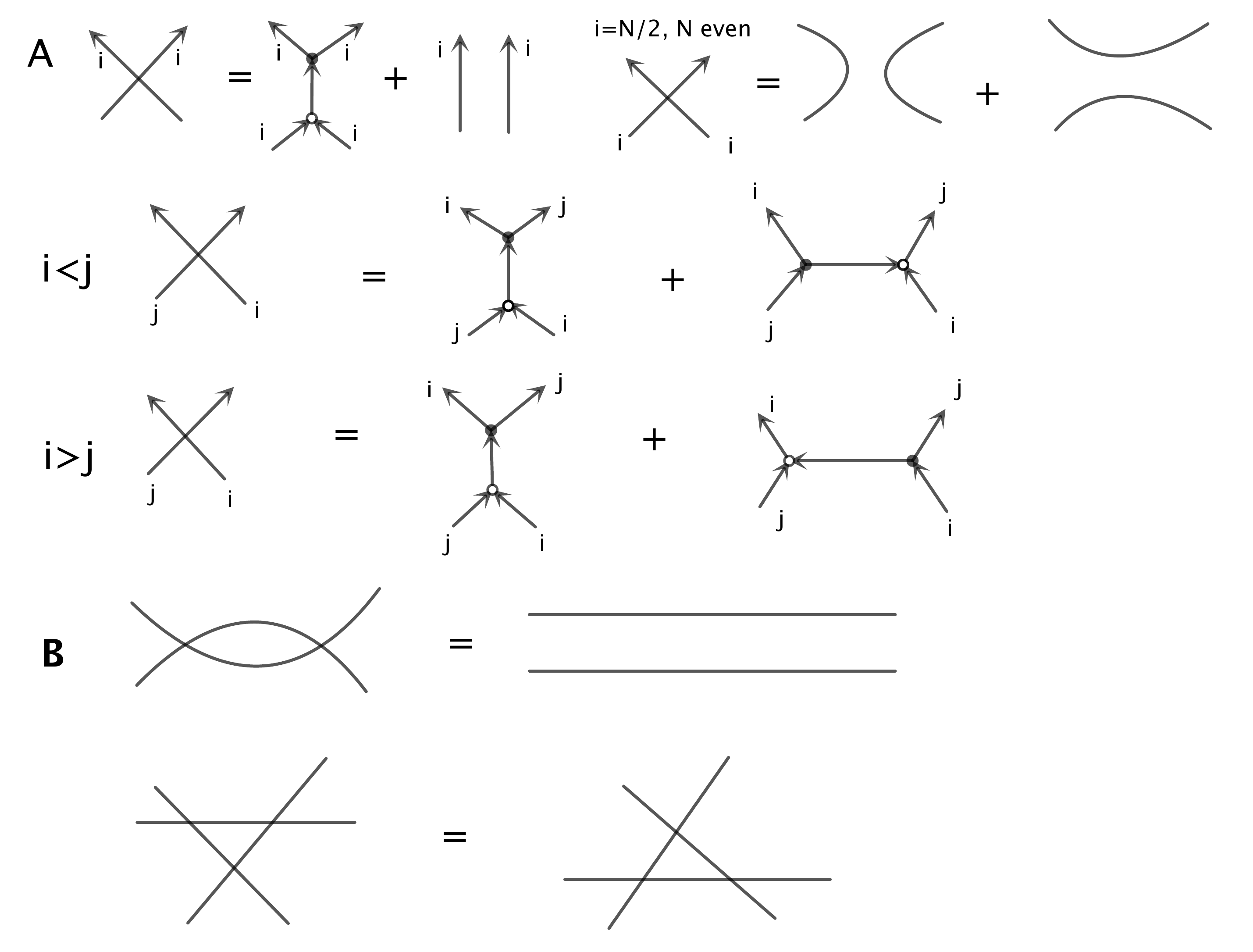}
\caption{Skein relations for general higher rank theory, here we assume $i+j<{N\over 2}$ to draw the orientation for the new generated legs.}
\label{highSkein}
\end{figure}

\begin{figure}[htbp]
\small
\centering
\includegraphics[width=8cm]{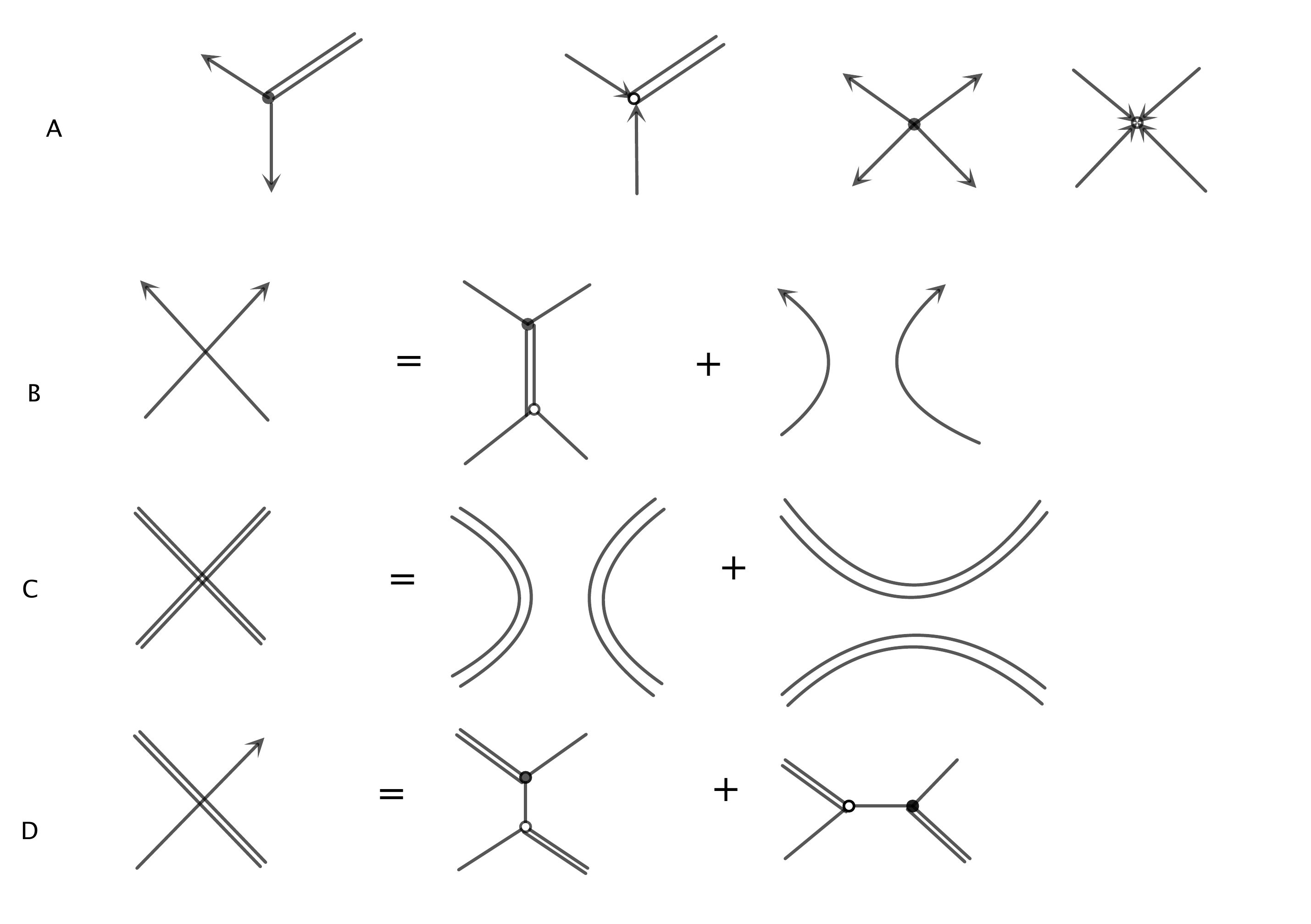}
\caption{The Skein relation for resolving the crossing for $\text{SU}(4)$ theory.}
\label{sl4}
\end{figure}

\begin{figure}[htbp]
\small
\centering
\includegraphics[width=10cm]{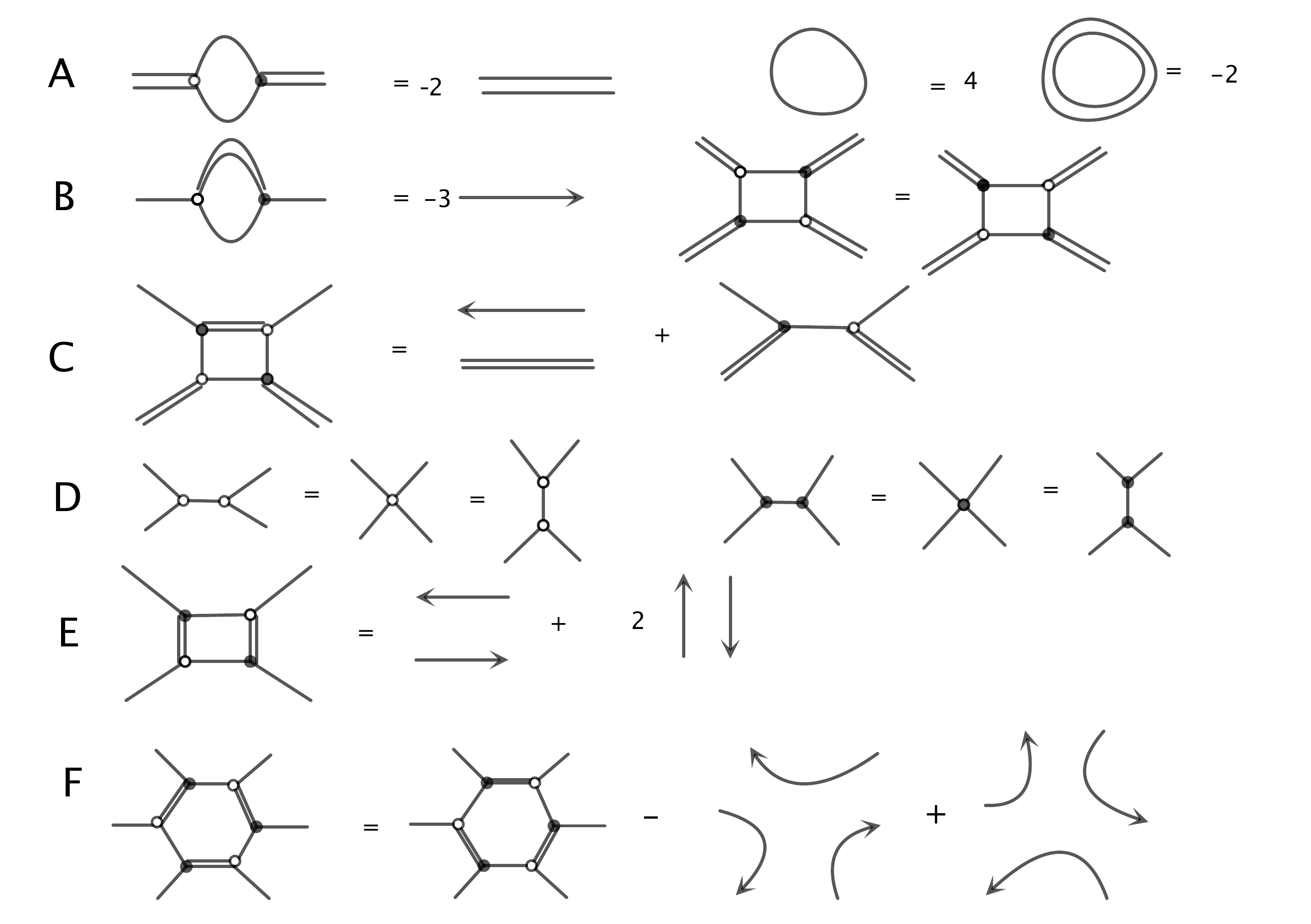}
\caption{The full Skein relations of $\text{SU}(4)$ theory.}
\label{sl4further}
\end{figure}

The  Skein relation between two oriented lines  can be easily found by doing the OPE between the ordinary 
Wilson and t' Hoof operators in various fundamental representations, and the result is summarized in figure.\ref{highSkein}.  
 There are more Skein relations involving bubbles,  squares, and hexagons which are derived from  obvious equivalence relations shown in figure.\ref{highSkein}.
A full Skein relations can be written down for $N=4$ case, since there is actually just one type of junction with three legs labeled by $(1,1,2)$. Instead of labeling 
the edge with  numbers, we use double line to represent the legs with label $2$, and there
is no need to specify its orientation since the corresponding representation is real. The two types of three junction are shown in figure. \ref{sl4}A, moreover,
we have the four junctions which can be formed by combining the three junctions. 
The Skein relation for resolving the cross of two lines are shown in figure. \ref{sl4further}.  
The details  for the general cases are tedious  and  we leave it to the interested reader.

Again, one can start with arbitrary tropical $a$ coordinates and construct the webs using a similar reconstruction procedure as
the one developed for $A_2$ theory. One the other hand, one can start with an irreducible webs and find its tropical $a$ coordinates. Our 
conjecture is that there is a one-to-one correspondence between the geometric and the algebraic description.

\newpage
\section{General bordered Riemann surface}
We discovered the basic ingredients such as line segments and 
three junctions with their coordinates on the dual triangle from the study of line operators on once punctured torus; we also find  Skein relations which are very powerful in finding the OPE.
The above elements are completely local and do not depend on the global structure of Riemann surface, therefore 
they should still be valid on any bordered Riemann surface. We have the following conjecture: 4d line 
operator is built as an irreducible bipartite web on the Riemann surface, and the OPE can be found using  Skein relations, etc. Other things 
like  tropical $a$ coordinates and canonical map are very similar.
In the following, we are going to study the webs on several interesting bordered Riemann surfaces, and we use $A_2$ theory 
as the main examples (the other cases are very similar).

\subsection{$T_N$ theory}
$T_N$ theory is defined by a sphere with three full punctures, and is a strongly coupled SCFT with $\text{SU(N)}^3$ flavor symmetry. 
There is no nontrivial closed cycle for  three punctured sphere, which means that there is no weakly coupled gauge theory description, so there is 
no ordinary Wilson line.

\begin{figure}[htbp]
\small
\centering
\includegraphics[width=12cm]{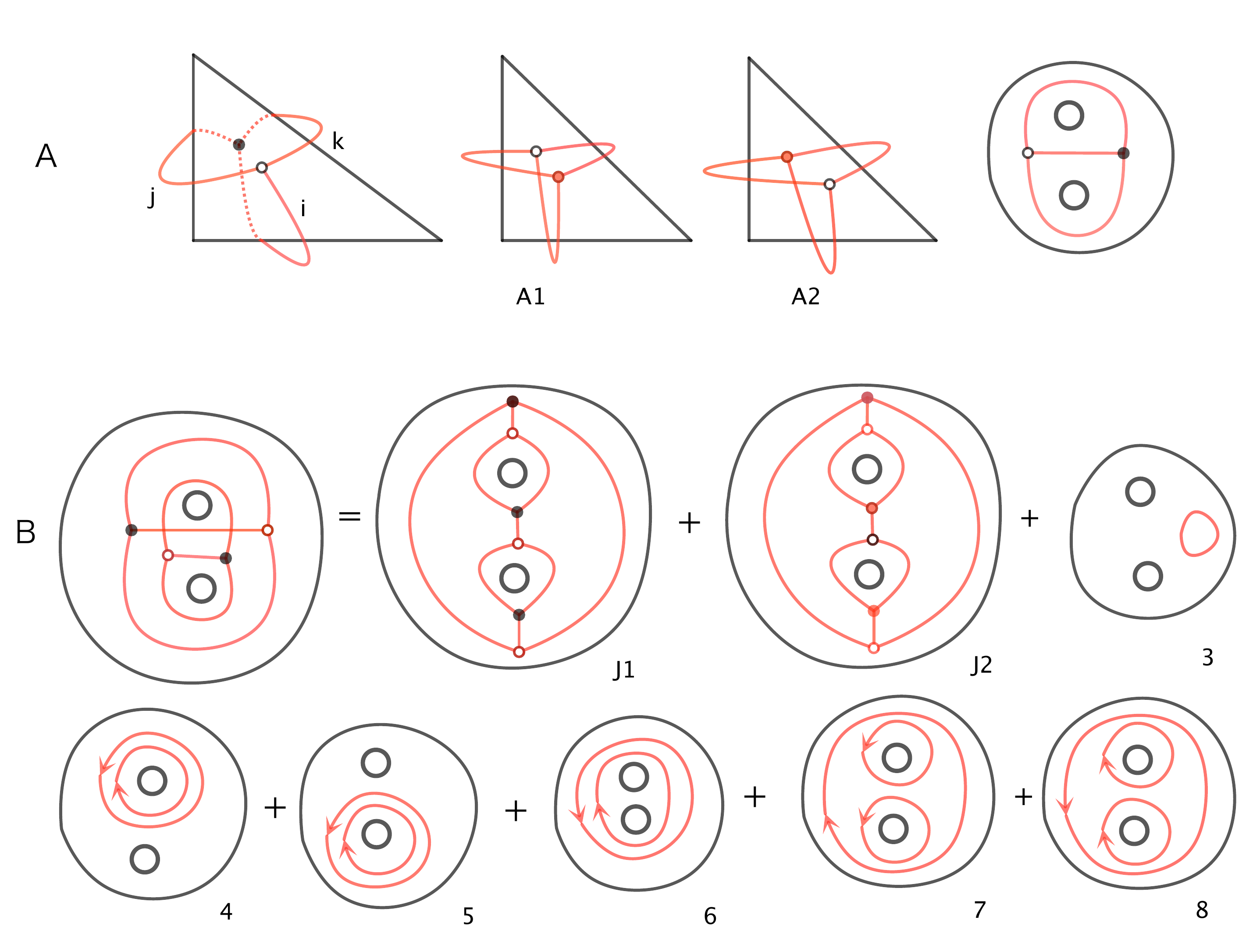}
\caption{A: the simplest webs on three punctured sphere. B: the OPE between above two webs are found using Skein relations, and there are two non-trivial webs in the OPE.}
\label{TN1}
\end{figure}

There are ${(N-1)(N-2)\over 2}$ Coulomb branch parameters and 
we should have non-trivial line operators represented by the webs using  three junctions. The simplest one is built by two junctions:
for each black junction with $i+j+k=N$, one naturally has a white junction on the back of the sphere to 
form a closed web, see figure. \ref{TN1}A. There are a  total of $(N-1)(N-2)$ of such webs as one can easily count.
In figure. \ref{TN1}A, we show the simplest web operators built from two junctions for $T_3$ theory, which are
denoted as $A_1$ and $A_2$. They can also be represented by the web on a plane graph in which the puncture is represented by a hole without 
any marked point, see figure. \ref{TN1}A.

 In figure. \ref{TN1}B, we use 
Skein relations to derive the OPE of $A_1$ and $A_2$, and two nontrivial line operators appear which are denoted as $J_1$ and $J_2$.
 We can further do OPE between $A_i$ and $J_i$, and find four more new line operators formed by eight three junctions, etc. 
Therefore, our conjecture is that the space of line operator is the same as the space of irreducible bipartite webs on three punctured sphere.

\begin{figure}[htbp]
\small
\centering
\includegraphics[width=10cm]{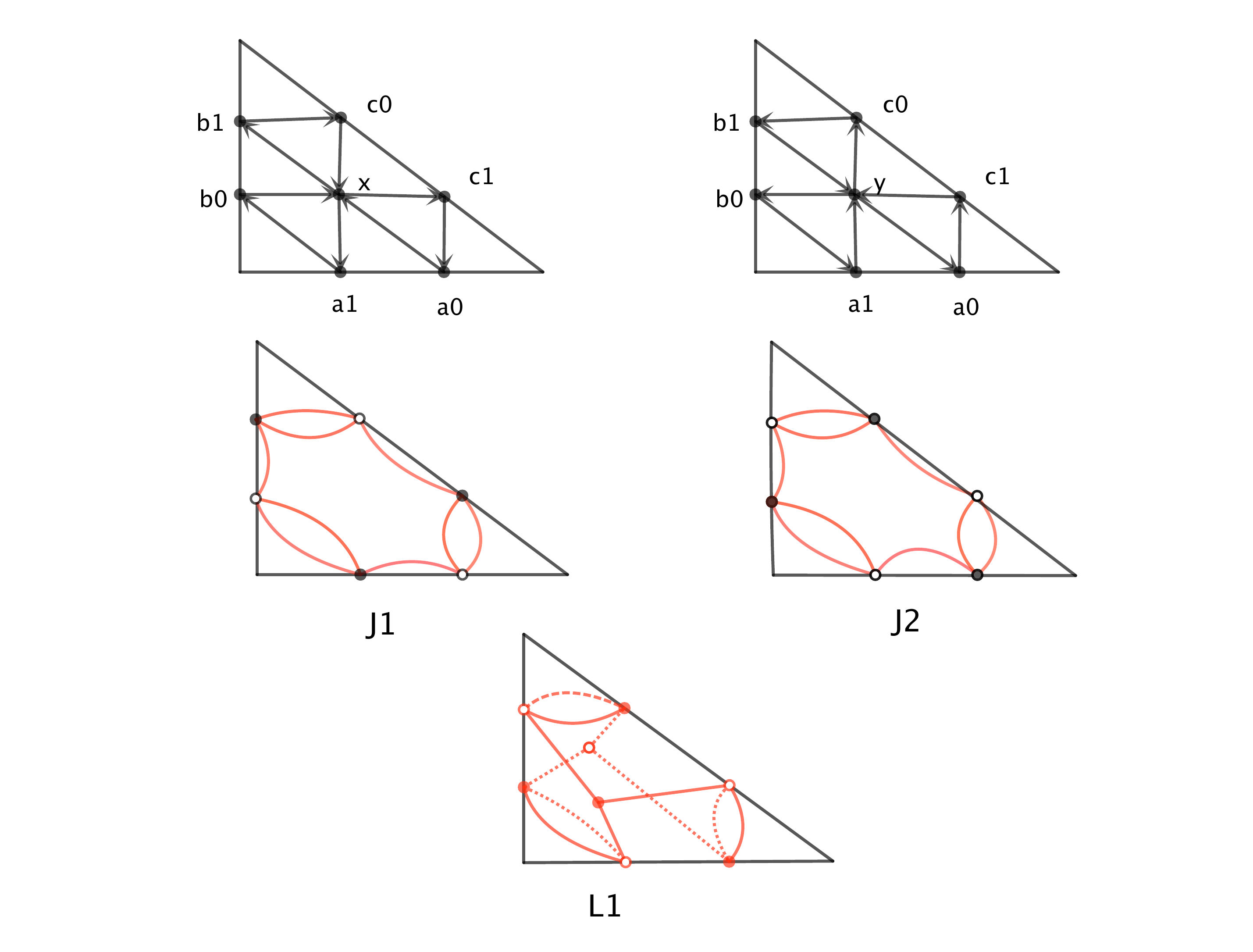}
\caption{Top: the cluster coordinates for $T_3$ theory. Middle: $J_1$ and $J_2$ webs. Bottom: another web found in the OPE between $J_1$ and $A_1$.}
\label{TN2}
\end{figure}

Given a triangulation and the cluster coordinates \cite{Xie:2012dw}, the tropical $a$ coordinates of the web can be
 found using the basic rule for three junction. The canonical map for them is nontrivial since here we can not use the monodromy of the ordinary Wilson line operator and 
OPE. We do know the leading order term which can be read from the tropical $a$ coordinates of the junctions, and the sub-leading order term is derived using the OPE:
\begin{align}
&I(A_1)=a_0^{2/3}a_1^{1/3}b_0^{2/3}b_1^{1/3}c_0^{2/3}c_1^{1/3}xy+a_0^{2/3}a_1^{1/3}b_0^{2/3}b_1^{1/3}c_0^{2/3}c_1^{1/3}x\ldots \nonumber\\
&I(A_2)=a_0^{1/3}a_1^{2/3}b_0^{1/3}b_1^{2/3}c_0^{1/3}c_1^{2/3}xy+a_0^{2/3}a_1^{1/3}b_0^{2/3}b_1^{1/3}c_0^{2/3}c_1^{1/3}y\ldots
\end{align}
It would be interesting to find the full canonical map for these two basic line operators. We also draw the webs $J_1$ and $J_2$ in the triangulation, and a web which would appear in 
the OPE between $J_1$ and $A_1$, see figure. \ref{TN2}. It is also not too hard to find that $A_1, A_2, J_1, J_2$ form closed Poisson brackets even though we do not know 
the full canonical map. Because the leading component of $A_1*A_2$ and 
$J_1*J_2$ have trivial dual $x$ coordinates, we find two constraints and two independent coordinates, which matches with the dimension of 
phase space of Hitchin integrable system. The explicit Poisson brackets and the Hamiltonian will be presented elsewhere.

\subsection{Fourth punctured sphere}
\begin{figure}[htbp]
\small
\centering
\includegraphics[width=10cm]{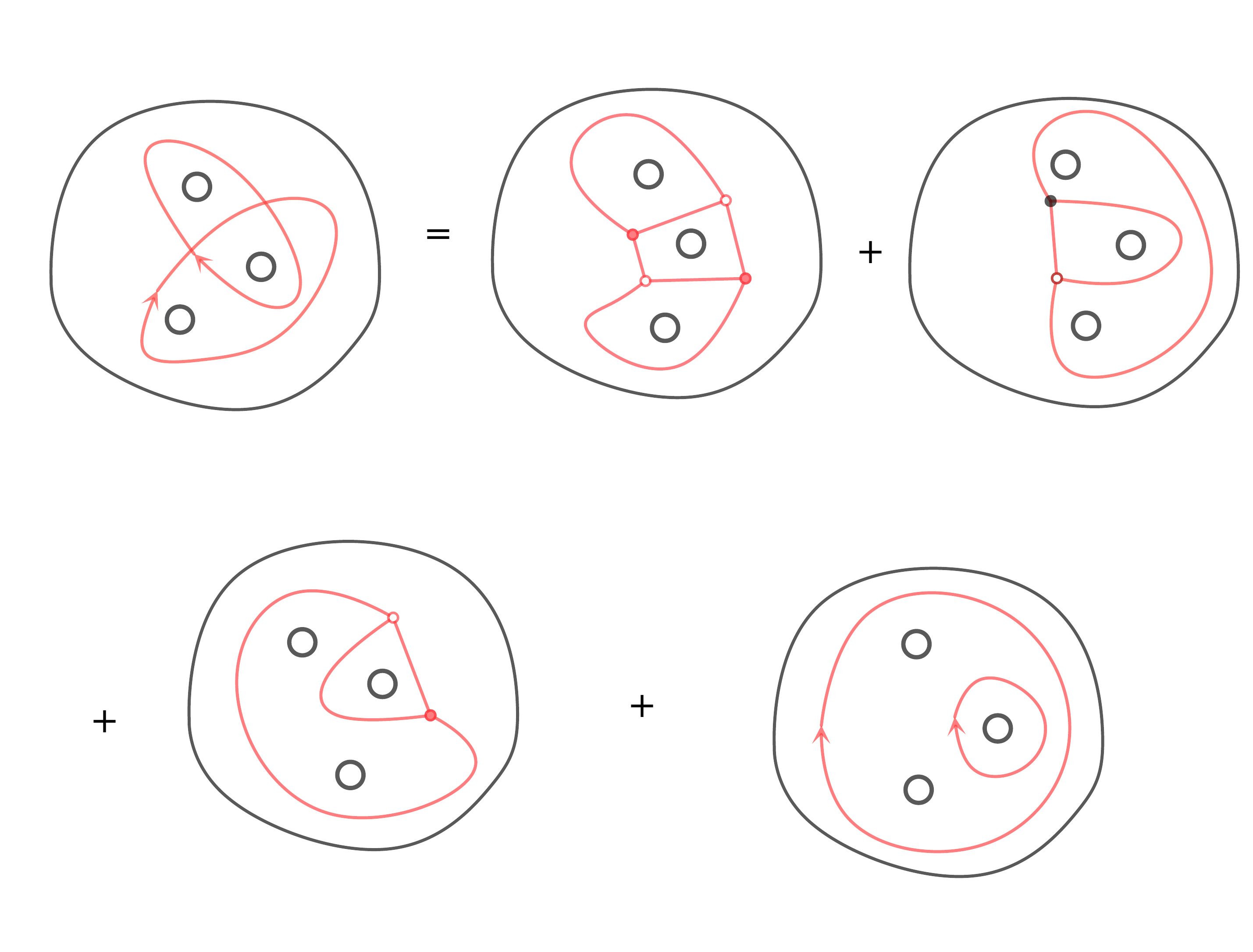}
\caption{The OPE between line operators on fourth punctured sphere using the Skein relations.}
\label{4pun}
\end{figure}

Let's consider  4d theory defined by a sphere with four full punctures. This theory is described by a $\text{SU}(N)$ gauge group coupled with two $T_N$ theories in one duality frame. 
The geometric picture of line operators for this theory is pretty similar: they are represented by irreducible bipartite webs. 
The OPE between two Wilson lines are derived using the Skein relations,  see figure. \ref{4pun}.
  One can find the tropical $a$ coordinates of these web operators by taking a triangulation, and  
  the canonical map of  various web operators can be found too. We leave the details to the interested reader.

\begin{figure}[htbp]
\small
\centering
\includegraphics[width=8cm]{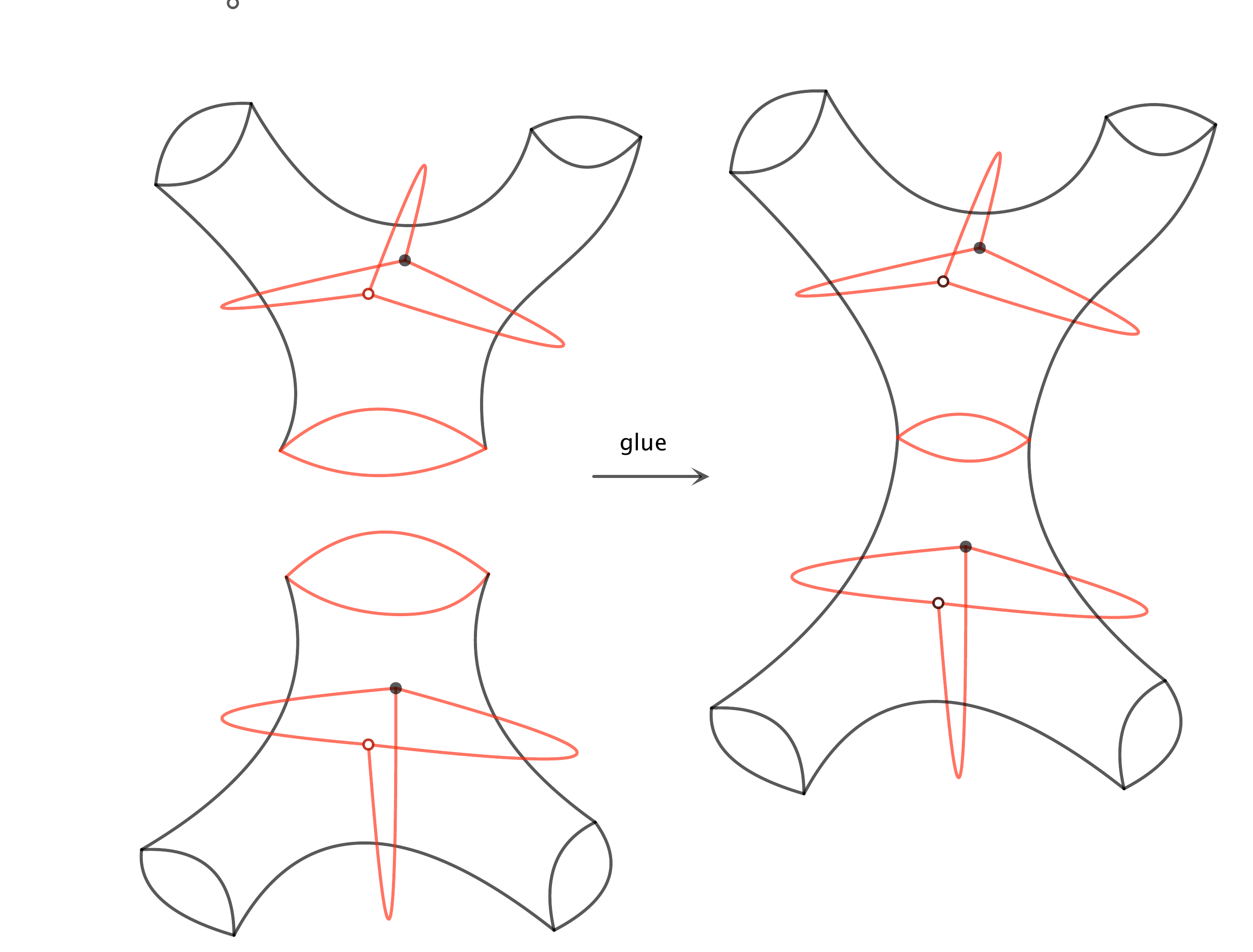}
\caption{The gluing of two three punctured spheres is achieved by using the line operator around the puncture.}
\label{glue}
\end{figure}

We would like to make a simple remark about the correlation function of 2d Toda field theory using our result (using the 2d/4d correspondence,
the line operator on the Riemann surface will also be the observables of 2d Toda field theory.). 
Since the fourth punctured sphere can be derived by gluing two three punctured spheres, and the gluing only uses the line operator
around the puncture, see figure. \ref{glue}, and  internal web operator is not used in gluing. 
In the 2d CFT interpretation, this means that the correlation function can be built from three point function, and the gluing is simple, which is consistent with 2d OPE.
The nontrivial part of  Toda field theory (counterpart of higher rank field theory) is that the three point function is complicated, as is manifested by the non-trivial line operators.
The nice thing about the gluing picture is that once we
 understand the quantization of  three punctured sphere, we should be able to understand the quantization of Toda field theory defined on any Riemann surface.

\subsection{Disc with marked points}
Let's consider a disc with marked points, which describes 4d Argyres-Douglas theory \cite{Xie:2012jd}. 
The special requirement  is: one need to ensure that the tropical $a$ coordinates of  boundary quiver nodes to be zero, as 
we only counter the internal quiver nodes in studying of 4d $\mathcal{N}=2$ theory. Another important 
rule is that for the curve connected with two boundary points, the canonical map is defined as
\begin{equation}
\text{I}(l)=\begin{pmatrix}
1 &0&0\\
\end{pmatrix}*
M*
\begin{pmatrix}
1 \\
0\\
0
\end{pmatrix},
\end{equation}
here $M$ is the monodromy around the path \footnote{ There are choices
on which edge we should start or end in the monodromy graph, here we start and end with the edge which have the same orientations
as the curve.} and we set the boundary coordinates to be one in the end, so the answer does not depend on 
the boundary coordinates.

\begin{figure}[htbp]
\small
\centering
\includegraphics[width=8cm]{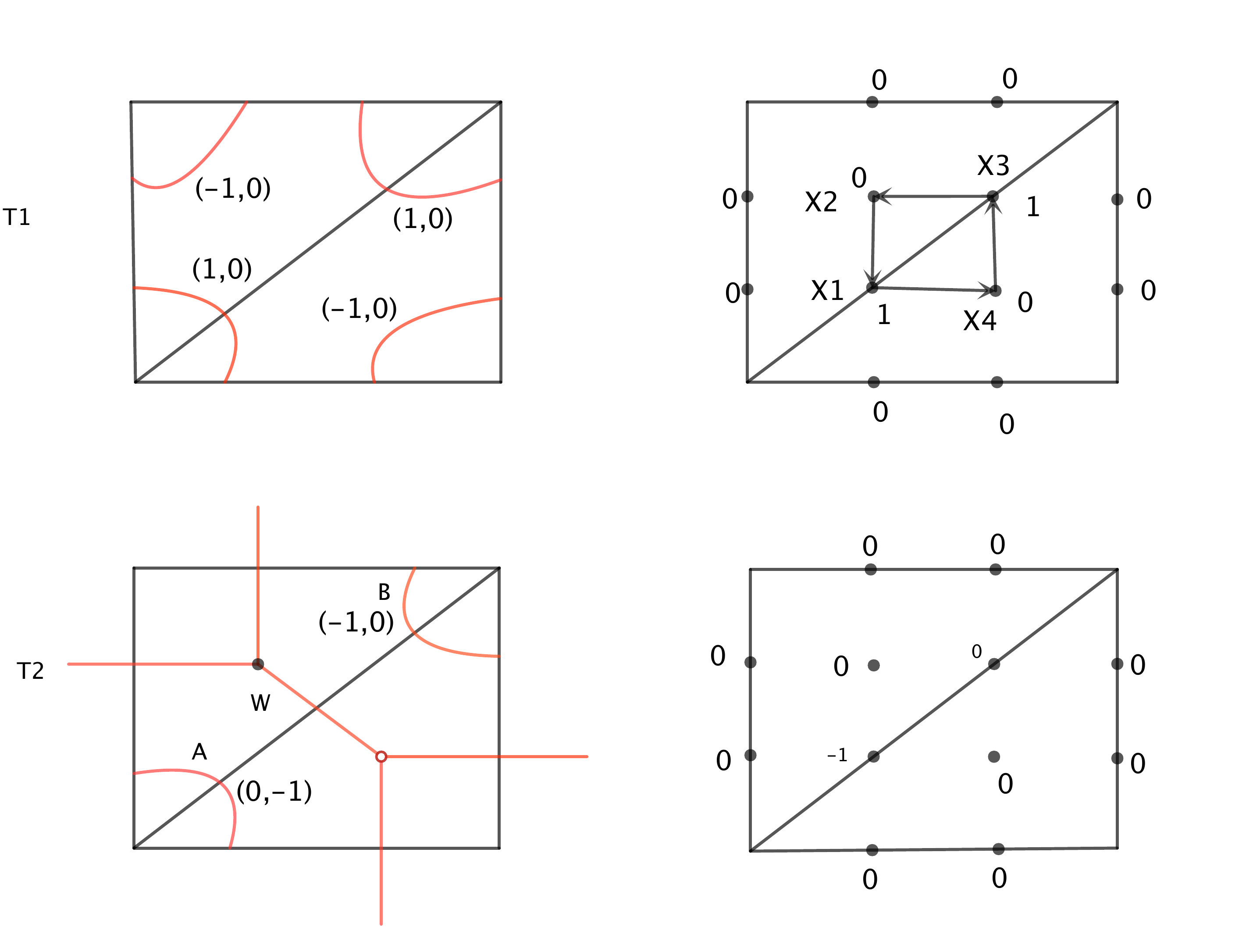}
\caption{Left: two laminations on fourth punctured disc of $A_2$ theory. Right: The tropical $a$ coordinates of these two laminations.}
\label{Disc}
\end{figure}

For simplicity, let's consider $A_2$ theory on a disc with four marked points, which  actually represents $D_4$ Argyres-Douglas theory \cite{Xie:2012jd}. A lamination 
satisfying the above constraint is shown in figure. \ref{Disc}, and its tropical $a$ coordinates is also shown. The canonical map is very simple 
$\text{I}(T_1)=X_1X_3$, which is actually the central element in the Poisson structure (consider only the quiver nodes insider the quadrilateral). Another 
lamination using  web is shown in figure. \ref{Disc}, and the canonical map of $W$ and the other two curves are \footnote{One first find the canonical map of 
 curve connecting two opposite boundaries using the monodromy, and then do the OPE to find  canonical map of web operator.}
\begin{align}
&\text{I}(W)=\frac{X_1^{1/3} X_2}{X_3^{1/3}}+X_1^{1/3} X_2 X_3^{2/3}+X_1^{1/3} X_2 X_3^{2/3} X_4, \nonumber\\
&\text{I}(A)=X_1^{-2/3}X_3^{-1/3}  X_2^{-1/3}X_4^{-1/3} ,~~~~~~\text{I}(B)=X_1^{-2/3}X_3^{-1/3} X_2^{-2/3}X_4^{-2/3},
\end{align}
so the canonical map of this lamination is
\begin{equation}
\text{I}(T_2)=\text{I}(W)*\text{I}(A)*\text{I}(B)=\frac{1}{\text{X}_1}+\frac{1}{\text{X}_1 \text{X}_4}+\frac{1}{\text{X}_1 \text{X}_3 \text{X}_4}.
\end{equation}

\begin{figure}[htbp]
\small
\centering
\includegraphics[width=8cm]{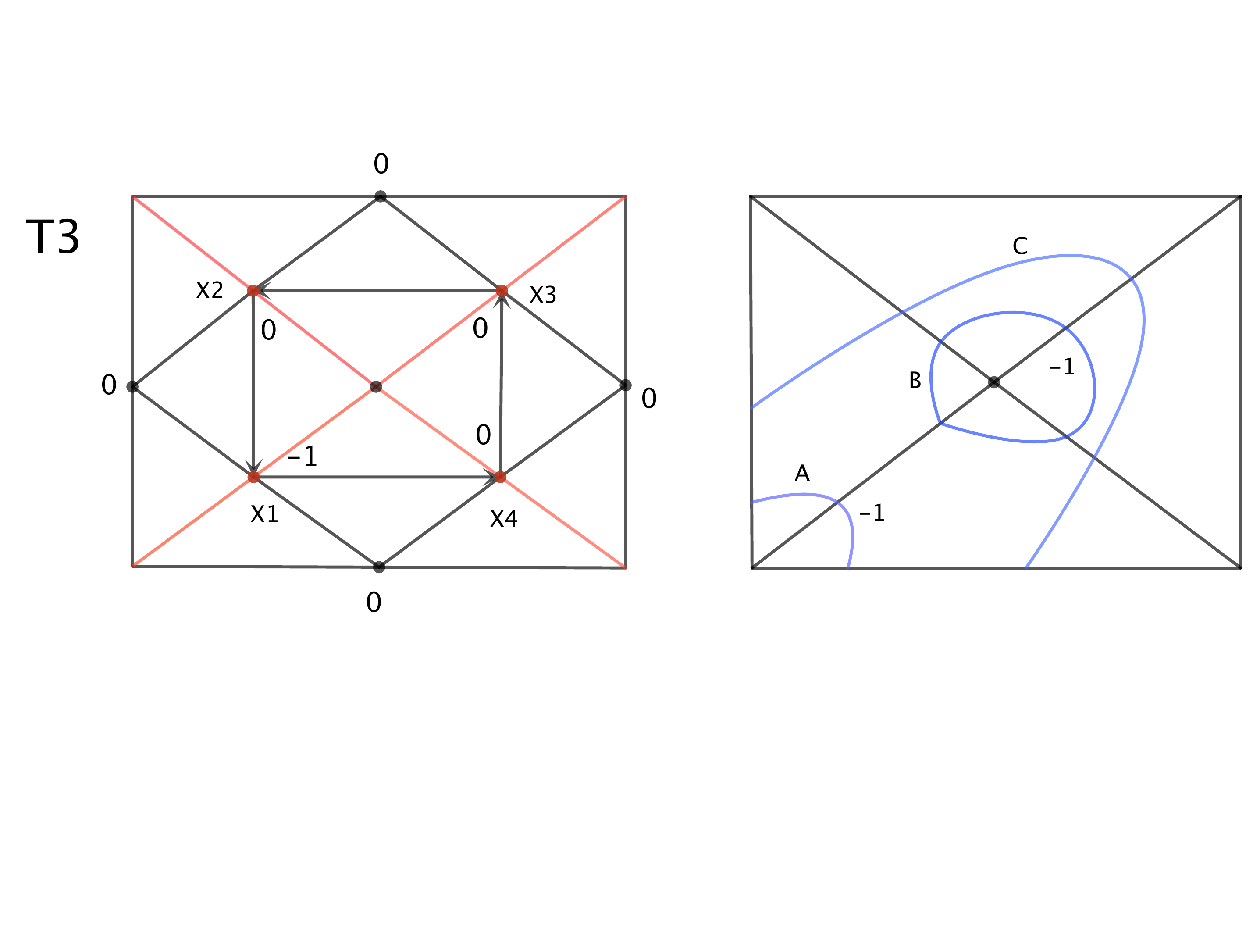}
\caption{A $A_1$ lamination which gives the same tropical $a$ coordinates as the lamination $T_2$.}
\label{disc2}
\end{figure}

The same theory can be constructed using  $A_1$ theory on a disc with four marked points and one bulk puncture, and the triangulation
which give the same quiver as  $A_2$ triangulations are shown in figure. \ref{disc2}.  The lamination with  the same tropical $a$ coordinates of 
$T_2$ is found using the reconstruction procedure described in section 2, see figure. \ref{disc2}. The canonical map of this lamination $T_3$ is found by using the 
monodromy representation of the $\text{SU}(2)$ local system described in section 2: 
\begin{equation}
\text{I}(\text{A})=X_1^{-{1\over2}},~~~\text{I}(\text{B})=X_1^{-{1\over2}}X_2^{-{1\over2}}X_3^{-{1\over2}}X_4^{-{1\over2}},~~\text{I}(\text{C})=X_2^{{1\over2}}X_3^{{1\over2}}X_4^{{1\over2}}+X_2^{{1\over2}}X_3^{{1\over2}}X_4^{-{1\over2}}+X_2^{{1\over2}}X_3^{-{1\over2}}X_4^{-{1\over2}},
\end{equation}
and the canonical map for this lamination is 
\begin{equation}
\text{I}(T_3)=\text{I}(\text{A})*\text{I}(\text{B})*\text{I}(\text{C})=\frac{1}{\text{X}_1}+\frac{1}{\text{X}_1 \text{X}_4}+\frac{1}{\text{X}_1 \text{X}_3 \text{X}_4}.
\end{equation}
Remarkably, this is equal to the answer found using  $\text{SU}(3)$ representation. This result is clearly a very strong evidence that our generalization of lamination to 
higher rank theory is correct.

\section{Conclusion}
In this paper, we initiated a systematical study of  half-BPS line operators of  4d $\mathcal{N}=2$ theory (theory of class ${\cal S}$) defined using full defects.
Although  the basic building blocks like line segments, three junctions are described for any $N$, we mainly focused on $A_2$ and $A_3$ theories, it
would be interesting to further study general cases, i.e.
giving the full Skein relations, finding the construction and reconstruction procedure as we did for  $A_2$ case, etc. 
There are many other open questions and further generalizations: 
\begin{itemize}
\item We found the canonical map of  line operators using  OPE, which is straightforward but tedious. It would be very useful if the answer can be  directly found
 using  web picture. This is important  for $T_N$ theory as the OPE method is not applicable.
 \item It would be interesting to further study the OPE and general properties
of the integer coefficients like its support and value. These results are useful in checking the conjectures given in \cite{Fock:math0311245}. It is also interesting to compare our results with the OPE of  Wilson-'t Hooft operators
 studied in the context of $\mathcal{N}=4$ SYM \cite{Kapustin:2006pk,Kapustin:2007wm,Saulina:2011qr,Moraru:2012nu} and three dimensional theory \cite{Kapustin:2013hpk}.
\item The non-full puncture case is worth to pursue,  since in this case there are theories with Lagrangian descriptions and  many field theory 
results are available \cite{Kapustin:2006hi,Pestun:2007rz,Gomis:2011pf,Rey:2010ry} (see also some results using Toda field theory \cite{Gomis:2010kv,Passerini:2010pr}).
The basic tools like junctions, Skein relations should still be valid, and the cluster coordinates are given in \cite{Xie:2012dw,Xie:2012jd} which 
can be used to parameterize the line operators, etc; The main difficulty is that there are no monodromy calculation tools available, so
we don't know how to get the canonical map, currently we are trying to  
deal with this problem \cite{xie2013}; 
\item Finally, it would be interesting to study various physical quantities related to the line operators, i.e. 
Superconformal index (see \cite{Gang:2012yr} for some $A_1$ cases), framed BPS states, $S^4$ partition function in the presence of line operators (see \cite{Drukker:2009id,Alday:2009fs} for $A_1$ theory with Lagrangian description) etc;  
it is also interesting to study how the line operator labels change under the 
pants decomposition, which would be a nice justification of  $S$ duality. 
\end{itemize}

One actually need a quantum version of line operators to answer many questions raised above. 
The quantum deformations of the line operator can  be studied using quantum cluster algebra \cite{Fock:math0311245}, for the $A_1$ case, see \cite{fock-2003, Gaiotto:2010be,Dimofte:2011jd}. 
A complete treatment about these quantum line operators of  higher rank case will 
be given in a future publication \cite{xie2013tt}.

In this paper, we mainly focus on  $A$ lamination and do not touch  $X$ lamination. The main difference 
is  the webs of  $X$ lamination can end on bulk punctures. Such $X$ lamination is important in representing 4d BPS states in the $A_1$ case \cite{Xie:2012gd}.
Naively, we expect the higher rank $X$ lamination should also give the 4d BPS states, and such interpretation matches perfectly with
the expectation from the BPS quiver approach in some special cases we check, and we hope to come to this 
point in the near future.

The three junctions play a crucial role in our construction, and our derivation is based on the canonical map, cluster coordinates, etc.
 It would be interesting to see whether it can be realized as the one quarter BPS surface operator of $(2,0)$ theory. Similar web-like 
 objects in M theory have been studied in \cite{Lee:2006gqa,Bolognesi:2011rq,Gaiotto:2012rg}, it is interesting to see if there is any relations between the junctions (superficially it seems that the number of junctions in our case is smaller than theirs.),
 see also the quarter BPS supergravity solutions  related to M2 branes in \cite{Lunin:2008tf}.
If these webs can indeed be realized as  quarter BPS surface operators, we expect them to be important in understanding the dynamics of  $(2,0)$ theory.

\begin{figure}[htbp]
\small
\centering
\includegraphics[width=12cm]{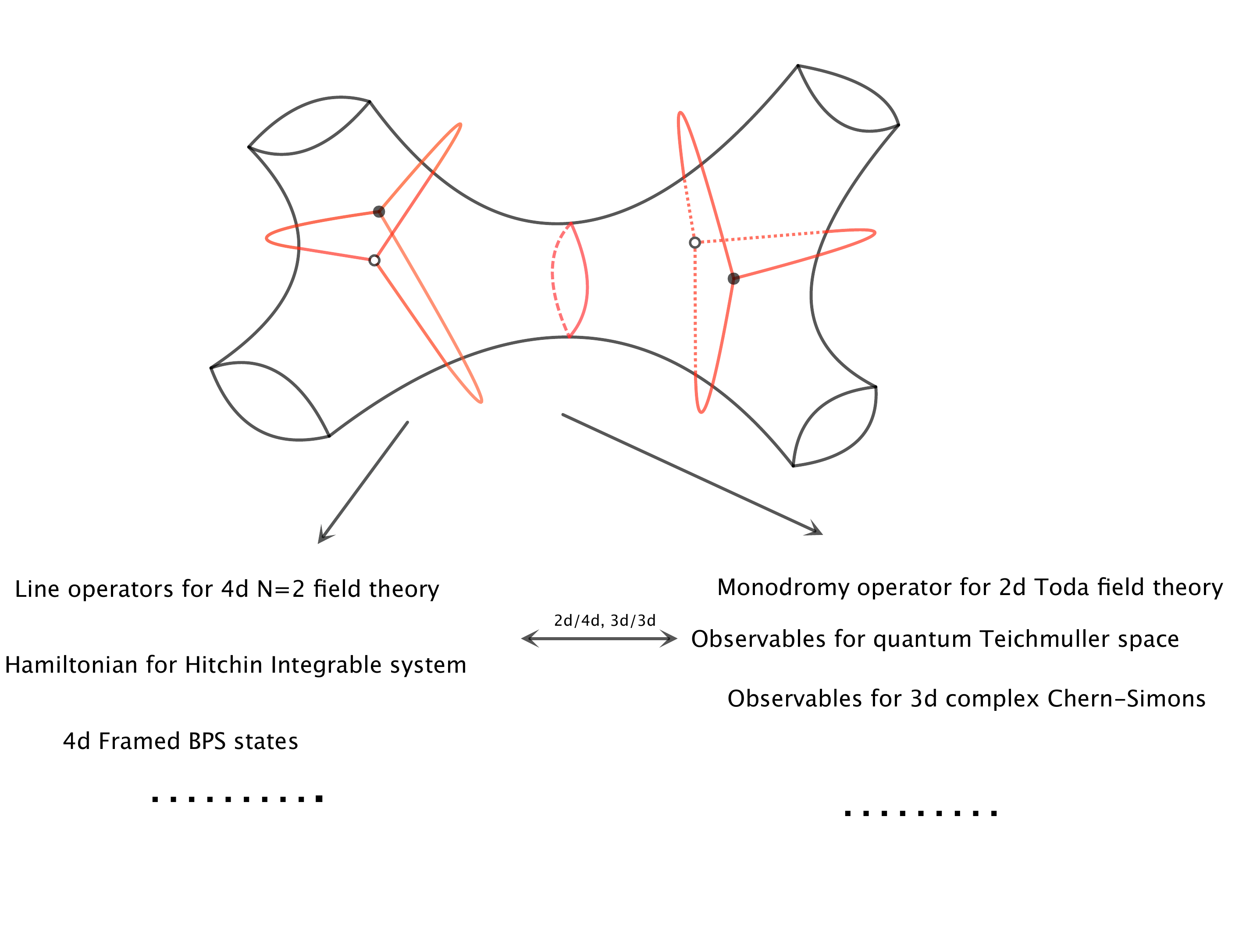}
\caption{The use of line (web) operators.}
\label{Con}
\end{figure}
These line operators would have a lot of applications in various physical contexts due to the magical six dimensional construction, see figure. \ref{Con}.
It seems that all of them can be pursued using our result, and it would be exciting to do some  concrete calculations. 

Our work also has a lot of direct relations to mathematics. Firstly our web structure seems to be the  geometric picture for the higher lamination proposed in 
\cite{fock-2003,Fock:math0311245} (see also recent work on 
geometric realization of higher lamination using  affine Grassmannia \cite{le2012higher}); The web diagrams studied by us
are also called tensor diagrams which are used to study  cluster algebra in \cite{fomin2012tensor}. Secondly it is amusing to note that our 
Skein relations coincide with the one used in the study of  Khovanov homology, see \cite{kuperberg1996spiders,khovanov2004sl}, it would be 
interesting to see if there is any deep connection. Thirdly the OPE of Wilson and 
t'Hooft loops \cite{Kapustin:2006pk} and quantization of Hitchin integrable system \cite{belindson}  play an important role in the study of  Geometric Langlands duality, 
Our results are both related to the OPE of line operators and  quantization of Hitchin integrable system, and it is 
interesting to see if there are any potential applications.

\newpage
\begin{flushleft}
\textbf{Acknowledgments}
\end{flushleft}

We thank Tudor Dimorfte, Davide Gaiotto, Juan Maldacena, Edward Witten, Yuji Tachikawa, Masahito Yamazaki, Piljin Yi and Peng Zhao for help discussions. 
This research is supported in part by Zurich Financial services membership and by the U.S. Department of Energy, grant DE- FG02-90ER40542 (DX). 

\appendix

\section{Calculating monodromy }
\begin{figure}[htbp]
\small
\centering
\includegraphics[width=10cm]{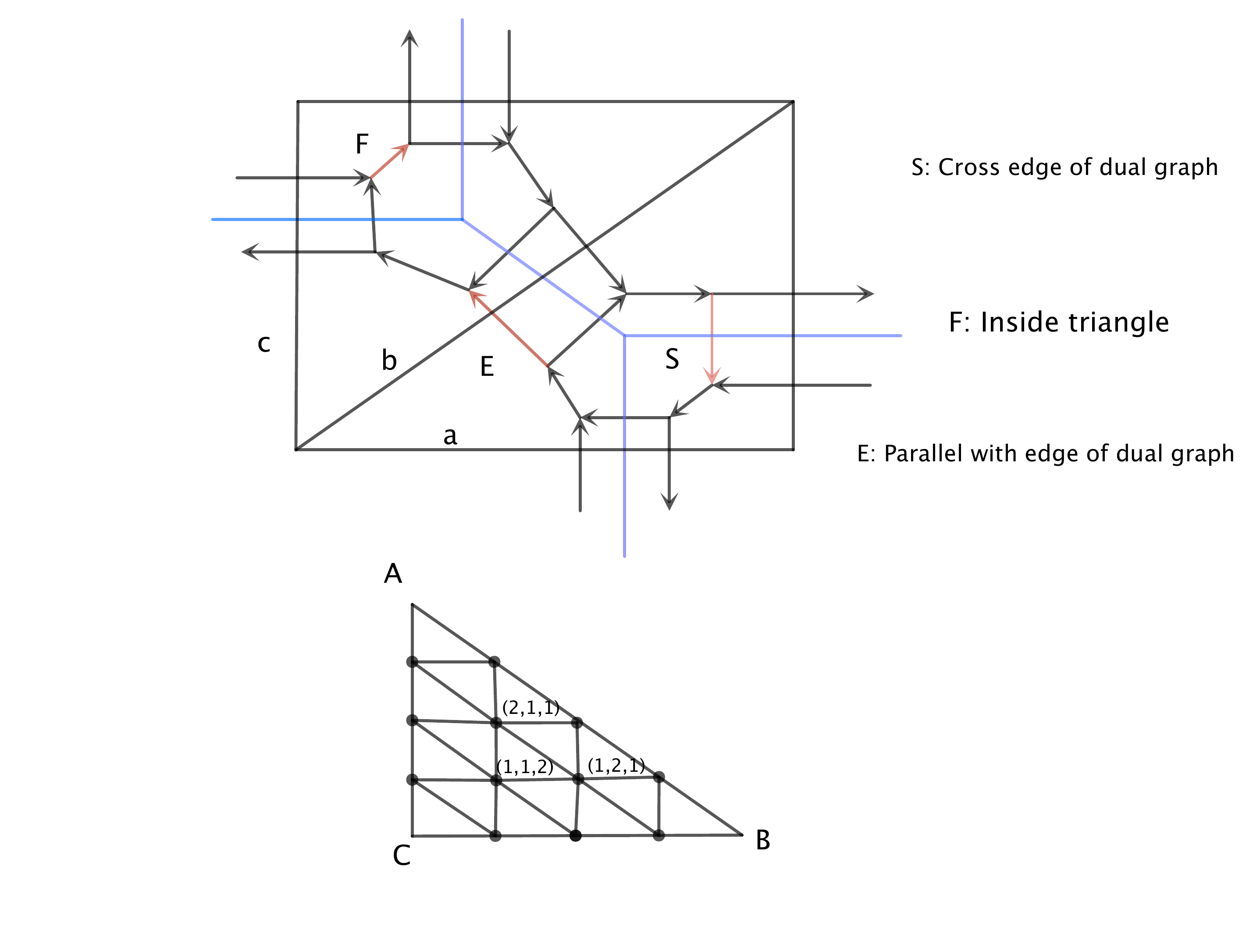}
\caption{ Top: the monodromy graph. Bottom: the label for coordinates inside a triangle, here the first index means the distance to the dual edge of vertex $A$.}
\label{appen}
\end{figure}

The cluster coordinates for higher rank theory defined using full punctures are found in \cite{fock-2003}. It was conjectured that they are the BPS quiver for  corresponding
$\mathcal{N}=2$ theory in \cite{Xie:2012dw} which also gave a brane construction. One start with an ideal triangulation of bordered Riemann surface, and 
then further triangulate  triangles in the triangulation. There are ${(N-1)(N-2)\over 2}$ new quiver nodes inside the triangle with three vertices $ABC$, and each internal quiver node can be labeled by 
three positive integers satisfying the condition 
\begin{equation}
a+b+c=N,
\end{equation}
here $a$ is defined as the distance to the dual side of the vertices $A$, etc, see figure. \ref{appen}.

Let's now summarize the ingredients for calculating the monodromy of higher rank local system following Fock and Goncharov \cite{fock-2003}. 
For each triangulation, one can construct a dual graph: put one trivalent vertex inside each triangle and connect those 
trivalent vertices together to form a dual graph. We form a hexagon around each vertex and a square around each edge in the dual graph.
We call  dual graph together with the hexagons and squares as the monodromy graph.
There are three types of edges in monodromy graph: S  edge which intersects with the dual graph,  E  edge 
which is parallel with the edge of dual graph, and F  edge which is inside the triangle, see figure. \ref{appen}.
We choose an orientation for edges in monodromy graph: the orientation is taken to be anti-clockwise for edges circling around the punctures,
and all the edges in hexagon is taken to be in clockwise orientation. 

\begin{figure}[htbp]
\small
\centering
\includegraphics[width=8cm]{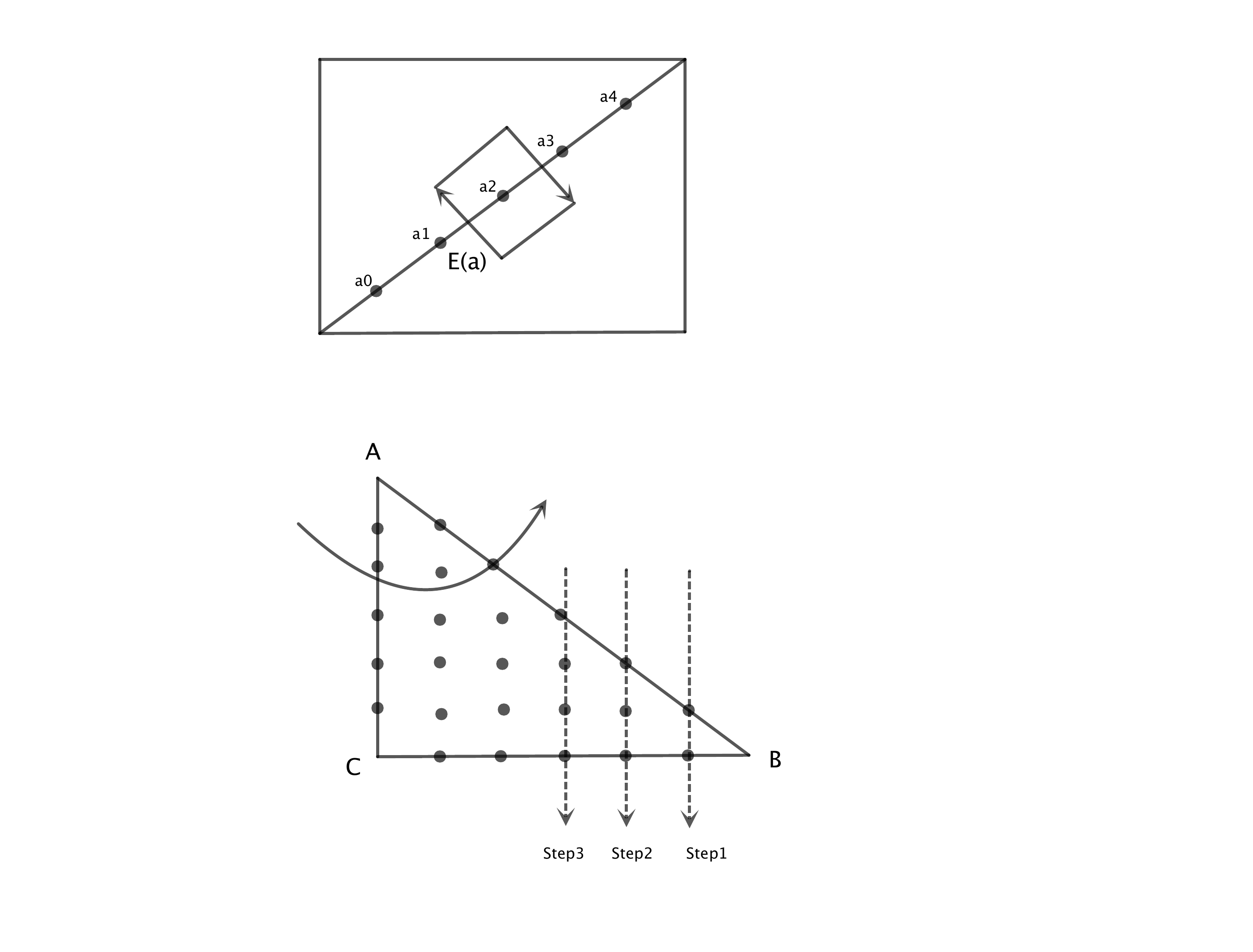}
\caption{Top: the labeling for the coordinates on the edge with which $E$ intersects, and this labeling is used 
in the calculation of the monodromy. Bottom: the rule of remembering the formula for calculating the F matrix going from $AB$ to $AC$.}
\label{keep}
\end{figure}

 The monodromy is calculated by assigning a $N\times N$ matrix to each edge in monodromy graph.
The E type is a diagonal matrix depending on the coordinates on the edge of the triangulation with which edge E intersects:
\begin{equation}
E(\bold{a})[N]=\text{diag}(a_0 a_1 a_2...a_{N-1}, a_1 a_2...a_{N-1},\ldots,a_{N-1},1),
\end{equation}
where the starting point for the label is taken as the vertex where edge $E$ circles around, see figure. \ref{keep}. Notice that 
the matrix for other edge parallel with $E$ is denoted as  $E(\bar{a})$ since we have different labeling for the coordinates.
The coordinates $\bar{\bold{a}}$ are related to the $a$ coordinates as 
\begin{equation}
\bar{\bold{a}}_0=a_{N-1},\ldots, \bar{\bold{a}}_j=a_{N-j-1},\ldots ,\bar{\bold{a}}_{N-1}=a_0.    
\end{equation}
The $S$ Matrix also has a simple form:
\begin{equation}
S_{i,j}[N]=\begin{cases}
(-1)^{i-1}~~j=N+1-i \\
0
\end{cases}
\end{equation}
The F(ace) matrix is more complicated and it is composed by $F_i$ and $H_i$ matrix. The $H_i$ matrix is a diagonal one which 
has the form depending on position of the internal quiver node:
\begin{equation}
H_i(x)=\text{diag}(\underbrace{x,x,x}_i,1,\ldots,1),
\end{equation}
and $F_i$ matrix is defined as 
\begin{equation}
(F_i)_{m,n}=\begin{cases}
1~\text{if}~ m=n,~ \text{or}~ m=i~ n=i-1 \\
0
\end{cases}
\end{equation}
So $F_i$ matrix is a  lower triangular matrix with entry $one$ on the diagonal. Using the above two building blocks, the 
F(ace) matrix  is expressed as
\begin{equation}
F[x,N]=\prod_{j=N-1}^1(\prod_{i=j-1}^{N-2} (H_{i+1}(X_{N-i-2,i-j,j-1})F_i)F_{N-1}).
\end{equation}
Notice that here an internal quiver node is labeled by three integers satisfying $a+b+c=N-3$, which is given by subtracting one from our
previous labeling.  For small $N$, the matrices are 
\begin{align}
&F[x,2]=F_1,~~~F[x,3]=F_2H_2(X_{0,0,0})F_1F_2, \nonumber \\
&F[x,4]=F_3H_3(X_{0,1,0})F_2F_3H_2(X_{1,0,0})F_1H_3(X_{0,0,1})F_2F_3.
\end{align}
There is an easy way of remembering this sequence by looking at the quiver diagram in figure. \ref{appen}. If the monodromy path is 
going from edge $AB$ to $AC$, then the above product is decomposed into $N-1$ steps. In each step, each quiver node 
on the line parallel to edge $AC$ contribute a factor $H_{i+1}F_{i}$, where $i$ is the distance to vertex $A$ of that quiver node. In each step,
the sequence of product is to start from top to bottom, see figure. \ref{keep} for the arrows indicating the order of product; finally one 
need to add a matrix $F_{N-1}$ at the end in each step. 

One should be really careful about the fact that the labeling of quiver nodes are quite different if the path is going through the other vertex
of the triangle, so it is better to use  different letters for the face matrix inside the same triangle, say $X, \bar{X}, \bar{\bar{X}}$, and 
finally one should relate three set of coordinates by looking at their positions, say $X_{a,b,c}=\bar{X}_{b,a,c}$, etc.

The above matrices are the basic ingredients for calculating the monodromy, and if a path goes in opposite direction as the chosen orientation, we 
just use the inverse of above matrices. The monodromy around the square and the hexagon is an identity matrix up to a constant scalar, therefore the monodromy
only depends on the homotopy class up to a complex  scalar, and it defines a $\text{PGL}(N,C)$ local system. The monodromy for the $\text{SL}(N,C)$
local system is defined as
\begin{equation}
I(l)=|\text{Det}(M)^{-1/N}\text{Tr}(M)|.
\end{equation}
here we take the absolute value to make sure that all the expressions are positive.

\section{Collection of canonical maps}
For  $A_2$ theory defined 
on once punctured torus, there are eight nontrivial generators. 
 The quivers and the geometric representations for these eight line operators are 
shown in figure. \ref{A2} and \ref{junction}. From the quiver, one can read the Poisson brackets for  cluster coordinates, and one then use
them to calculate the Poisson structure between the line operators combining the explicit expression of canonical map.
We collect canonical maps for these eight generators, and 
the interested reader can do all kinds of calculations presented in the paper.

\begin{align}
&\mbox{I}(\mbox{L}_1)= {1\over{x^{2/3} y^{1/3} a_0^{1/3} a_1^{2/3} b_0^{1/3} b_1^{2/3}}}+{{b_1^{1/3}}\over{x^{2/3} y^{1/3} a_0^{1/3} a_1^{2/3} b_0^{1/3}}}+\frac{x^{1/3} b_1^{1/3}}{y^{1/3} a_0^{1/3} a_1^{2/3} b_0^{1/3}} 
+\frac{x^{1/3} a_1^{1/3} b_1^{1/3}}{y^{1/3} a_0^{1/3} b_0^{1/3}}+\nonumber\\
&\frac{x^{1/3} b_0^{2/3} b_1^{1/3}}{y^{1/3} a_0^{1/3} a_1^{2/3}}+\frac{x^{1/3} a_1^{1/3} b_0^{2/3} b_1^{1/3}}{y^{1/3} a_0^{1/3}}
+\frac{x^{1/3} y^{2/3} a_1^{1/3} b_0^{2/3} b_1^{1/3}}{a_0^{1/3}}+x^{1/3} y^{2/3} a_0^{2/3} a_1^{1/3} b_0^{2/3} b_1^{1/3}
\end{align}

\begin{align}
&\mbox{I}(\mbox{L}_2)=\frac{1}{x^{1/3} y^{2/3} a_0^{2/3} a_1^{1/3} b_0^{2/3} b_1^{1/3}}+\frac{b_0^{1/3}}{x^{1/3} y^{2/3} a_0^{2/3} a_1^{1/3} b_1^{1/3}}+\frac{y^{1/3} b_0^{1/3}}{x^{1/3} a_0^{2/3} a_1^{1/3} b_1^{1/3}}+\frac{y^{1/3} a_0^{1/3} b_0^{1/3}}{x^{1/3} a_1^{1/3} b_1^{1/3}}+\nonumber\\
&\frac{y^{1/3} b_0^{1/3} b_1^{2/3}}{x^{1/3} a_0^{2/3} a_1^{1/3}}+\frac{y^{1/3} a_0^{1/3} b_0^{1/3} b_1^{2/3}}{x^{1/3} a_1^{1/3}}+\frac{x^{2/3} y^{1/3} a_0^{1/3} b_0^{1/3} b_1^{2/3}}{a_1^{1/3}}+x^{2/3} y^{1/3} a_0^{1/3} a_1^{2/3} b_0^{1/3} b_1^{2/3}
\end{align}
\begin{align}
&\mbox{I}(\mbox{M}_1)=\frac{1}{x^{1/3} y^{2/3} b_0^{1/3} b_1^{2/3} c_0^{1/3} c_1^{2/3}}+\frac{c_1^{1/3}}{x^{1/3} y^{2/3} b_0^{1/3} b_1^{2/3} c_0^{1/3}}+\frac{y^{1/3} c_1^{1/3}}{x^{1/3} b_0^{1/3} b_1^{2/3} c_0^{1/3}}+\frac{y^{1/3} b_1^{1/3} c_1^{1/3}}{x^{1/3} b_0^{1/3} c_0^{1/3}}\nonumber\\
&+\frac{y^{1/3} c_0^{2/3} c_1^{1/3}}{x^{1/3} b_0^{1/3} b_1^{2/3}}+\frac{y^{1/3} b_1^{1/3} c_0^{2/3} c_1^{1/3}}{x^{1/3} b_0^{1/3}}+\frac{x^{2/3} y^{1/3} b_1^{1/3} c_0^{2/3} c_1^{1/3}}{b_0^{1/3}}+x^{2/3} y^{1/3} b_0^{2/3} b_1^{1/3} c_0^{2/3} c_1^{1/3}
\end{align}
\begin{align}
&\mbox{I}(\mbox{M}_2)=\frac{1}{x^{2/3} y^{1/3} b_0^{2/3} b_1^{1/3} c_0^{2/3} c_1^{1/3}}+\frac{c_0^{1/3}}{x^{2/3} y^{1/3} b_0^{2/3} b_1^{1/3} c_1^{1/3}}+\frac{x^{1/3} c_0^{1/3}}{y^{1/3} b_0^{2/3} b_1^{1/3} c_1^{1/3}}+\frac{x^{1/3} b_0^{1/3} c_0^{1/3}}{y^{1/3} b_1^{1/3} c_1^{1/3}}\nonumber\\
&+\frac{x^{1/3} c_0^{1/3} c_1^{2/3}}{y^{1/3} b_0^{2/3} b_1^{1/3}}+\frac{x^{1/3} b_0^{1/3} c_0^{1/3} c_1^{2/3}}{y^{1/3} b_1^{1/3}}+\frac{x^{1/3} y^{2/3} b_0^{1/3} c_0^{1/3} c_1^{2/3}}{b_1^{1/3}}+x^{1/3} y^{2/3} b_0^{1/3} b_1^{2/3} c_0^{1/3} c_1^{2/3}
\end{align}
\begin{align}
&\mbox{I}(\mbox{C}_1)=\frac{1}{x^{1/3} y^{2/3} a_0^{1/3} a_1^{2/3} c_0^{2/3} c_1^{1/3}}+\frac{a_1^{1/3}}{x^{1/3} y^{2/3} a_0^{1/3} c_0^{2/3} c_1^{1/3}}+\frac{y^{1/3} a_1^{1/3}}{x^{1/3} a_0^{1/3} c_0^{2/3} c_1^{1/3}}+\frac{y^{1/3} a_0^{2/3} a_1^{1/3}}{x^{1/3} c_0^{2/3} c_1^{1/3}}\nonumber\\
&+\frac{y^{1/3} a_1^{1/3} c_0^{1/3}}{x^{1/3} a_0^{1/3} c_1^{1/3}}+\frac{y^{1/3} a_0^{2/3} a_1^{1/3} c_0^{1/3}}{x^{1/3} c_1^{1/3}}+\frac{x^{2/3} y^{1/3} a_0^{2/3} a_1^{1/3} c_0^{1/3}}{c_1^{1/3}}+x^{2/3} y^{1/3} a_0^{2/3} a_1^{1/3} c_0^{1/3} c_1^{2/3}
\end{align}
\begin{align}
&\mbox{I}(\mbox{C}_2)=\frac{1}{x^{2/3} y^{1/3} a_0^{2/3} a_1^{1/3} c_0^{1/3} c_1^{2/3}}+\frac{a_0^{1/3}}{x^{2/3} y^{1/3} a_1^{1/3} c_0^{1/3} c_1^{2/3}}+\frac{x^{1/3} a_0^{1/3}}{y^{1/3} a_1^{1/3} c_0^{1/3} c_1^{2/3}}+\frac{x^{1/3} a_0^{1/3} a_1^{2/3}}{y^{1/3} c_0^{1/3} c_1^{2/3}}\nonumber\\
&+\frac{x^{1/3} a_0^{1/3} c_1^{1/3}}{y^{1/3} a_1^{1/3} c_0^{1/3}}+\frac{x^{1/3} a_0^{1/3} a_1^{2/3} c_1^{1/3}}{y^{1/3} c_0^{1/3}}+\frac{x^{1/3} y^{2/3} a_0^{1/3} a_1^{2/3} c_1^{1/3}}{c_0^{1/3}}+x^{1/3} y^{2/3} a_0^{1/3} a_1^{2/3} c_0^{2/3} c_1^{1/3}
\end{align}
\begin{align}
&\mbox{I}(I_1)=\frac{a_1^{1/3} b_0^{1/3}}{a_0^{1/3} b_1^{1/3} c_0^{1/3} c_1^{2/3}}+\frac{a_0^{2/3} a_1^{1/3} b_0^{1/3}}{b_1^{1/3} c_0^{1/3} c_1^{2/3}}+\frac{a_1^{1/3} c_1^{1/3}}{a_0^{1/3} b_0^{2/3} b_1^{1/3} c_0^{1/3}}+\frac{b_0^{1/3} c_1^{1/3}}{a_0^{1/3} a_1^{2/3} b_1^{1/3} c_0^{1/3}} \nonumber\\
&+\frac{2 a_1^{1/3} b_0^{1/3} c_1^{1/3}}{a_0^{1/3} b_1^{1/3} c_0^{1/3}}+\frac{a_0^{2/3} a_1^{1/3} b_0^{1/3} c_1^{1/3}}{b_1^{1/3} c_0^{1/3}}+\frac{b_0^{1/3} b_1^{2/3} c_1^{1/3}}{a_0^{1/3} a_1^{2/3} c_0^{1/3}}+\frac{a_1^{1/3} b_0^{1/3} b_1^{2/3} c_1^{1/3}}{a_0^{1/3} c_0^{1/3}} \nonumber\\
&+\frac{a_1^{1/3} c_0^{2/3} c_1^{1/3}}{a_0^{1/3} b_0^{2/3} b_1^{1/3}}+\frac{a_1^{1/3} b_0^{1/3} c_0^{2/3} c_1^{1/3}}{a_0^{1/3} b_1^{1/3}}+\frac{1}{a_0^{1/3} a_1^{2/3} b_0^{2/3} b_1^{1/3} c_0^{1/3} c_1^{2/3} y}+\frac{a_1^{1/3}}{a_0^{1/3} b_0^{2/3} b_1^{1/3} c_0^{1/3} c_1^{2/3} y} \nonumber\\
&+\frac{b_0^{1/3}}{a_0^{1/3} a_1^{2/3} b_1^{1/3} c_0^{1/3} c_1^{2/3} y}+\frac{a_1^{1/3} b_0^{1/3}}{a_0^{1/3} b_1^{1/3} c_0^{1/3} c_1^{2/3} y}+\frac{c_1^{1/3}}{a_0^{1/3} a_1^{2/3} b_0^{2/3} b_1^{1/3} c_0^{1/3} y}+\frac{a_1^{1/3} c_1^{1/3}}{a_0^{1/3} b_0^{2/3} b_1^{1/3} c_0^{1/3} y} \nonumber\\
&+\frac{b_0^{1/3} c_1^{1/3}}{a_0^{1/3} a_1^{2/3} b_1^{1/3} c_0^{1/3} y}+\frac{a_1^{1/3} b_0^{1/3} c_1^{1/3}}{a_0^{1/3} b_1^{1/3} c_0^{1/3} y}+\frac{1}{a_0^{1/3} a_1^{2/3} b_0^{2/3} b_1^{1/3} c_0^{1/3} c_1^{2/3} x y}+\frac{a_1^{1/3} b_0^{1/3} c_1^{1/3} y}{a_0^{1/3} b_1^{1/3} c_0^{1/3}} \nonumber\\
&+\frac{a_0^{2/3} a_1^{1/3} b_0^{1/3} c_1^{1/3} y}{b_1^{1/3} c_0^{1/3}}+\frac{a_1^{1/3} b_0^{1/3} b_1^{2/3} c_1^{1/3} y}{a_0^{1/3} c_0^{1/3}}+\frac{a_0^{2/3} a_1^{1/3} b_0^{1/3} b_1^{2/3} c_1^{1/3} y}{c_0^{1/3}}+\frac{a_1^{1/3} b_0^{1/3} c_0^{2/3} c_1^{1/3} y}{a_0^{1/3} b_1^{1/3}} \nonumber\\
&+\frac{a_0^{2/3} a_1^{1/3} b_0^{1/3} c_0^{2/3} c_1^{1/3} y}{b_1^{1/3}}+\frac{a_1^{1/3} b_0^{1/3} b_1^{2/3} c_0^{2/3} c_1^{1/3} y}{a_0^{1/3}}+a_0^{2/3} a_1^{1/3} b_0^{1/3} b_1^{2/3} c_0^{2/3} c_1^{1/3} y+a_0^{2/3} a_1^{1/3} b_0^{1/3} b_1^{2/3} c_0^{2/3} c_1^{1/3} x y
\end{align}
\begin{align}
&\mbox{I}(I_2)=\nonumber \\
&\frac{1}{x a_0^{2/3} a_1^{1/3} b_0^{1/3} b_1^{2/3} c_0^{2/3} c_1^{1/3}}+\frac{1}{x y a_0^{2/3} a_1^{1/3} b_0^{1/3} b_1^{2/3} c_0^{2/3} c_1^{1/3}}+\frac{a_0^{1/3}}{x a_1^{1/3} b_0^{1/3} b_1^{2/3} c_0^{2/3} c_1^{1/3}}+\frac{b_1^{1/3}}{x a_0^{2/3} a_1^{1/3} b_0^{1/3} c_0^{2/3} c_1^{1/3}}\nonumber\\
&+\frac{a_0^{1/3} b_1^{1/3}}{a_1^{1/3} b_0^{1/3} c_0^{2/3} c_1^{1/3}}+\frac{a_0^{1/3} b_1^{1/3}}{x a_1^{1/3} b_0^{1/3} c_0^{2/3} c_1^{1/3}}+\frac{a_0^{1/3} a_1^{2/3} b_1^{1/3}}{b_0^{1/3} c_0^{2/3} c_1^{1/3}}+\frac{c_0^{1/3}}{x a_0^{2/3} a_1^{1/3} b_0^{1/3} b_1^{2/3} c_1^{1/3}}\nonumber\\
&+\frac{a_0^{1/3} c_0^{1/3}}{a_1^{1/3} b_0^{1/3} b_1^{2/3} c_1^{1/3}}+\frac{a_0^{1/3} c_0^{1/3}}{x a_1^{1/3} b_0^{1/3} b_1^{2/3} c_1^{1/3}}+\frac{b_1^{1/3} c_0^{1/3}}{a_0^{2/3} a_1^{1/3} b_0^{1/3} c_1^{1/3}}+\frac{b_1^{1/3} c_0^{1/3}}{x a_0^{2/3} a_1^{1/3} b_0^{1/3} c_1^{1/3}}\nonumber\\
&+\frac{2 a_0^{1/3} b_1^{1/3} c_0^{1/3}}{a_1^{1/3} b_0^{1/3} c_1^{1/3}}+\frac{a_0^{1/3} b_1^{1/3} c_0^{1/3}}{x a_1^{1/3} b_0^{1/3} c_1^{1/3}}+\frac{x a_0^{1/3} b_1^{1/3} c_0^{1/3}}{a_1^{1/3} b_0^{1/3} c_1^{1/3}}+\frac{a_0^{1/3} a_1^{2/3} b_1^{1/3} c_0^{1/3}}{b_0^{1/3} c_1^{1/3}}\nonumber\\
&+\frac{x a_0^{1/3} a_1^{2/3} b_1^{1/3} c_0^{1/3}}{b_0^{1/3} c_1^{1/3}}+\frac{b_0^{2/3} b_1^{1/3} c_0^{1/3}}{a_0^{2/3} a_1^{1/3} c_1^{1/3}}+\frac{a_0^{1/3} b_0^{2/3} b_1^{1/3} c_0^{1/3}}{a_1^{1/3} c_1^{1/3}}+\frac{x a_0^{1/3} b_0^{2/3} b_1^{1/3} c_0^{1/3}}{a_1^{1/3} c_1^{1/3}}\nonumber\\
&+\frac{x a_0^{1/3} a_1^{2/3} b_0^{2/3} b_1^{1/3} c_0^{1/3}}{c_1^{1/3}}+\frac{a_0^{1/3} c_0^{1/3} c_1^{2/3}}{a_1^{1/3} b_0^{1/3} b_1^{2/3}}+\frac{a_0^{1/3} b_1^{1/3} c_0^{1/3} c_1^{2/3}}{a_1^{1/3} b_0^{1/3}}+\frac{x a_0^{1/3} b_1^{1/3} c_0^{1/3} c_1^{2/3}}{a_1^{1/3} b_0^{1/3}}\nonumber\\
&+\frac{x a_0^{1/3} a_1^{2/3} b_1^{1/3} c_0^{1/3} c_1^{2/3}}{b_0^{1/3}}+\frac{x a_0^{1/3} b_0^{2/3} b_1^{1/3} c_0^{1/3} c_1^{2/3}}{a_1^{1/3}}+x a_0^{1/3} a_1^{2/3} b_0^{2/3} b_1^{1/3} c_0^{1/3} c_1^{2/3}+x y a_0^{1/3} a_1^{2/3} b_0^{2/3} b_1^{1/3} c_0^{1/3} c_1^{2/3}
\end{align}
\begin{align}
&\mbox{I}(2\omega_1)  \nonumber\\
&=\frac{1}{x^{4/3} y^{2/3} a_0^{2/3} a_1^{4/3} b_0^{2/3} b_1^{4/3}}+\frac{2}{x^{4/3} y^{2/3} a_0^{2/3} a_1^{4/3} b_0^{2/3} b_1^{1/3}}+\frac{2}{x^{1/3} y^{2/3} a_0^{2/3} a_1^{4/3} b_0^{2/3} b_1^{1/3}}+\frac{2 b_0^{1/3}}{x^{1/3} y^{2/3} a_0^{2/3} a_1^{4/3} b_1^{1/3}} \nonumber\\
&+\frac{b_1^{2/3}}{x^{4/3} y^{2/3} a_0^{2/3} a_1^{4/3} b_0^{2/3}}
+\frac{2 b_1^{2/3}}{x^{1/3} y^{2/3} a_0^{2/3} a_1^{4/3} b_0^{2/3}}+\frac{x^{2/3} b_1^{2/3}}{y^{2/3} a_0^{2/3} a_1^{4/3} b_0^{2/3}}+\frac{2 b_1^{2/3}}{x^{1/3} y^{2/3} a_0^{2/3} a_1^{1/3} b_0^{2/3}} \nonumber\\
&+\frac{2 x^{2/3} b_1^{2/3}}{y^{2/3} a_0^{2/3} a_1^{1/3} b_0^{2/3}}+\frac{x^{2/3} a_1^{2/3} b_1^{2/3}}{y^{2/3} a_0^{2/3} b_0^{2/3}}+\frac{2 b_0^{1/3} b_1^{2/3}}{x^{1/3} y^{2/3} a_0^{2/3} a_1^{4/3}} 
+\frac{2 x^{2/3} b_0^{1/3} b_1^{2/3}}{y^{2/3} a_0^{2/3} a_1^{4/3}}
+\frac{2 b_0^{1/3} b_1^{2/3}}{x^{1/3} y^{2/3} a_0^{2/3} a_1^{1/3}} \nonumber\\
&+\frac{4 x^{2/3} b_0^{1/3} b_1^{2/3}}{y^{2/3} a_0^{2/3} a_1^{1/3}}+\frac{2 x^{2/3} y^{1/3} b_0^{1/3} b_1^{2/3}}{a_0^{2/3} a_1^{1/3}}
\frac{2 x^{2/3} a_1^{2/3} b_0^{1/3} b_1^{2/3}}{y^{2/3} a_0^{2/3}} +\frac{2 x^{2/3} y^{1/3} a_1^{2/3} b_0^{1/3} b_1^{2/3}}{a_0^{2/3}}+ \nonumber \\
&\frac{x^{2/3} b_0^{4/3} b_1^{2/3}}{y^{2/3} a_0^{2/3} a_1^{4/3}}+\frac{2 x^{2/3} b_0^{4/3} b_1^{2/3}}{y^{2/3} a_0^{2/3} a_1^{1/3}}+\frac{2 x^{2/3} y^{1/3} b_0^{4/3} b_1^{2/3}}{a_0^{2/3} a_1^{1/3}}+\frac{2 x^{2/3} y^{1/3} a_0^{1/3} b_0^{4/3} b_1^{2/3}}{a_1^{1/3}} \nonumber\\
&+\frac{x^{2/3} a_1^{2/3} b_0^{4/3} b_1^{2/3}}{y^{2/3} a_0^{2/3}}
+\frac{2 x^{2/3} y^{1/3} a_1^{2/3} b_0^{4/3} b_1^{2/3}}{a_0^{2/3}}
+\frac{x^{2/3} y^{4/3} a_1^{2/3} b_0^{4/3} b_1^{2/3}}{a_0^{2/3}} \nonumber\\
&+2 x^{2/3} y^{1/3} a_0^{1/3} a_1^{2/3} b_0^{4/3} b_1^{2/3}
+2 x^{2/3} y^{4/3} a_0^{1/3} a_1^{2/3} b_0^{4/3} b_1^{2/3} +x^{2/3} y^{4/3} a_0^{4/3} a_1^{2/3} b_0^{4/3} b_1^{2/3}
\end{align}

\bibliographystyle{utphys} 
 \bibliography{PLforRS}    

\end{document}